\documentclass[journal,compsoc]{IEEEtran}
\pdfoutput=1

\usepackage{bm}
\usepackage{bbm}
\usepackage[pdftex]{graphics}
\usepackage{enumerate}
\usepackage{caption}
\usepackage{algorithm}
\usepackage{algorithmic}
\usepackage{color}
\usepackage{subfigure}
\usepackage[pdftex]{graphicx}
\usepackage{epstopdf}
\usepackage{stfloats}
\usepackage{float}
\usepackage{amsthm}
\usepackage{amsmath}
\usepackage{booktabs}
\usepackage{multirow}

\usepackage[square, comma, sort&compress, numbers]{natbib}

\usepackage{color}
\usepackage{pifont}
\newtheorem{defi}{Definition}
\newtheorem{thm}{Theorem}

\begin{document}

\title{A Differentially Private Framework for Spatial Crowdsourcing with Historical Data Learning}

\author{Shun~Zhang,
	Benfei~Duan,
	Zhili~Chen$^\ast$,
Hong~Zhong,
	and~Qizhi~Yu
\IEEEcompsocitemizethanks{\IEEEcompsocthanksitem S. Zhang, B. Duan, Z. Chen and H. Zhong are with School of Computer Science and Technology, Anhui University, Hefei 230601, China\protect\\
E-mail: szhang@ahu.edu.cn (S. Zhang), dbf97@stu.ahu.edu.cn (B. Duan), zlchen@ahu.edu.cn (Z. Chen), zhongh@ahu.edu.cn (H. Zhong)
\IEEEcompsocthanksitem Q. Yu is with Zhejiang Lab, Hangzhou 311121, China\protect\\
E-mail: yuqz@zhejianglab.com
\IEEEcompsocthanksitem $^\ast$ Corresponding author (Zhili Chen)}
\thanks{Manuscript received Month xx, 2020; revised Month xx, 2020.}}

% The paper headers
\markboth{Journal of \LaTeX\ Class Files,~Vol.~xx, No.~x, Month~20xx}%
{Shell \MakeLowercase{\textit{et al.}}: Bare Demo of IEEEtran.cls for Computer Society Journals}
% The only time the second header will appear is for the odd numbered pages
% after the title page when using the twoside option.
%
% *** Note that you probably will NOT want to include the author's ***
% *** name in the headers of peer review papers.                   ***
% You can use \ifCLASSOPTIONpeerreview for conditional compilation here if
% you desire.

% The publisher's ID mark at the bottom of the page is less important with
% Computer Society journal papers as those publications place the marks
% outside of the main text columns and, therefore, unlike regular IEEE
% journals, the available text space is not reduced by their presence.
% If you want to put a publisher's ID mark on the page you can do it like
% this:
%\IEEEpubid{0000--0000/00\$00.00~\copyright~2015 IEEE}
% or like this to get the Computer Society new two part style.
%\IEEEpubid{\makebox[\columnwidth]{\hfill 0000--0000/00/\$00.00~\copyright~2015 IEEE}%
%\hspace{\columnsep}\makebox[\columnwidth]{Published by the IEEE Computer Society\hfill}}
% Remember, if you use this you must call \IEEEpubidadjcol in the second
% column for its text to clear the IEEEpubid mark (Computer Society jorunal
% papers don't need this extra clearance.)

% use for special paper notices
%\IEEEspecialpapernotice{(Invited Paper)}

% for Computer Society papers, we must declare the abstract and index terms
% PRIOR to the title within the \IEEEtitleabstractindextext IEEEtran
% command as these need to go into the title area created by \maketitle.
% As a general rule, do not put math, special symbols or citations
% in the abstract or keywords.
\IEEEtitleabstractindextext{
\begin{abstract}
Spatial crowdsourcing (SC) is an increasing popular category of crowdsourcing in the era of mobile Internet and sharing economy. It requires workers to arrive at a particular location for task fulfillment. Effective protection of location privacy is essential for workers' enthusiasm and valid task assignment. However, existing SC models with differential privacy usually perturb real-time location data for both partition and data publication. Such a way may produce large perturbations to counting queries that affect assignment success rate and allocation accuracy. This paper proposes a framework (R-HT) for protecting location privacy of workers taking advantage of both real-time and historical data. We simulate locations by sampling the probability distribution learned from historical data, use them for grid partition, and then publish real-time data under this partitioning with differential privacy. This realizes that most privacy budget is allocated to the worker count of each cell and yields an improved Private Spatial Decomposition approach. Moreover, we introduce some strategies for \emph{geocast region} construction, including quality scoring function and local maximum geocast radius. A series of experimental results on real-world datasets shows that R-HT attains a stable success rate of task assignment, saves performance overhead and fits for dynamic assignment on crowdsourcing platforms.
\end{abstract}

% Note that keywords are not normally used for peerreview papers.
}

% make the title area
\maketitle

% To allow for easy dual compilation without having to reenter the
% abstract/keywords data, the \IEEEtitleabstractindextext text will
% not be used in maketitle, but will appear (i.e., to be "transported")
% here as \IEEEdisplaynontitleabstractindextext when the compsoc
% or transmag modes are not selected <OR> if conference mode is selected
% - because all conference papers position the abstract like regular
% papers do.
\IEEEdisplaynontitleabstractindextext
% \IEEEdisplaynontitleabstractindextext has no effect when using
% compsoc or transmag under a non-conference mode.

% For peer review papers, you can put extra information on the cover
% page as needed:
% \ifCLASSOPTIONpeerreview
% \begin{center} \bfseries EDICS Category: 3-BBND \end{center}
% \fi
%
% For peerreview papers, this IEEEtran command inserts a page break and
% creates the second title. It will be ignored for other modes.
\IEEEpeerreviewmaketitle

\section{Introduction}
\IEEEPARstart{S}{patial} crowdsourcing (SC) is a new platform that harnesses the potential of the crowd to perform real-world tasks including collecting and analyzing environmental, social, and other spatiotemporal information. SC has been applied in various domains such as smart cities, environmental sensing, and journalism. However, disclosing individual locations has
serious privacy implications. Many mobile users do not agree to engage in SC if their privacy is violated. Thus, ensuring location privacy is an important aspect of SC.

In the last decade, several papers have been published on location privacy in SC, see the surveys \cite{TS18,TZZ20} and papers cited there. Usually workers send
their locations to a trusted \emph{Cellular Service Provider (CSP)} which collects
updates and releases a \emph{Private Spatial Decomposition (PSD)}.
Usually a PSD partitions the domain into
smaller cells and reports statistics on the locations within
each cell.
 To determine a PSD, some methods have been adopted such as
 kd-tree based partitioning  \cite{XXY10}, \emph{uniform grid (UG)} method \cite{QYL13} and \emph{adaptive grids (AG)} approach  \cite{QYL13,TGS14}. A framework was proposed by To et al. \cite{TGF17} for protecting location privacy of workers involved in SC. They achieved privacy protection by building PSDs based on an extended AG approach, which creates sanitized data releases using noisy real-time data at the CSP. However, since the privacy budget is sequentially divided into three parts, the large scale of noise greatly affects framework performance. In particular, the success rate of task assignment failed, at most cases, to reach the \emph{expected utility (EU)} as is one of the specific challenges identified therein.

Similar to \cite{TGF17}, we focus on protecting privacy of worker locations in SC.
We utilize historical data for learning the probability distribution of real-time locations, and build a differentially private SC framework that protects location privacy of workers using a minor fraction of privacy budget for grid partitioning.
Historical data learning by linear regression predicts the probability distribution of real-time locations, and simulated locations randomly sampled from this distribution are used for level-2 partition.
Afterwards, the real-time data are imported into grids where noise is added directly to the count of workers in each level-2 cell. Since the simulated points are independent of the real-time locations, this results in that most privacy budget can be allocated to the publication of real-time counts, and essentially improves the performance of privacy framework.

The main contributions are as follows:
\begin{enumerate}[(i)]
\item To avoid excessive noise additions in SC with differential privacy, we propose R-HT scheme allocating most of the privacy budget to location counting publication and the remaining minor budget to grid partition. To our knowledge, we are the first to propose the strategy that employs historical data learning in building PSD, which opens a new connection between differentially private location protection framework and machine learning methods. For constructing the continuous region $GR$, we first build PSD by random sampling simulation with historical data learning. Such sampling points by simulation for partition is completely independent of real-time data. In view of the privacy framework, our scheme improves the efficiency of system operations and ensures stably high success rate of task assignment.

\item We investigate some techniques across local cell selections to fuse those cells with negative noisy counts. Our selection of each newly added cell depends on a refined \emph{quality scoring function} involving the area of cells, instead of its utility.

\item We carry out extensive experiments on three real-world datasets which demonstrate that our R-HT scheme achieves stable success rate of task assignment and shows better performance in most aspects.

\end{enumerate}

The remainder of this paper is structured as follows. In Section \ref{sec:rela_work}, we conduct a survey of related work. Section \ref{sec:backg} introduces some necessary backgrounds. Section \ref{sec:framework} describes the proposed privacy framework. Section \ref{sec:skill} discusses some new techniques used in our framework. Experimental results are presented in Section \ref{sec:evaluation}. Finally, we conclude this
paper in Section \ref{sec:conclusion}.

\section{Related work}\label{sec:rela_work}
\textbf{Location Privacy Model.}
With the rapid development of smart mobile devices, more and more mobile applications provide (spatial) crowdsourcing services, which greatly facilitates our life. However, disclosing individual locations brings
many privacy implications. Leaked locations often lead to a breach of sensitive
information such as personal health, political and religious preferences.
Traditional methods of location privacy protection mainly include $k$-anonymity, expected distance error and cryptography. Wang et al. \cite{WLT17} used $k$-anonymity method to generate $k$-1 proper and dummy points and perform $k$ indistinguishable queries to the service provider, using the real location and dummy locations. However, only using anonymous method cannot offer good protection to a wide range of data and is vulnerable to background knowledge attack
 \cite{YXS18,ZLZ17}.
  Expected distance error reflects the accuracy degree to which the adversary can guess the real location by observing the obfuscated location and using available side-information \cite{STT12}. Explicitly, the obfuscation mechanisms are defined by a given prior, representing the adversary's side information \cite{ABC13}.
  Cryptography is suitable for multiple parties, completely protects data privacy and
prevents the leakage of data in the process of location service \cite{LCZ17}, however it normally results in
high computational costs and the availability of data will decrease significantly \cite{YXS18}.  \emph{Differential privacy (DP)} is a new and promising privacy model, which is completely independent of attacker's
background knowledge and currently a popular research topic in both academia and industry \cite{ZLZ17,ABC13}.

Differential privacy has been proven effective in sensitive data release. Particularly, many authors have attempted to bring differential privacy into location data protection for \emph{spatial crowdsourcing (SC)}.
Spatiotemporal information of workers, tasks, and intermediate results needs to be properly transformed to avoid privacy leakage while allowing efficient
information processing including task assignment. There are some recent works devoted to
balance between the strength of privacy protection and the efficiency
of other operations in various SC scenarios.
To et al. \cite{TGF17} divided the whole data domain into indexed units of grids (PSD) at the CSP. After receiving a task, the untrusted SC-server queries the PSD
to determine a \emph{geocast region (GR)} for task assignment. Any adversary cannot identify worker's location from the published $GR$. However, this scheme adds noises to all grids layer by layer, which reduces the efficiency of privacy budget and affects the performance of the framework. Xiong et al. \cite{XZZ17} proposed a new SC model based on reward, and adopts a reward based allocation strategy to ensure the task assignment success rate. Based on geo-indistinguishability, To et al. \cite{TSX18} presented a framework for protecting location privacy of both workers and tasks during the
tasking phase without relying on any trusted entity, in which techniques were devised to quantify the probability of reachability between a task and a worker. Wang et al. \cite{WPC19} proposed a novel distributed agent-based privacy-preserving framework that introduces a new level of multiple agent between users and the untrusted server and realizes the $w$-event $\epsilon$-DP for real-time crowd-sourced statistical data publishing with the untrusted server. Recently, Wei et al. \cite{WLY19} constructed two sets of PSDs to achieve task allocation with high data utility and simultaneously protect task and worker locations.

\textbf{Private Spatial Decomposition.}
To create sanitized data releases, PSD structures are often constructed in SC system. These partition the domain into
smaller regions and report noisy statistics on the locations within each region.
A suitable PSD can often improve the success rate of task assignment and reduce system overhead. Previously, PSDs are usually based on tree, especially kd-tree and quadtree  \cite{XXY10, LWZ18}, and the result is a deep tree. The typical simple method is \emph{Uniform Grid method (UG)}, which treats all dense and sparse regions
 equally in the domain \cite{QYL13}. Alternatively, \emph{Adaptive Grids approach (AG)} was proposed \cite{QYL13}. At the first level, AG creates a coarse-grained, equally spaced $m_1 \times m_1$ grid over the data domain. Then, each level-1 cell is partitioned into $m_2 \times m_2$ level-2 cells with $m_2$ chosen adaptively. This partition method emphasizes cell's difference in sparseness brought by UG, and it can be applied to various cases of data distribution. Later, The partition granularity was optimized with good universality in \cite{TGF17}, and such a granularity arrangement is utilized in our proposed scheme. Gong et al. \cite{GZF18} proposed a partitioning method (R-PSD) based on  reputation and location, where reputation is regarded as a new data dimension in building PSD by AG method and each R-PSD is composed of several sub-PSDs with different reputation levels.

In the traditional two-layer AG method, the privacy budget is divided into three parts for worker counting in the whole domain, level-1 cells, and level-2 cells, respectively. It is worth noting that partitioning does not depend on the noisy counting in level-2 cells. In this paper we deploy historical data learning to perform the second level partition, in which the real-time distribution of locations is simulated by random sampling (i.i.d.) with probability determined by the predicted proportion of location counts among level-1 cells.

%and achieve the separation of the original (historical and real-time) data, so that the privacy budget are assigned in parallel at PSD and $GR$ construction stages.

\textbf{Prediction by Historical Data Learning.}
When real-time data is unavailable or difficult to obtain, researchers often use historical data instead. Indeed, using historical data to make predictions by learning methods can often generate good results reflecting the real-time case. As for track of whereabouts of moving users, there are many examples of using historical locations or historical trajectories to predict real-time locations. Xu et al. \cite{XWJ17} proposed a real-time road traffic state prediction based on ARIMA model and Kalman filter, with using historical traffic data. Liu et al. \cite{LCZ18} predicted user's movement trajectory and position at the next moment by collecting locations and historical check-in information on social network.

 Real-time location prediction using historical data has also been extended to the field of SC. Jiang et al. \cite{JHC18} predicted worker positions at the next moment by analyzing their historical movement trajectory, and assigned tasks to those workers who were willing to go to or able to physically move to the position of the task on time with a high probability. On the aspect of location distribution on grids, To et al. \cite{TGF17} performed random perturbation to simulate subsequent distribution using historical positions, while only updating the counts in level-2 cells without changing AG structure. Chen et al. \cite{CKZ19} resamples the data at regular intervals to update the counts in the fixed grid structure. However, these above methods do not use historical data to update the partition structure, but renew only the data in fixed grids.
In contrast, we investigate linear regression method to perform domain partitioning with historical locations. Indeed, there are many advanced prediction methods, such as ARIMA model mentioned above, while multiple noise adding on statistical counting weakens the reliability of the system performance on prediction precision.

In this paper, we first propose to build a PSD with historical data learning based on AG method \cite{TGF17}. In particular, the level-2 partition is performed on sampling data (i.i.d.) generated by a probability distribution learned from historical data. This avoids privacy leakage caused by the strong correlation between historical and real-time data.
Then we can allocate privacy budget mainly to adding real-time counting noise in each level-2 cell. Moreover, we develop a quality scoring function derived from exponential mechanism instead of the utility function, for optimizing selections of neighboring cells, which involves the factors of cell's area and distance for $GR$ construction. This promotes significantly the performance in various aspects of the system.

\section{Background}\label{sec:backg}

In this section, we introduce some notations and initial definitions,
and review spatial crowdsourcing,
differential privacy, and linear regression method.

\subsection{Spatial Crowdsourcing}\label{sec:SC}

\emph{Spatial Crowdsourcing (SC)} is a new avenue of crowdsourcing related to
real-world scenarios involving physical locations, which requires workers to physically move to a particular location to perform tasks. The roles of components involved in SC are tasks, workers and the platform (mainly SC-server). The SC-server publishes or assigns the spatial task after receiving request and finally one or more workers accept and finish the task. Usually there are two categories of task assignment modes based
on how workers are matched to tasks in SC \cite{KS12}. In the \emph{Worker Selected Tasks (WST)} mode, SC-server is only responsible for the release of tasks, and workers autonomously select suitable tasks according to their own locations, without reporting their locations to the SC-server. In \emph{Server Assigned Tasks (SAT)} mode, SC-server collects worker locations and runs a complex optimization matching algorithm to assign the task to one or more workers, who decide whether to accept the task or not.

The quality of spatial tasks in SC is the main criterion to determine whether the tasks are assigned effectively \cite{TZZ20}. It is usually evaluated by reliability, which is formalized as the probability that over $50$ percent of workers correctly answer the task \cite{KSC13}, or the chance that as least one worker completes the task successfully \cite{ZYL16}. The former is generally used for spatial tasks that require qualified answers, such as spatial data collection related to pictures and videos. For such spatial tasks, the main challenge for SC is how to verify the validity of the results provided by untrusted workers. Since malicious workers may upload some incorrect information, the number of tasks correctly completed needs to be maximized \cite{KSC13}.
The latter is generally used in the case that a spatial task should be finished by a single worker, such as taxi calling service and fast food delivery. In such situations, the two factors, the worker-task distance and task expiration time, should be taken into
account, to ensure that at least one assigned worker can correctly finish the task \cite{ZYL16}.
Similarly, the framework proposed by To et al. \cite{TGF17} does not guarantee that the task are disseminated to enough workers, since the SC-server assigns tasks only by the sanitized PSD. Then the method of effectiveness evaluation is to compute the probability that among those assigned to the task at least one worker is willing to accept the task.

%Therefore, it is very important to develop an appropriate noise addition strategy and worker selection strategy to improve the success rate of task assignment, which will be the focus of this paper.

\subsection{Differential Privacy}\label{sec:DP}
\emph{Differential privacy (DP)} has emerged as the \emph{de facto} standard privacy notion for privacy-preservation research on data analysis and publishing. It makes that the probability of any output is equally likely from all nearly identical input datasets, so that it is unable to infer any sensitive information of an individual. Afterwards, any adversary cannot conclude with high confidence whether a particular individual is involved in the query result or not.

For applying DP, a crucial choice is the condition under which the datasets $D$ and $D^{\prime}$ are considered to be neighboring. The notion of \emph{Unbounded DP} is used in our framework, which means that two datasets $D$ and $D^{\prime}$ are neighboring if $D$ can be obtained from $D^{\prime}$ by adding or removing one element.

\begin{defi}[$\epsilon$-DP  \cite{Dwork06}]\label{def:DP}
Given any two neighboring datasets $D$ and $D^{\prime}$, for any set of outcomes $\Omega$, a randomized mechanism $M$ gives $\epsilon$-DP if the probability distribution of the mechanism output on $D$ and $D^{\prime}$ is bounded by:
\begin{equation}
  \frac{Pr(M(D) \in \Omega)}{Pr(M(D^{\prime}) \in \Omega )}\leq e^{\epsilon}.
\end{equation}
\end{defi}
The parameter $\epsilon$ is termed privacy budget that is specified by a data owner and represents the privacy level to be achieved. A lower budget means a higher privacy level.

\begin{defi}[Global Sensitivity  \cite{DR14}]\label{def:Sensitivity}
Let $D$ and $D^{\prime}$ denote any pair of neighboring datasets. The global sensitivity of a function $f$, denoted by $\Delta f$, is given as below,
\begin{equation}
\Delta f = \mathop {\rm max}\limits_{D,\;D^{\prime}} \left\| {f\left( D \right) - f\left( {D^{\prime}} \right)} \right\|,
\end{equation}
which represents the maximal change on the output of $f$ when deleting any record in $D$.
\end{defi}

\begin{defi}[Laplace Mechanism  \cite{Dwork06}]\label{def:Laplace}
Given a function $f:D\to {\mathbbm{R}}$, $\Delta f$ is the sensitivity of $f$. The mechanism is given by
\begin{equation}
M(D)=f(D)+Lap(\frac{{\Delta f}}{\epsilon }),
\end{equation}
where $Lap(\frac{{\Delta f}}{\epsilon })$ means randomly
distributed Laplace noise with scale $\frac{{\Delta f}}{\epsilon}$.
\end{defi}

The Laplace mechanism provides the $\epsilon$-DP \cite{LLS16}. The following important properties are related to composition of algorithms for preserving $\epsilon$-DP.

\begin{thm}[Sequential Composition  \cite{LLS16}]\label{thm:seq_comp}
$M_1, M_2, \ldots, M_m$ is a set of mechanisms, where $M_i$ provides $\epsilon_{i}$-DP. Let $M$ be a mechanism that executes $M_1(D), M_2(D), \ldots, M_m(D)$ using independent randomness for each $M_i$, and returns the vector of the outputs of these mechanisms. Then, $M$ satisfies $(\sum\limits_{i = 1}^m \epsilon_i)$-DP.
\end{thm}

\begin{thm}[Parallel Composition  \cite{LLS16}]\label{thm:par_comp}
Let $M_1,M_2,...,M_m$ be $m$ mechanisms that satisfy $\epsilon_1$-DP,\, $\epsilon_2$-DP,\, \ldots,\, $\epsilon_m$-DP, respectively. For a deterministic partitioning function $f$, let $D_1, D_2, \ldots, D_m$ be respectively the resulting partitions of excuting $f$ on dataset $D$. Then publishing the results of $M_1(D_1), M_2(D_2), \ldots, M_m(D_m)$ satisfies $(\rm max\{\epsilon_1, \epsilon_2, \ldots, \epsilon_m\})$-DP.
\end{thm}

\begin{thm}[Post-Processing  \cite{DR14}]\label{thm:pos-pro}
Given a randomized mechanism $M_1$ that satisfies $\epsilon$-DP, then for any (possibly randomized) algorithm  $M_2$ accessing only the output of $M_1$ and not its input, the composition $M_2(M_1)$ still satisfies $\epsilon$-DP.
\end{thm}

\subsection{Linear Regression and Sampling Simulation}
%Before constructing $GR$, we first need to determine the granularity of spatial partition by the noisy count of workers. In the previous schemes, two-layer AG method is usually used for data domain partition with noisy real-time numbers of workers' locations \cite{TGS14}.

In our setting, we add no noise to the counting of the historical locations in each level-1 cell, and use the linear regression method to predict the real-time count of workers in each cell. The sequence of predicted counts can be regarded as a proportion of sampling probability distribution by which locations of the total (noisy) real-time number are randomly (i.i.d.) generated so as to determine a simulated partition of the second level.

Linear regression is one of the basic learning methods in machine learning. The goal is to find a line, a plane, or even a higher-dimensional hyperplane that minimizes the error between the predicted and actual values. There are univariate and multiple linear regressions. Univariate linear regression means that only one factor is considered and the solution involves a linear equation. We use unitary (least squares) regression method to predict location counts.

\begin{figure}[htp]
	\centering
	\subfigure[Simulation, Ta.]{
		\includegraphics[height=1.2 in]{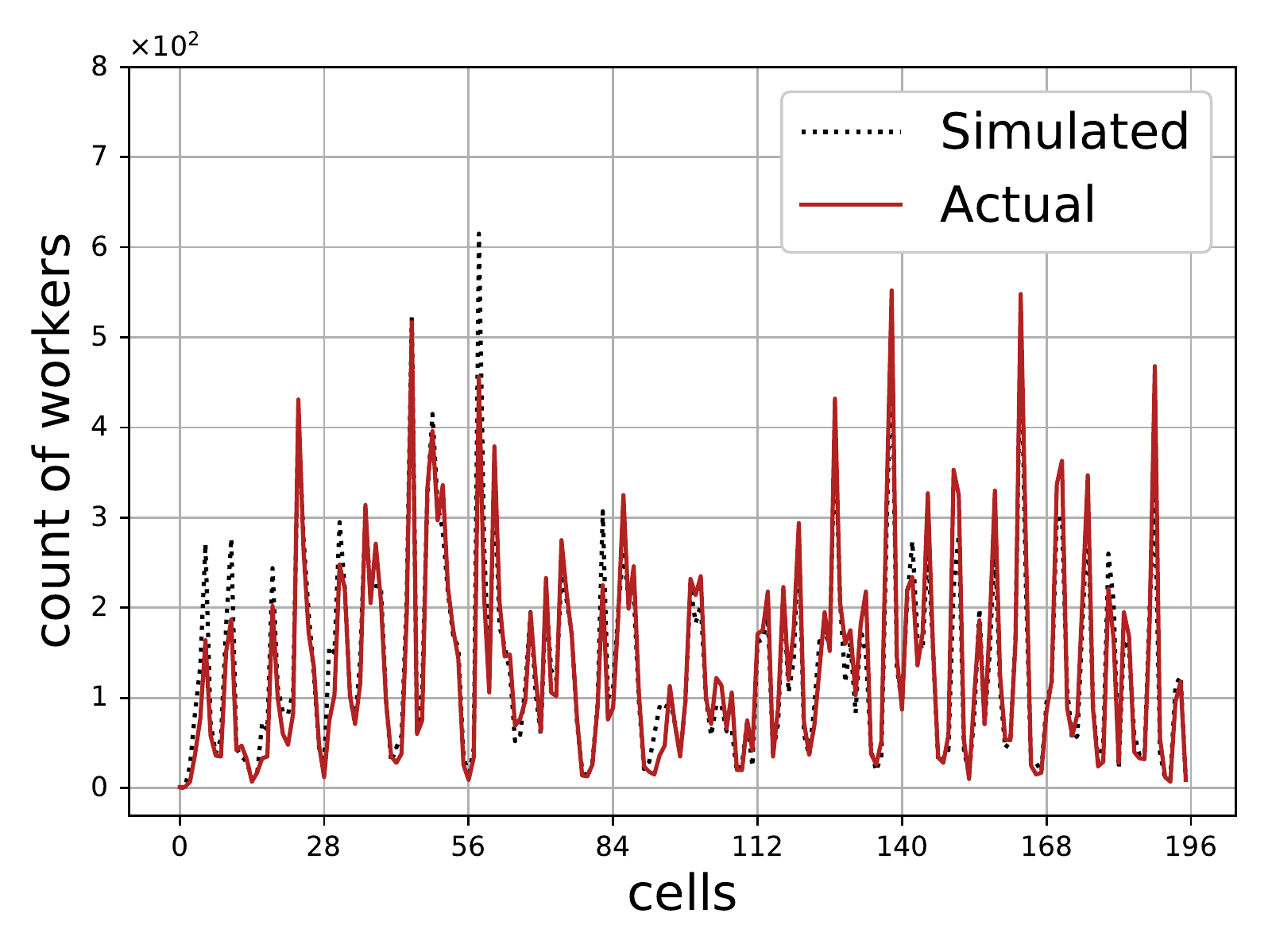}
	}
	\subfigure[Simulation, Go.]{
		\includegraphics[height=1.2 in]{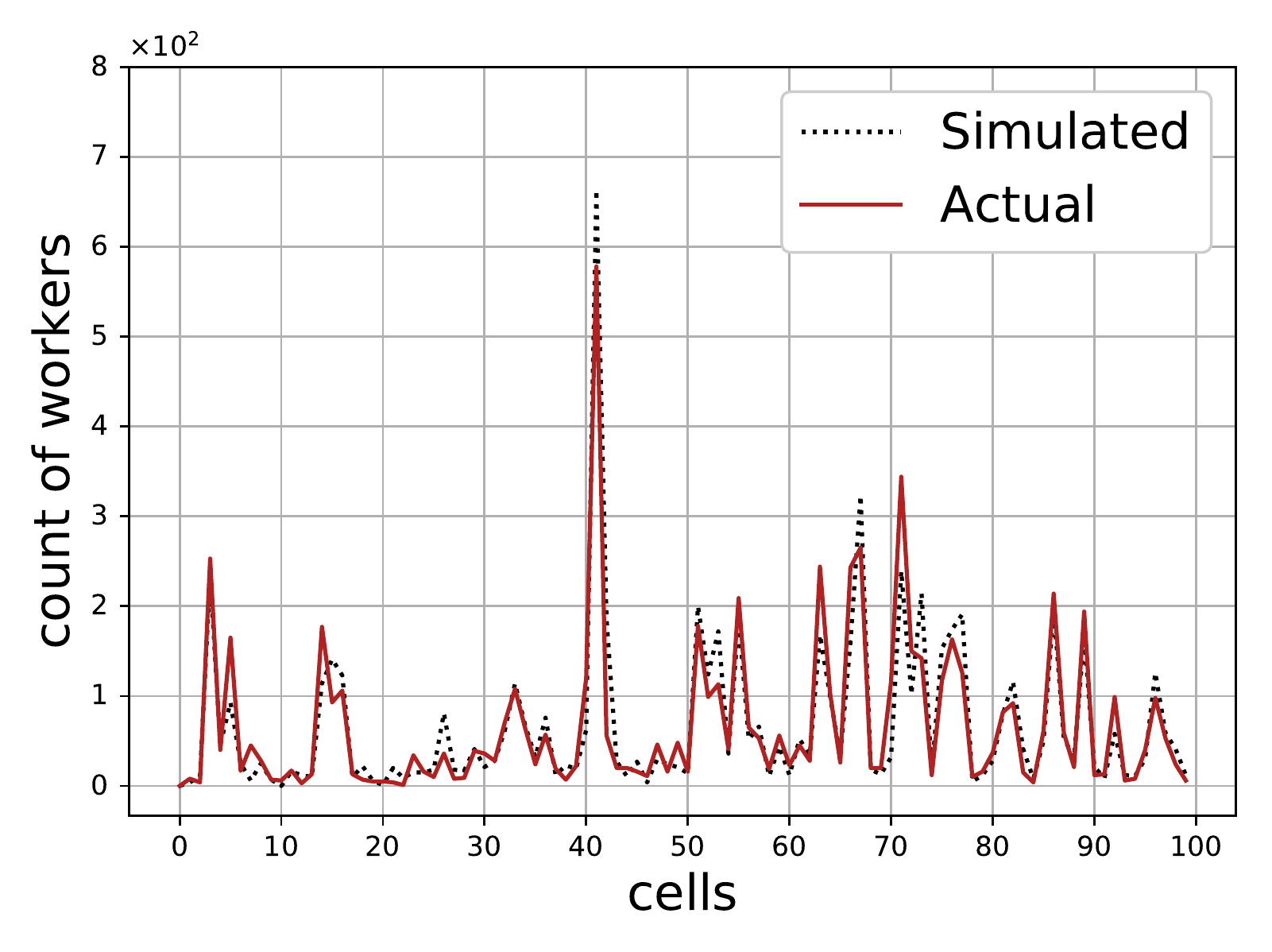}
	}
	\caption{Simulated and actual distributions of locations in the first-level cells}
    \label{fig:predict}
\end{figure}

We evaluate related schemes with three datasets, see Section \ref{sec:dataset} for details.  Valid historical locations of the previous 20 periods without noises in counting are used to make a prediction. This produces a probability distribution of locations by which we randomly (i.i.d.) sample a (noisy) total real-time number of points to simulate the real-time distribution. Fig. \ref{fig:predict} shows the comparison between the actual counts of level-1 cells and the simulated counts with noise (with sensitivity 1 and privacy budget 0.04) added to the total count in the real-time period on NYTaxi and Gowalla datasets, respectively. The average error rate $\gamma$ is calculated by,

\begin{equation}\label{eq:AER}
\gamma=\frac{\bar{E}}{\bar{A}},~\text{with}~\bar{E}=\frac{1}{n} \sum_{i=1}^{n}\mid P_i-A_i\mid,
\end{equation}

\noindent where $n$ represents the count of level-1 cells, $\bar{A}$ represents the actual average count of workers of level-1 cells, $P_i$ and $A_i$ represent the simulated and actual counts of workers in the $i$-th level-1 cells, respectively. The rates $\gamma$ of NYTaxi and Gowalla are $15.1\%$ and $30.2\%$, respectively. This shows that in the case of very low privacy budget, the noise interference is much, the predicted result still reflect roughly the distribution of actual data, and the simulated distribution of workers can be regarded as the actual distribution.

\subsection{Problem Statement}

%We assume a new problem model, a trusted third-party CSP continuously collects workers' location information, predicts real-time location distribution and generates PSD based on learning from noisy historical location data when SC-server forwards the task request and then import real-time noisy data for the server to query and construct GR.
%
%As SC server is not trusted, historical and real-time location data of workers are sensitive information, so we need to adapt to changes in real-time location distribution and dynamically update the PSD grid over time. In terms of differential privacy, data information adds a time dimension, which undoubtedly provides adversaries with more available information, especially the use of historical sensitive data to predict and attack workers' real-time location. On the other hand, how to reasonably allocate the privacy budget in parallel in the overall framework and build an optimized data noising mechanism on the dynamic grid to achieve efficient GR publishing and task allocation will be a more important challenge.
%
%To sum up, the problems statement in this paper are as follows: Given workers' historical and real-time locations and privacy budget, it is required to generate a dynamic PSD that protects privacy, the released GR can achieve efficient spatial task allocation especially the ASR can reaches EU stably, and there is no significant increase in system overhead.

%\new{Some more description.... like}

DP guarantees
that the probability of producing a given output does
not depend much on whether any record is present
in the input dataset or not. Random noise is added to each query result to preserve privacy, such that an adversary will not be able to deduce the privacy information of any user from the query outputs, regardless of its prior knowledge.

Existing SC models with differential privacy protection mainly harness real-time location data and generate PSD by performing noise additions and domain partitioning in a crossed way. They have to add noise to worker counts in grids at all levels, which tends to incur high error on many aspects and affects efficiency of SC system.

For this, we are intended to propose a new SC model. That is, given historical and real-time data of worker locations, and a privacy budget $\epsilon$, perturb their counts with necessary noise, generate a suitable PSD and then design a more efficient SC model with privacy protection. Specifically, with the simulated locations drawn from the probability distribution learned from historical data, a domain partitioning is performed with little privacy budget. On the other hand, based on the grids determined previously, the real-time data are directly imported into the bottom cells in which the counts are perturbed with almost the whole privacy budget. In addition, how to optimize task assignments in SC from various perspectives (particularly to achieve stable success rate of task assignment) based on PSD with perturbation is also considered in this paper.

\section{Basic Framework}\label{sec:framework}

This section mainly introduces the basic framework of the newly proposed privacy protection scheme, including system model, domain partitioning method, cell selection strategy in $GR$, and performance metrics for system design.

\subsection{System Model}\label{sec:model}
\begin{figure}[ht]
\centering
\includegraphics[scale=0.5]{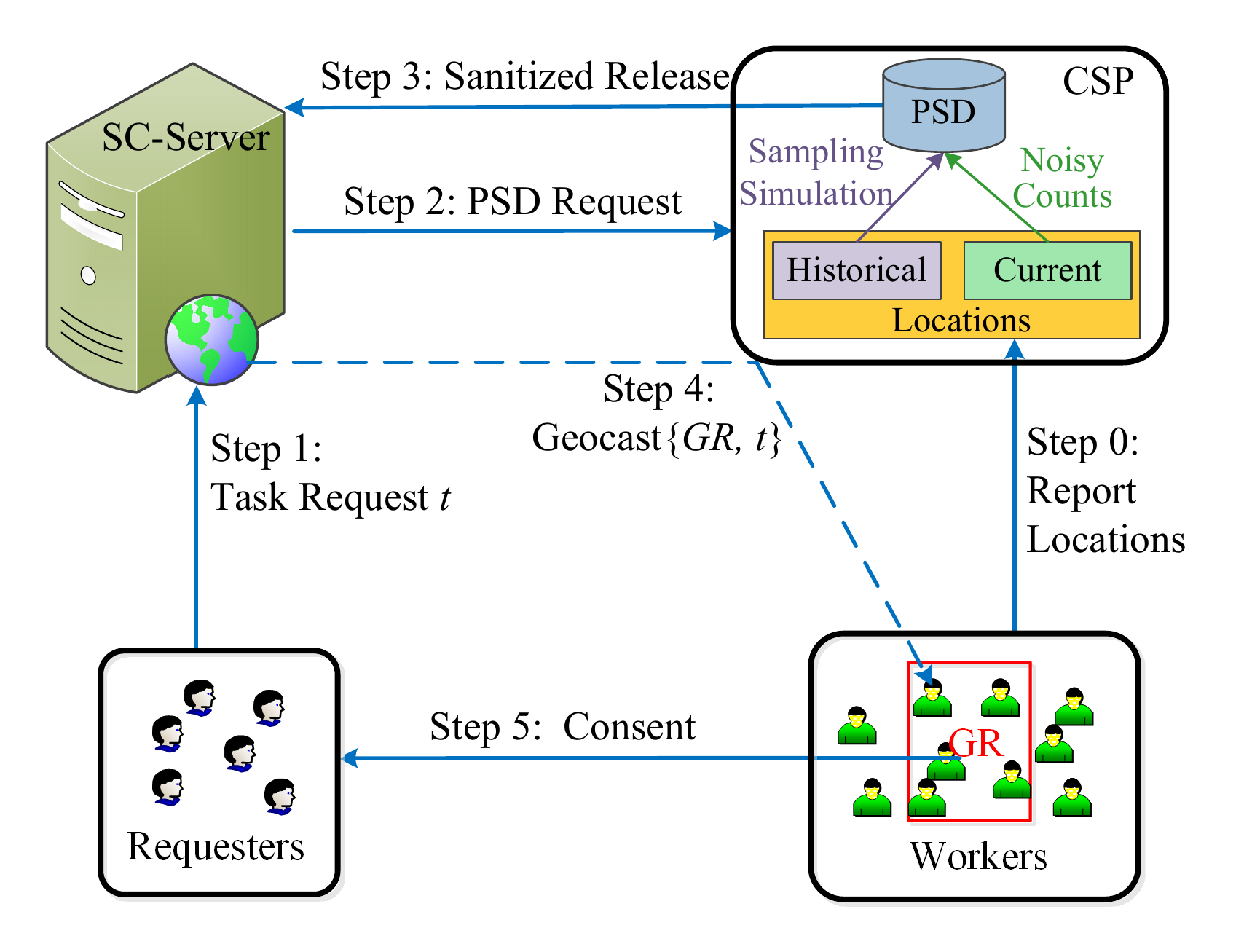}
\caption{Privacy framework for spatial crowdsourcing}
\label{fig:SC_model}
\end{figure}

 Like \cite{TGF17}, we consider the privacy protection problem of SC worker locations in the SAT mode. Fig. \ref{fig:SC_model} describes our proposed system framework that consists of four parts: CSP, SC-server, requesters and workers.
 Workers send their real location information to CSP. As a trusted third party, CSP collects locations reported by workers, and construct PSD with noises. Requesters submits tasks and exposed location information to SC-Server.
 After receiving a task request, the untrusted SC-server determines $GR$ by querying the PSD and initiates a geocast communication process.

 The SC-server is assumed to be semi-honest and intentionally deduces sensitive information of locations from the PSD. Under mutually-agreed upon rules with workers for data disclosure, CSP exposes PSD release (without identities and locations) to SC-servers and contacts workers in the $GR$. CSP plays mainly the role of communications, and computations to generate $GR$ are carried out at the SC-server part.
  Possible disclosure of worker location and identity after her consent to the task is outside our scope.

 To be specific, our work focuses on the design of PSD using historical data learning at CSP part. Based on historical locations, we adopt linear regression method to make prediction and simulate the real-time distribution of locations at the level-1 grid, which yields the partition together with adaptive grids (AG) method. Then CSP publishes the PSD with noisy real-time counts. Fig. \ref{fig:AG} describes the procedure of building PSD with learning.
 See Section \ref{sec: AG} for details.

As shown in Fig. \ref{fig:SC_model}, the whole scheme consists of the following steps.
\begin{enumerate}[Step 0:]
	\item[Step 0:] Workers report their real locations to the CSP, which will be divided into real-time and historical data.
	\item[Step 1:] A requester sends task $t$ to SC-server.
	\item[Step 2:] SC-server queries the PSD with the CSP.
	\item[Step 3:] According to the given privacy budget $\epsilon$, CSP partitions the data domain by learning historical locations reported by workers and imports noisy real-time counts into the generated grids to update the PSD for answering to SC-server.
	\item[Step 4:] SC-server determines $GR$ and initiates a geocast communication process in two ways (infrastructure-based or infrastructure-less mode) as in \cite{TGF17}.
	\item[Step 5:] If a worker in $GR$ accepts the task, she sends a consent message to the SC-server (or the requester) for confirming her availability.
	\end{enumerate}

The above scheme is mainly designed for a single-task assignment system model in dynamic scenarios where for each real-time task request, the CSP collects properly the latest real-time location data of workers and updates the PSD with newly confirmed historical data, and only a worker is required to complete the task. If worker locations change rapidly, our scheme allows fixing the grid structure and updating only real-time counting in cells, which is assigned the same privacy budget. This is a significant advantage of our scheme, in particular on running time, see Section \ref{sec:runtime}.

In the static scenario, SC-server deploys a fixed PSD in a period to release $GR$, and Steps 0, 2 and 3 can be skipped. In the case of multiple tasks, our system model is still valid under the assumption that the task assignments do not interfere with each other.

\subsection{Building PSD with Learning}\label{sec: AG}

The first stage in our framework consists of building
a PSD (at the CSP part), which determines the accuracy of released data and also affects performance metrics. Here we improve the extended \emph{Adaptive Grid method (AG)} developed in \cite{TGF17} with historical locations learning.
Table \ref{tlb:notations} summarizes the notations used in our description.
\begin{table}[htbp]
	\caption{Summary of notations}\label{tlb:notations}
	\begin{tabular}{ll}
		\toprule
		Symbol&Definition\\
		\midrule
$\epsilon$ & Total privacy budget\\
$N_p$ & Total count of workers by prediction\\
${N}_p^{ij}$& Predicted worker count in first-level cell $c_{ij}$\\
$m_{1}$& The first level grid granularity\\
$m_{2}^{ij}$& Second-level partitioning granularity at cell $c_{ij}$ \\
$\beta$ & Budget allocation parameter for prediction,~$\beta=0.04$\\
$\epsilon^{\prime}$ & Parameter for domain partitioning in  \cite{TGF17}, $\epsilon^{\prime}=0.5\epsilon$\\
		\bottomrule
	\end{tabular}
\end{table}

\begin{figure}[ht]
\centering
\includegraphics[scale=0.22]{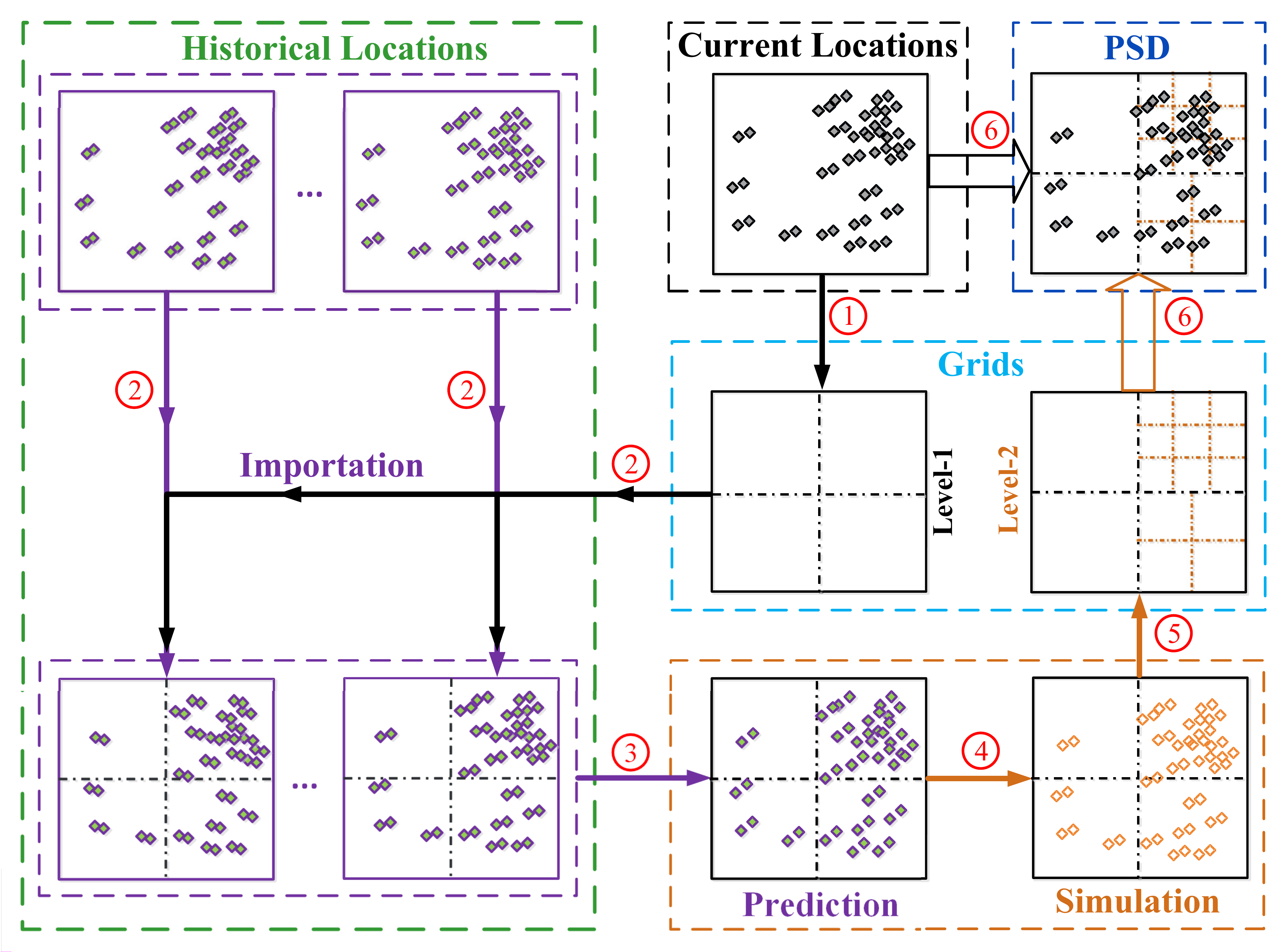}
\caption{Procedure of building PSD with learning}
\label{fig:AG}
\end{figure}

The work flow of our improved AG is shown in Fig. \ref{fig:AG}.
%AG reduces the non-uniformity error brought by UG
Compared to the extended AG proposed in  \cite{TGF17}, our PSD construction takes advantage of historical data learning. The detailed procedure is given as follows.

Firstly (Step \ding{172} in Fig. \ref{fig:AG}, level-1 partition),
as before, at the first level, AG creates a fixed-size $m_1 \times m_1$ grid over the data domain with the level-1 granularity $m_1$ chosen as
\begin{equation}\label{eq: m1}
m_1  = \max\left(10,\left\lceil \frac{1}{4}\sqrt{\frac{N^\prime\times\epsilon}{k_1}}\right\rceil\right),
\end{equation}
where $N^\prime$ is the noisy total number of locations computed by adding random Laplace noise with scale
 $1/(\beta \epsilon)$,
$\epsilon$ is used here just as a parameter for partitioning (privacy cost $\beta \epsilon$ is a small fraction of the total budget $\epsilon$) and $k_1=10$ as in \cite{TGF17}.

Secondly (Steps \ding{173}-\ding{174} in Fig. \ref{fig:AG}, prediction with historical data learning), historical data (without noise) are imported into the level-1 grid to make prediction of each cell's count.
The historical data divided by $n$ periods is denoted by $H=\{h_1,h_2,...,h_n\}$.
%, and the count of location data $h_k$ in the $k$-th period $N_k$.
 AG issues count queries in all level-1 cells for each $h_k$, denoted by $N_k^{ij}$. For each cell $c_{ij}$, the linear regression on $\{N_k^{ij} \}_{k=1}^n$ yields $N_{p}^{ij}$, a locally predicted count of the real-time locations in cell $c_{ij}$. Then, the probability distribution is evaluated with fractions $N_{p}^{ij}/\sum_{i,j}N_{p}^{ij}$ of predicted total locations.

Next (Steps \ding{175}-\ding{176} in Fig. \ref{fig:AG}, simulation and level-2 partition),
according to the probability distribution of locations on level-1 cells (as index nodes) predicted above, randomly generate $N^\prime$ points (independent and identically distributed, i.i.d.) to simulate the real-time distribution.
 Then level-2 granularity $m_2$ for each level-1 cell $c_{ij}$ is chosen as
\begin{equation}\label{eq: m2}
m_2^{ij}  = \left\lceil\sqrt{\frac{N^{ij}_{\rm sim}\times\epsilon^{\prime}}{k_2}}  \right\rceil,
\end{equation}
where $N^{ij}_{\rm sim}$ denotes the point count on cell $c_{ij}$ by sampling simulation and  $k_2=\sqrt{2}$ is selected as in \cite{TGF17} and $\epsilon^{\prime}$ is still a parameter with $\epsilon^{\prime}$ = 0.5 $\times\epsilon$ but not privacy budget.

%the historical data set divided by $n$ periods is denoted by $H=\{h_1,h_2,...,h_n\}$, and the count of location data $h_k$ in the $k$-th period $\widetilde{N_k}$. Adding Laplace noise to each $\widetilde{N_k}$ with scale $1/(\beta \epsilon)$ generates $\widetilde{N_k^{\prime}}$. The linear regression model on $\{\widetilde{N_k^{\prime}}\}_{k=1}^n$ is established to predict the count $N_p$ in the $(n+1)$-th period (real-time positions). Then level-1 granularity $m_1$ is chosen as

Afterwards (Step \ding{177} in Fig. \ref{fig:AG}, sanitized
countings), real-time locations are imported into the adaptive grids generated above, and the location count in each cell are
added by Laplace noise with scale $1/\epsilon$.
We mention that each noisy count is set zero immediately if negative. Then the PSD is completed as is processed in Fig. \ref{fig:AG}.

Indeed, we can use only real-time data even in the simulation for level-2 partition which costs no privacy budget. However, location distribution sometimes changes rapidly and irregularly while historical data learning provides the overall trend of distribution theoretically. Such a scheme using historical data allows us to fix grid structure and update only noisy real-time count in each cell for quick release of tasks while workers move rapidly.

\noindent

\subsection{Geocast Region Construction}\label{sec:GR}

In this section, we mainly introduce some new strategies for constructing $GR$. We investigate some techniques across local cell selections to fuse those cells with negative noisy counts and develop a quality scoring function involving the area of cells instead of  utility function for cell selections.

Currently we consider just the static case for task locations. The probability of a worker accepting a task is only related to the worker-task distance. We refer to the \emph{mean contribution distance (MCD)} proposed in  \cite{HG10}, where the MCD is computed as

\begin{equation}\label{eq:MCD}
MCD(w_i)=\sum_{j=1}^{n}\frac{d\left(L_{w_i},L_{c_j} \right)}{n},
\end{equation}
where $L_{w_i}$ is the location of worker $w_i$, and $L_{c_j}$ the locations of its $n$ contributions. Following  \cite{MG13}, we regard $90\%$ of $MCD$ as \emph{maximum travel distance (MTD)} that is the maximum distance a worker accept to travel to perform a task.

In contrast to the utility function proposed in \cite{TGF17}, we introduce a quality scoring function that considers noisy worker count and the cell-task distance as well as cell's area. For adding each neighboring cell to $GR$, the score $q_t^k$ of cell $c_k$ in the candidate heap $Q$ for task $t$ is defined as:
\begin{equation}\label{eq:score}
q_t^k=\frac{N_k}{f_s(S_k)\cdot f_d(D_t^k)},
\end{equation}
where $N_k$ represents the number of noisy workers in  candidate cell $c_k$, $S_k$ the area of cell $c_k$  and $D_t^k$ the cell-task distance computed by the average distance between the task and four corners of cell $c_k$. It is reasonable that the higher density of workers $N_k / S_k$ in cell $k$ or the shorter the cell-task distance $D_t^k$, the higher score $q_t^k$. The functions $f_s$ and $f_d$ are (positive and) monotonic on $S_k$ and $D_t^k$, respectively, and linear mappings to some interval $[a,b]$. Such functions are designed to control the influence of varying the factors, area and distance.

Our algorithm to construct $GR$ is shown in Algorithm \ref{alg:RHT}.
\begin{algorithm}[h]
	\caption{R-HT Algorithm}
	\label{alg:RHT}
	\begin{algorithmic}[1]
		\REQUIRE
		Maximum travel square $MTD$ centered at task $t$, expected utility $EU$
		\ENSURE
		Geocast region $GR$\\
		\STATE Init $GR=\{\}$, $Q=\{\}$, $U=0$
		\STATE Compute Local Maximum Geocast Region $LGR$ with Algorithm \ref{alg:R}\label{state:compute_R}
		\STATE $GR=GR\cup c_t$, $c_t$ is the level-2 cell that covers $t$
		\STATE Compute ${c_t}$\,'s utility $U_t$ by Eq. (\ref{eq:Uci})\label{state:compute_Utc1}
		\STATE If $U_t>0$ then $c=c_t$ and $U=U_t$
		\WHILE {$U< EU$}
		\STATE Find $neighbors=\{\{{c}\,'s~neighbors\}-GR\}\cap MTD \cap LGR$
		\STATE $Q=Q\cup neighbors$
		\STATE If $Q$ is $null$, return $GR$
		\STATE Update cell $c$ with the highest score in $Q$
		\STATE If cell $c$\,'s score $q_c\leq 0$, return $GR$\label{state:break}
		\STATE Remove $c$ from $Q$
		\STATE $GR=GR\cup c$
		\STATE Compute $c$\,'s utility $U_t^c$ by Eq. (\ref{eq:Uci})\label{state:compute_Utc2}
		\STATE Update $U$ by Eq. (\ref{eq:U})\label{state:compute_U}
		\ENDWHILE
	\end{algorithmic}
\end{algorithm}
	
In Line \ref{state:compute_R}, in order to improve the compactness of the $GR$,
 %and avoid excessive neglect of nearby workers,
 we set a restriction by \emph{Local Maximum Geocast Region (LGR)} centered at task $t$ during cell selection and the radius computation of $LGR$ is presented in Section \ref{sec:LGR}. In Line \ref{state:break}, we add a termination condition for
cyclic program in the case that all candidate cells include non-positive (and noisy) count of workers, which will be discussed in detail in Section \ref{sec:break}.

In Lines \ref{state:compute_Utc1} and \ref{state:compute_Utc2}, we calculate the utility value of some cell $c$ by, the utility function naturally defined in \cite{TGF17},
\begin{equation}\label{eq:Uci}
U_t^c=1-\left(1-AR_t^c\right)^{N_c}.
\end{equation}
Here, $N_c$ denotes the noisy count of workers in cell $c$, $AR_t^c$ represents the average task acceptance probability of each worker in cell $c$. All workers in the same cell are regarded to have the same $AR$ value. Then $\left(1-AR_t^c\right)^{N_c}$ gives the probability that no workers in $c$ are willing to perform the task $t$. The practical meaning of $U_t^c$ is the probability that at least one worker in cell $c$ is willing to finish task $t$. All possible candidate workers for each task are limited in the maximal distance $MTD$ from task $t$. The $AR$ value is assumed to decrease linearly with distance and defined as
\begin{equation}\label{eq:AR}
AR_t^c=(1-d / MTD)\cdot MAR,
\end{equation}
where $d$ represents the average distance between the task $t$ and the four corners of the cell $c$, and $MAR$ is defined as the maximum task acceptance rate of workers. Once the worker-task distance reaches $MTD$, the $AR$ vanishes.

In Line \ref{state:compute_U}, after adding $c$ to $GR$, we should update the $GR$ utility by

\begin{equation}\label{eq:U}
U=1-(1-U)(1-U_t^c).
\end{equation}

\begin{figure*}[htp]
	\centering
	\includegraphics[scale=0.5]{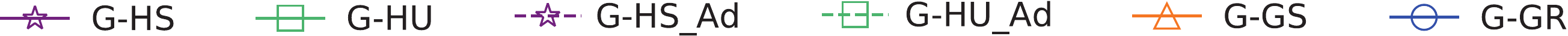}\\
	\subfigure[ASR, Ye.]{
		\includegraphics[height=0.97 in]{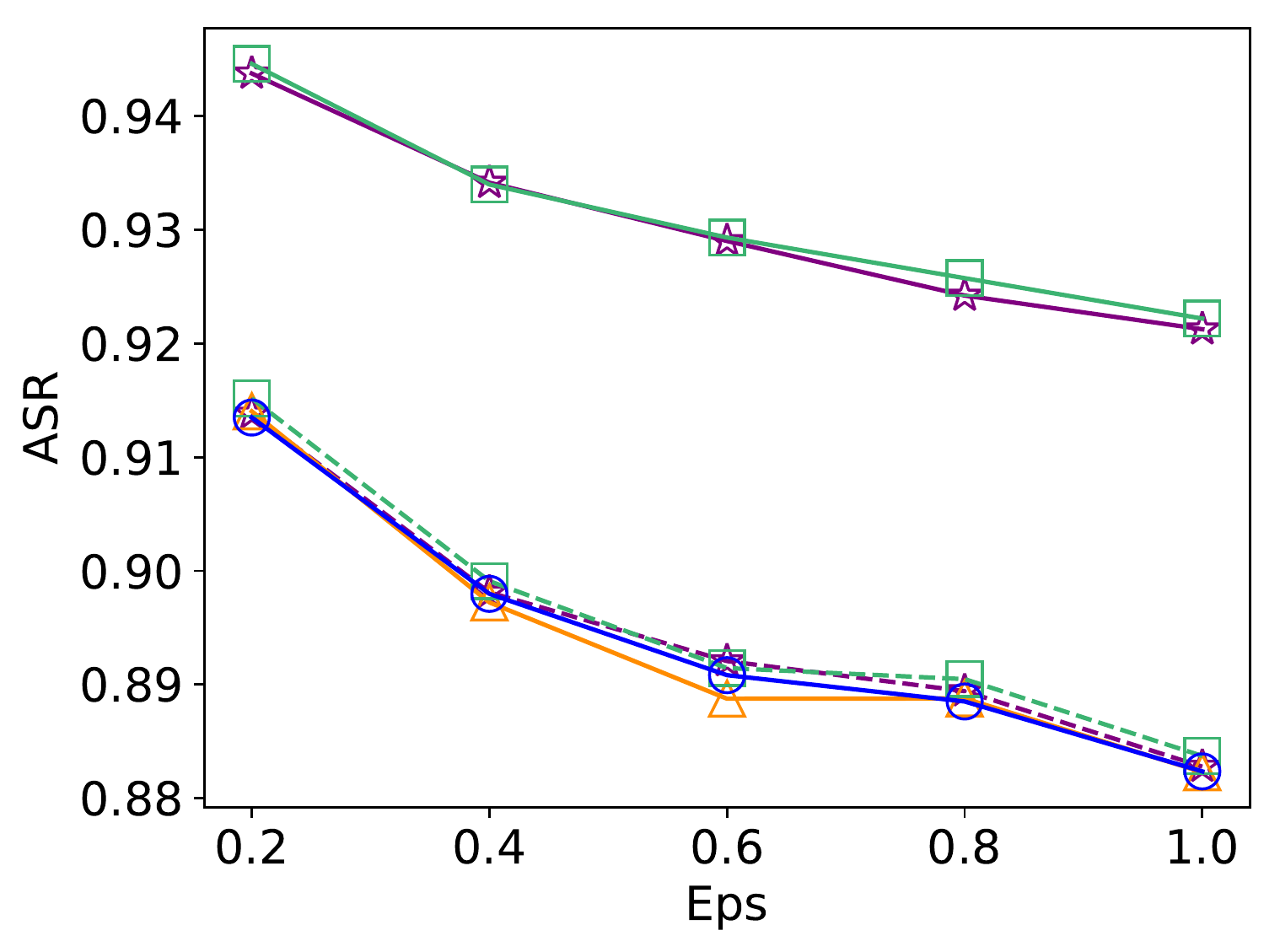}
	}
	\subfigure[WTD, Ye.]{
		\includegraphics[height=0.97 in]{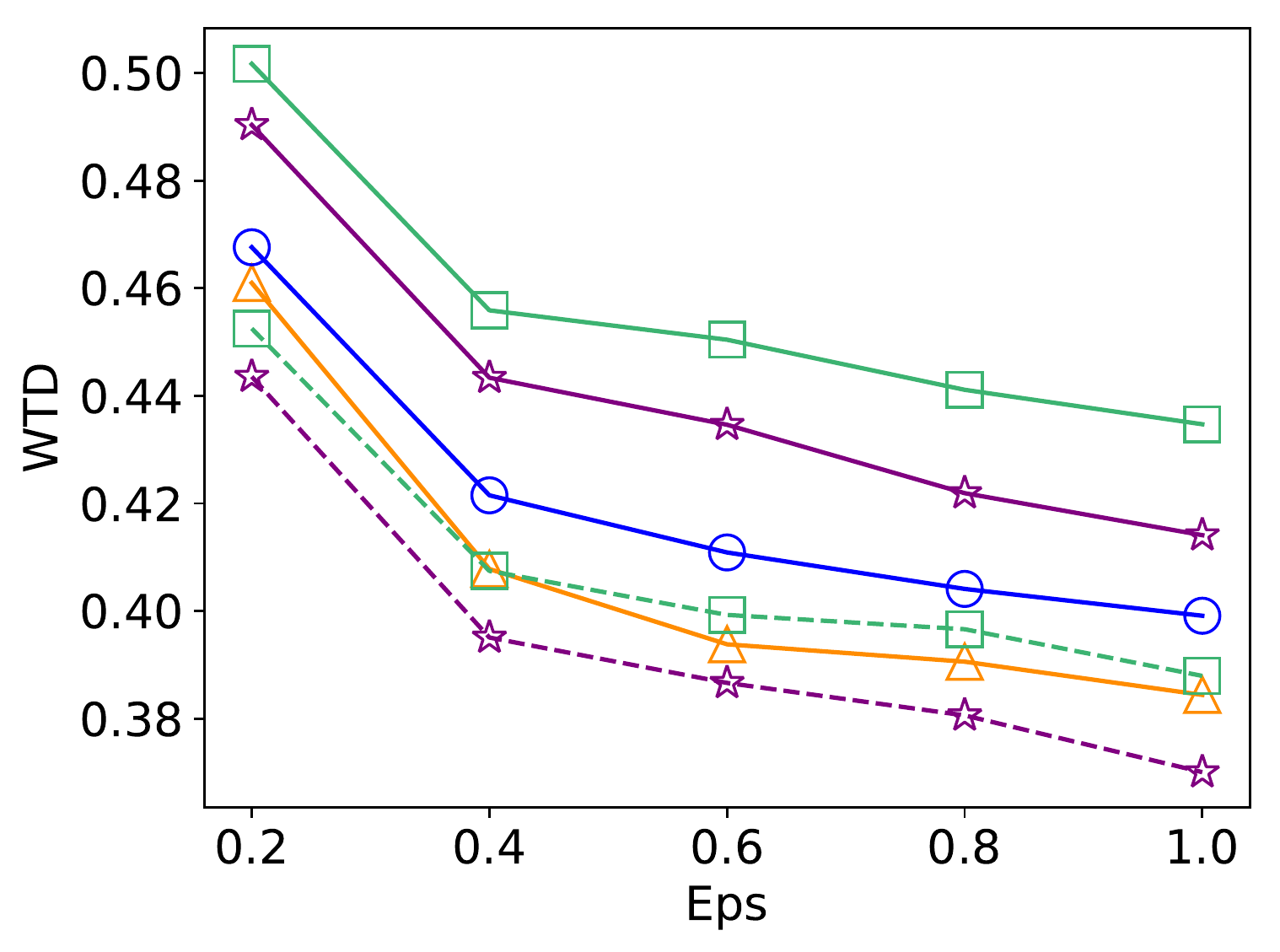}
	}
	\subfigure[HOP, Ye.]{
		\includegraphics[height=0.97 in]{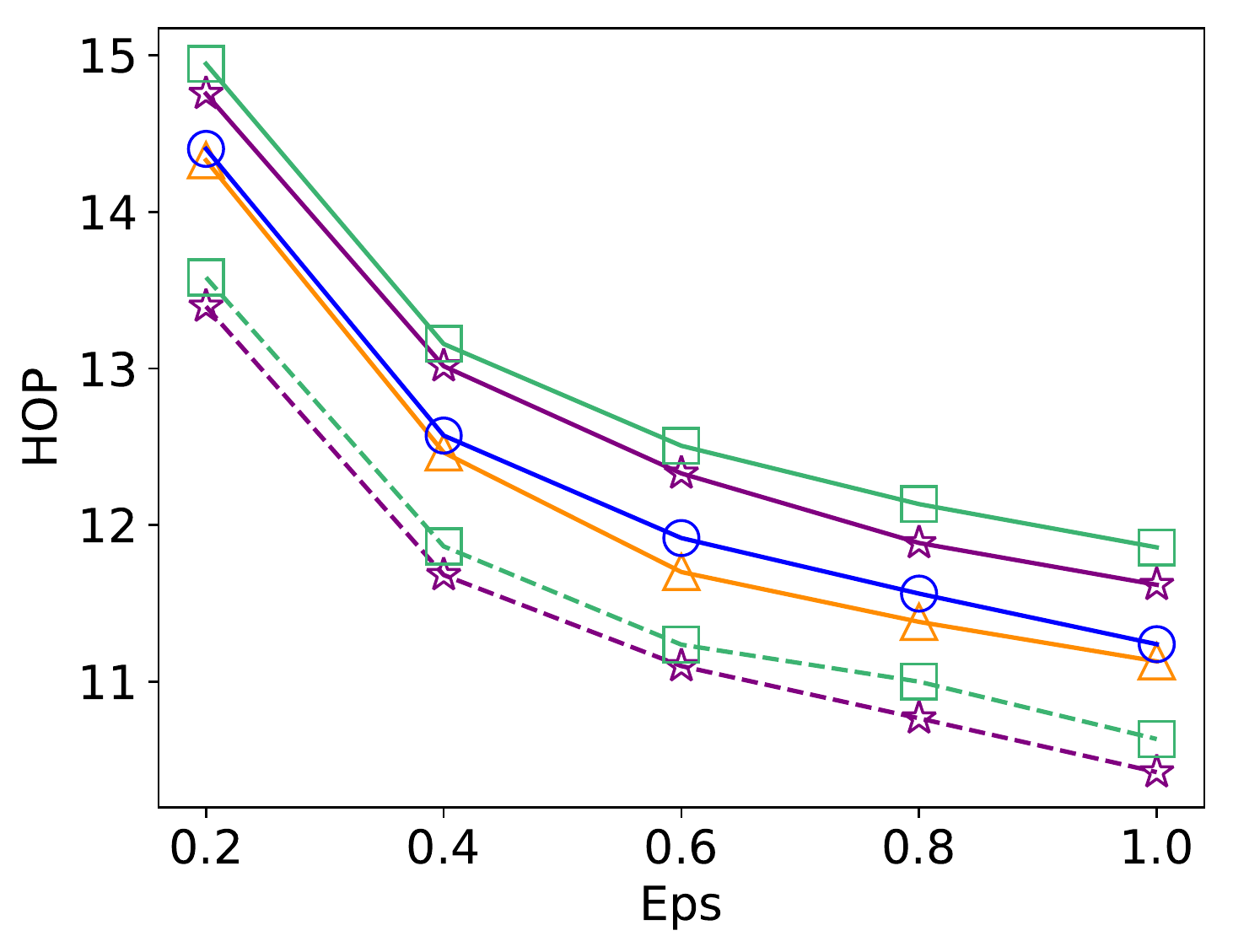}
	}
	\subfigure[ANW, Ye.]{
		\includegraphics[height=0.97 in]{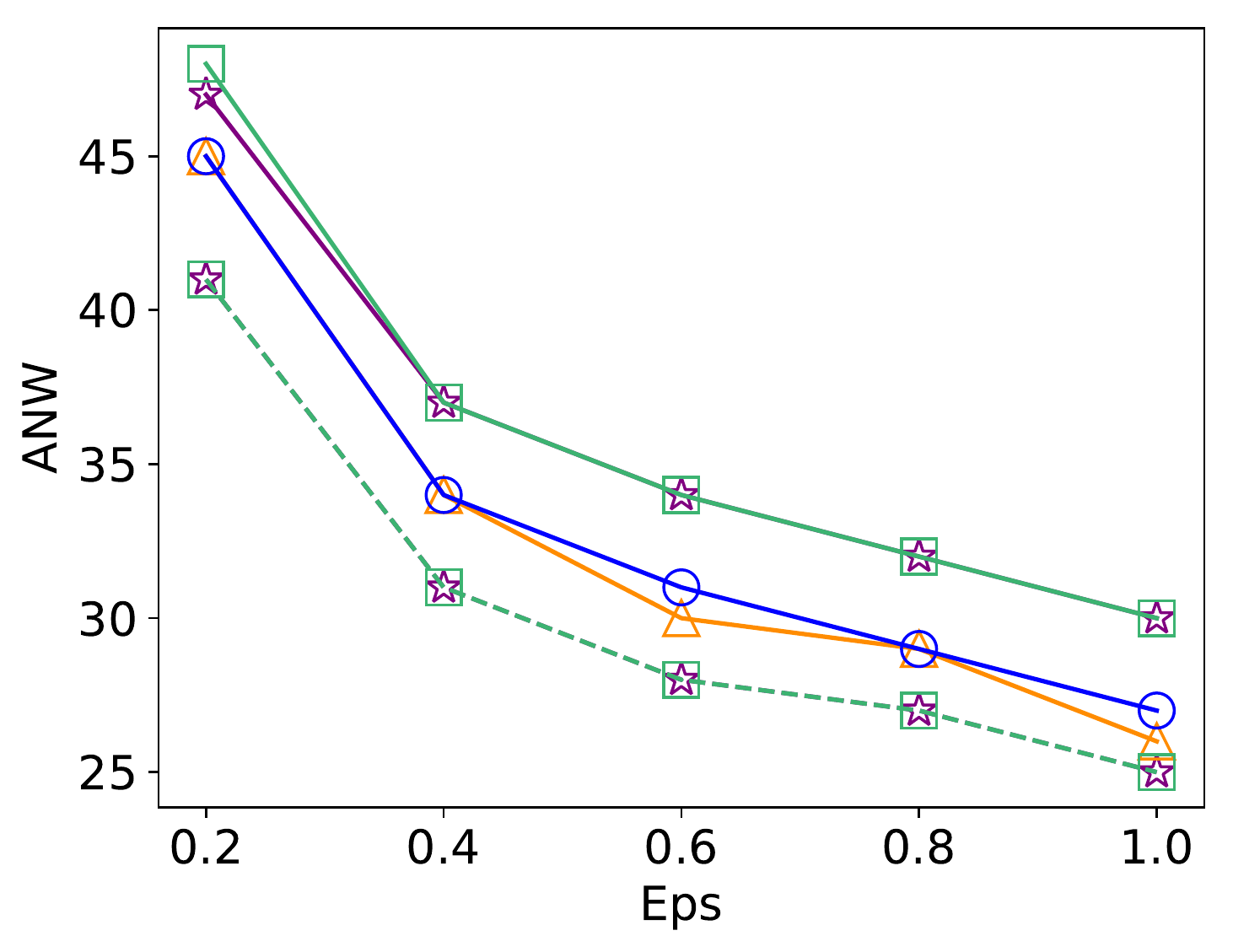}
	}
	\subfigure[DCM, Ye.]{
		\includegraphics[height=0.97 in]{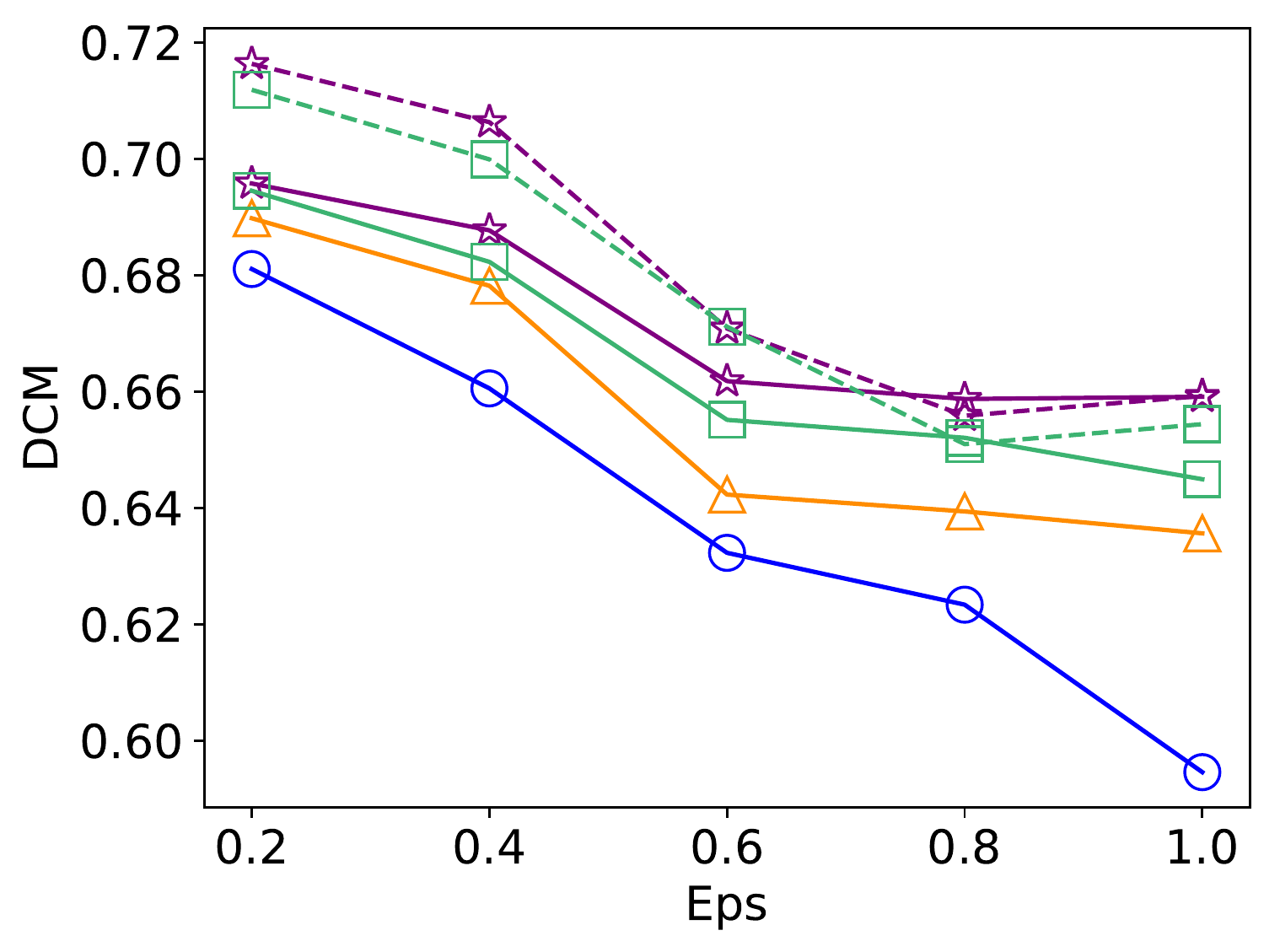}
	}
	\caption{Effect of quality scoring function on scheme performance}
	\label{fig:score}
\end{figure*}

Once $U$ reaches the expected utility $EU$, it returns $GR$ directly. Otherwise, we continue to select a new neighboring cell which ensures that the $GR$ is a continuous region.
The significant effect of using quality scoring function is shown as in Fig. \ref{fig:score}, see Section \ref{sec:scoring} for details.

\begin{figure*}[htp]
	\centering
	\includegraphics[scale=0.5]{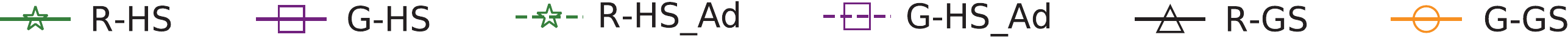}\\
	\subfigure[ASR, Ta.]{
		\includegraphics[height=0.98 in]{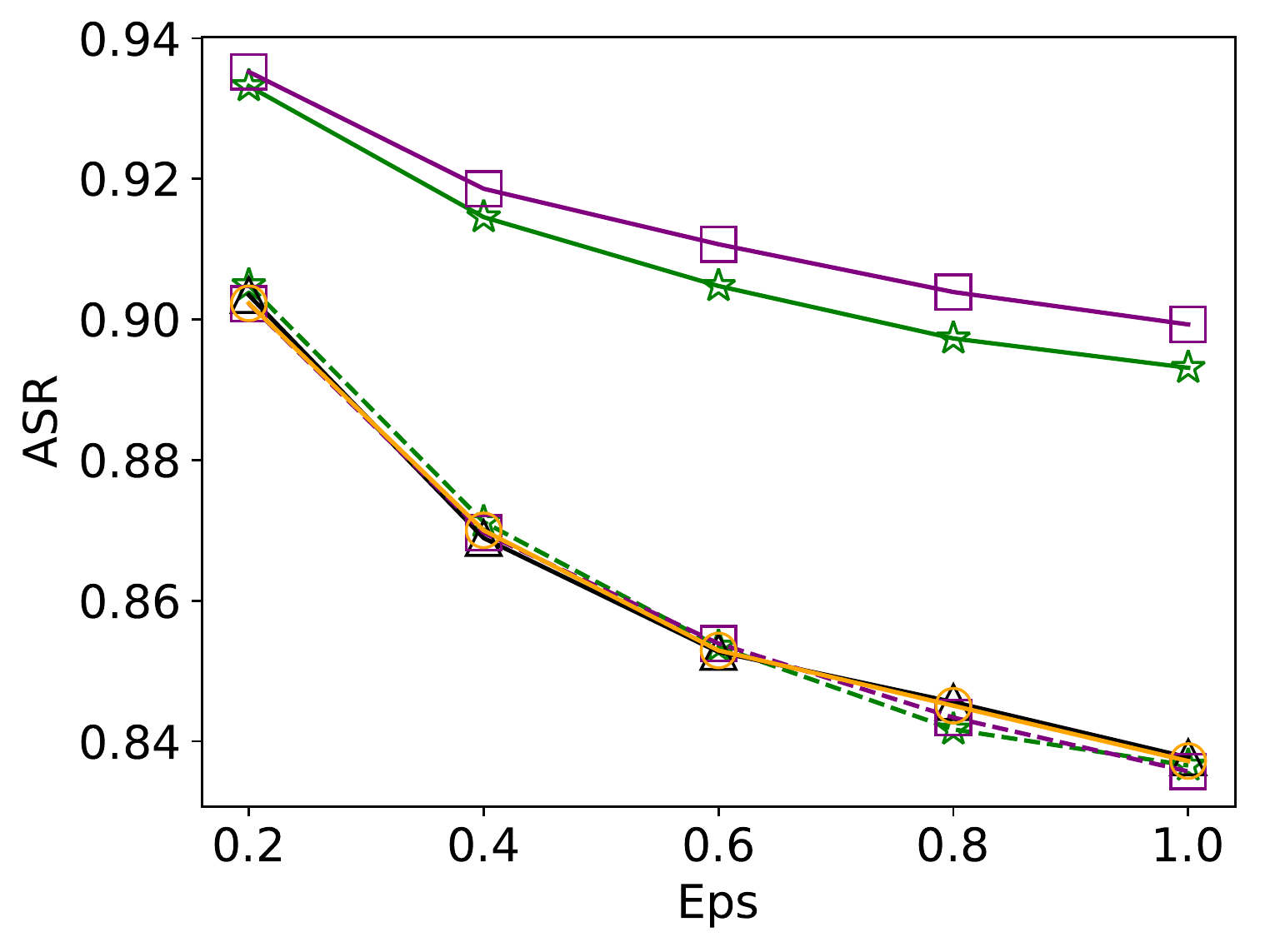}
	}
	\subfigure[WTD, Ta.]{
		\includegraphics[height=0.98 in]{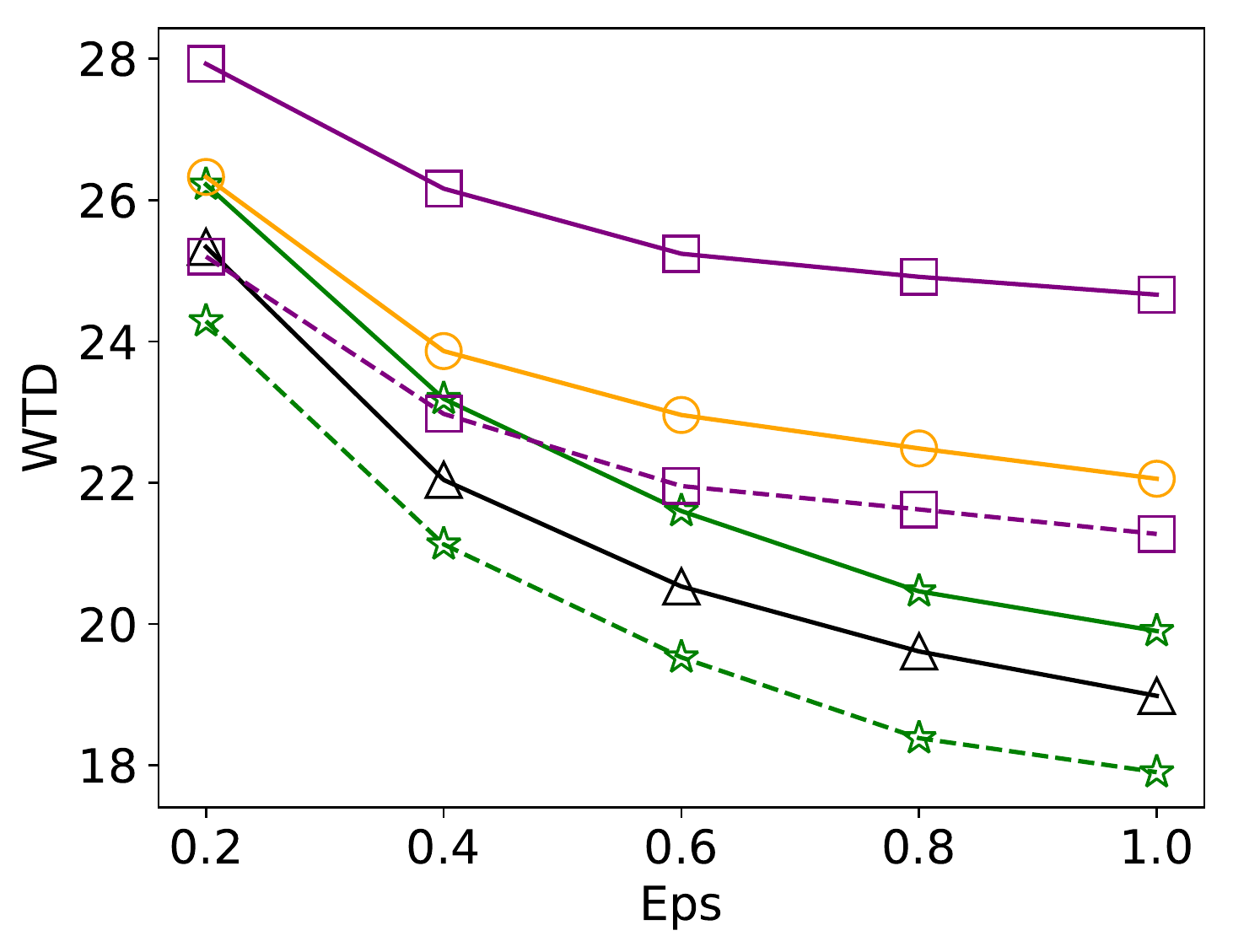}
	}
	\subfigure[HOP, Ta.]{
		\includegraphics[height=0.98 in]{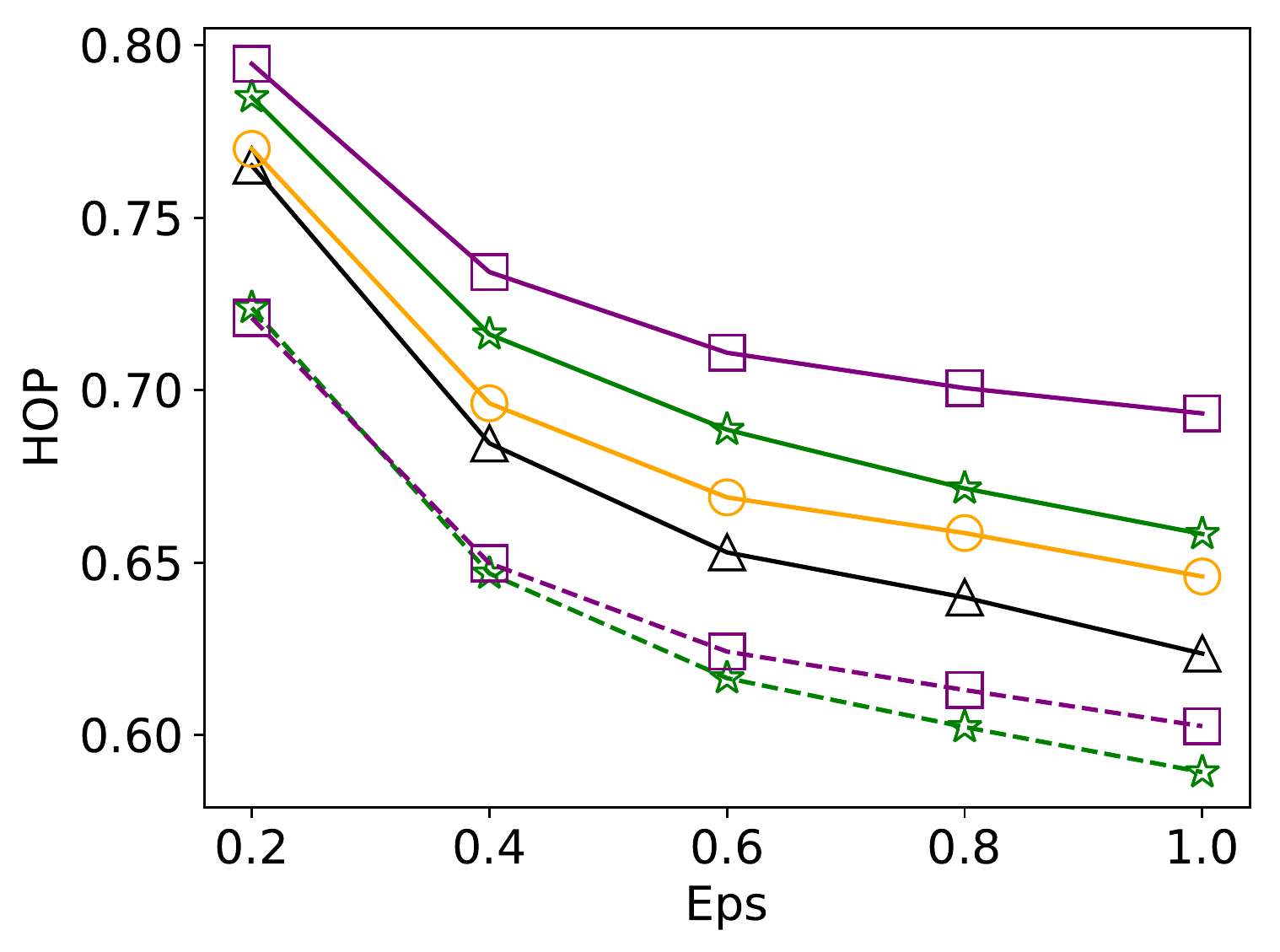}
	}
	\subfigure[ANW, Ta.]{
		\includegraphics[height=0.98 in]{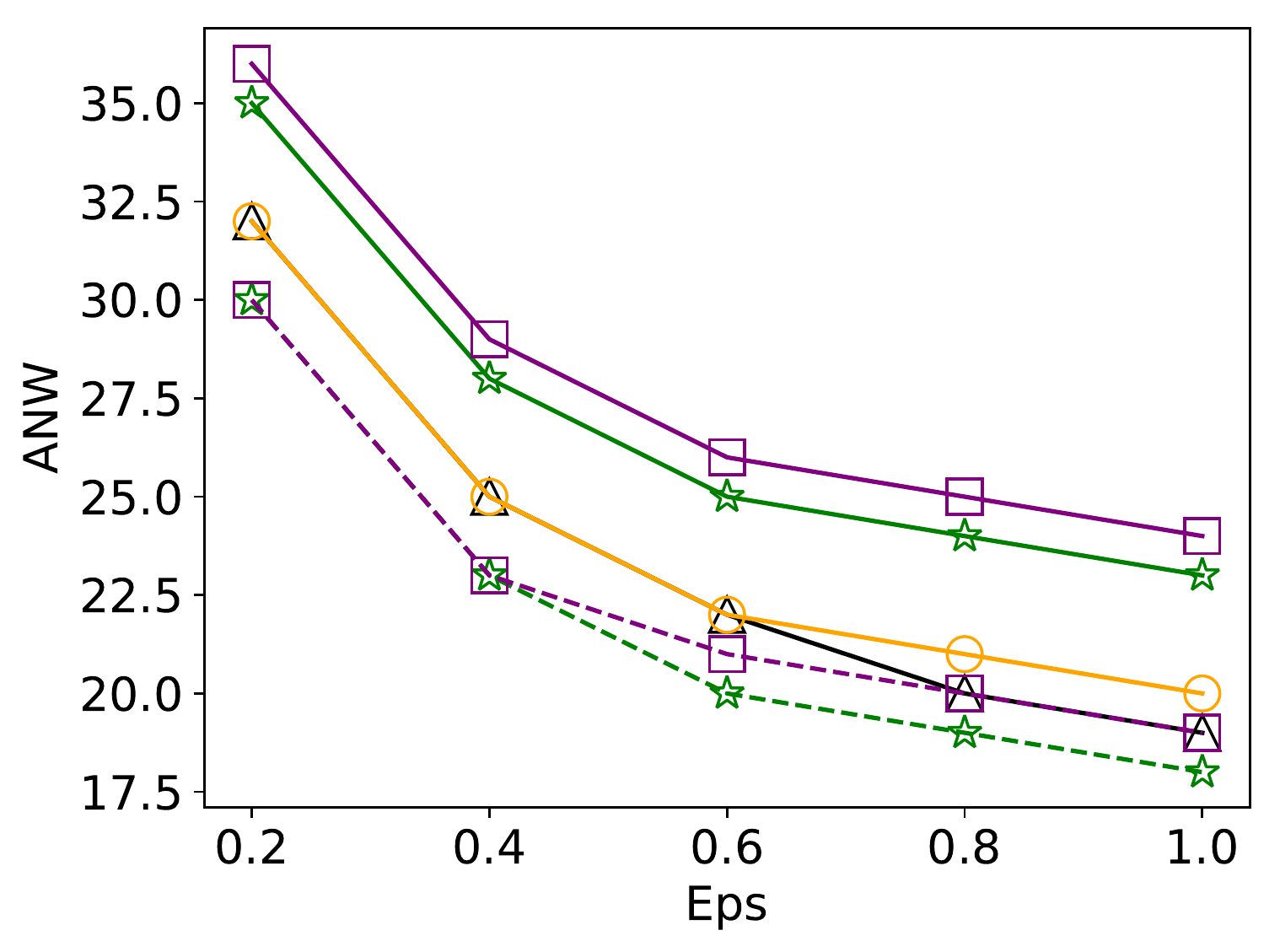}
	}
	\subfigure[DCM, Ta.]{
		\includegraphics[height=0.98 in]{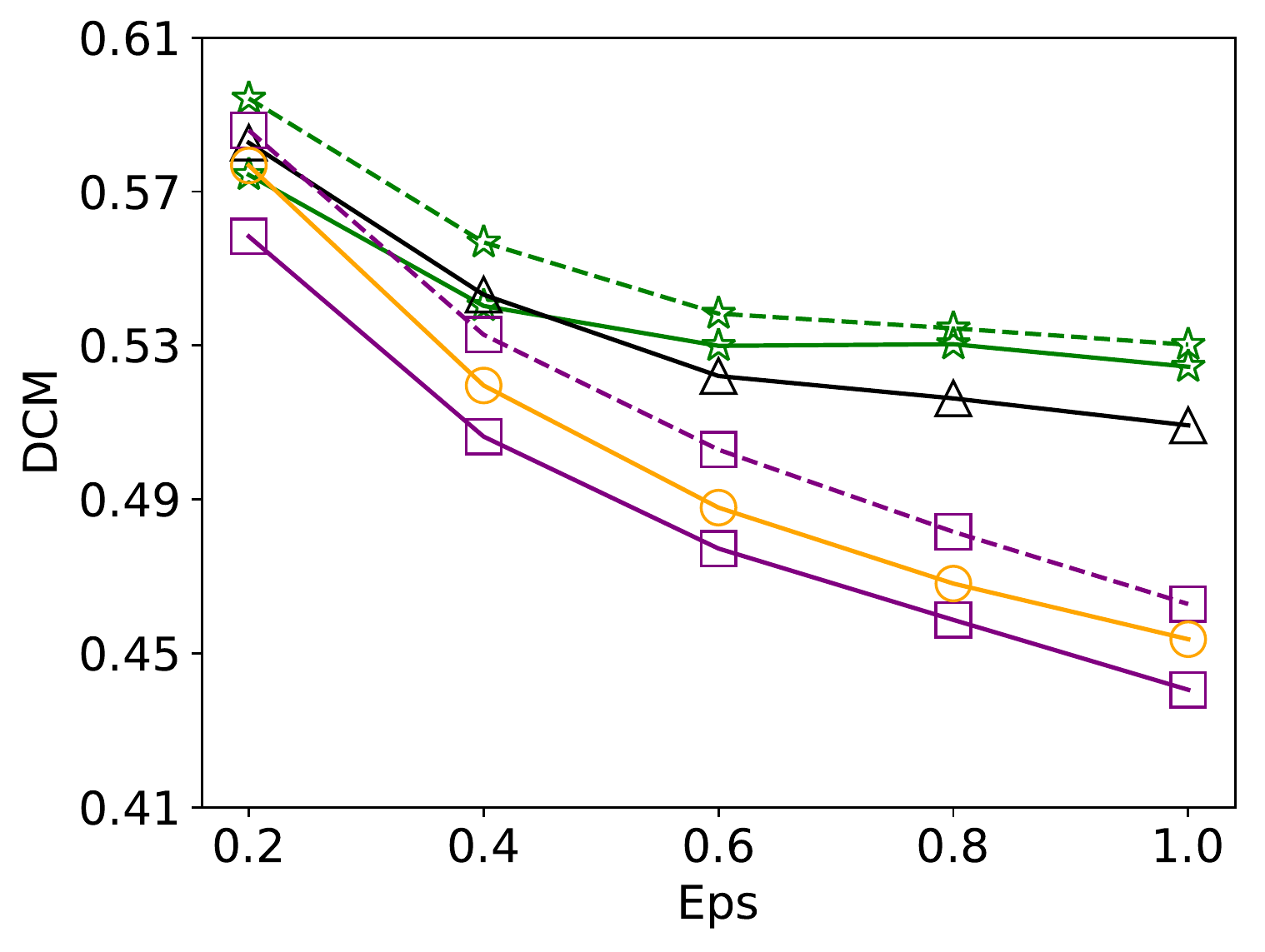}
	}
	\caption{ Effect of applying $LGR$ on scheme performance}
	\label{fig:R}

\end{figure*}
\begin{figure*}[htp]
	\centering
	\includegraphics[scale=0.5]{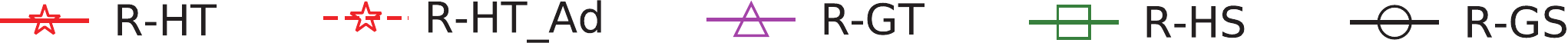}\\
	\subfigure[ASR, Go.]{
		\includegraphics[height=0.99 in]{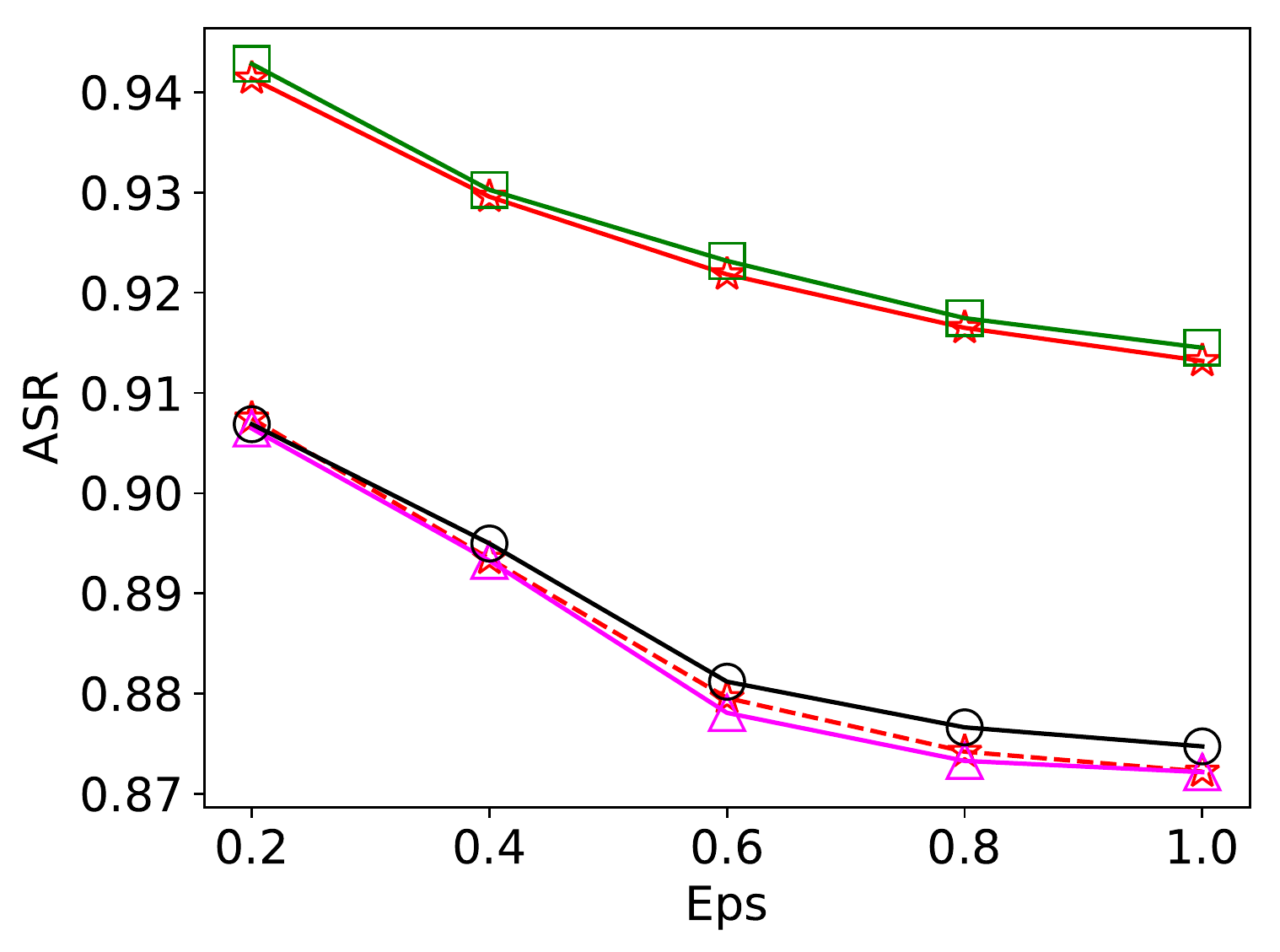}
	}
	\subfigure[WTD, Go.]{
		\includegraphics[height=0.99 in]{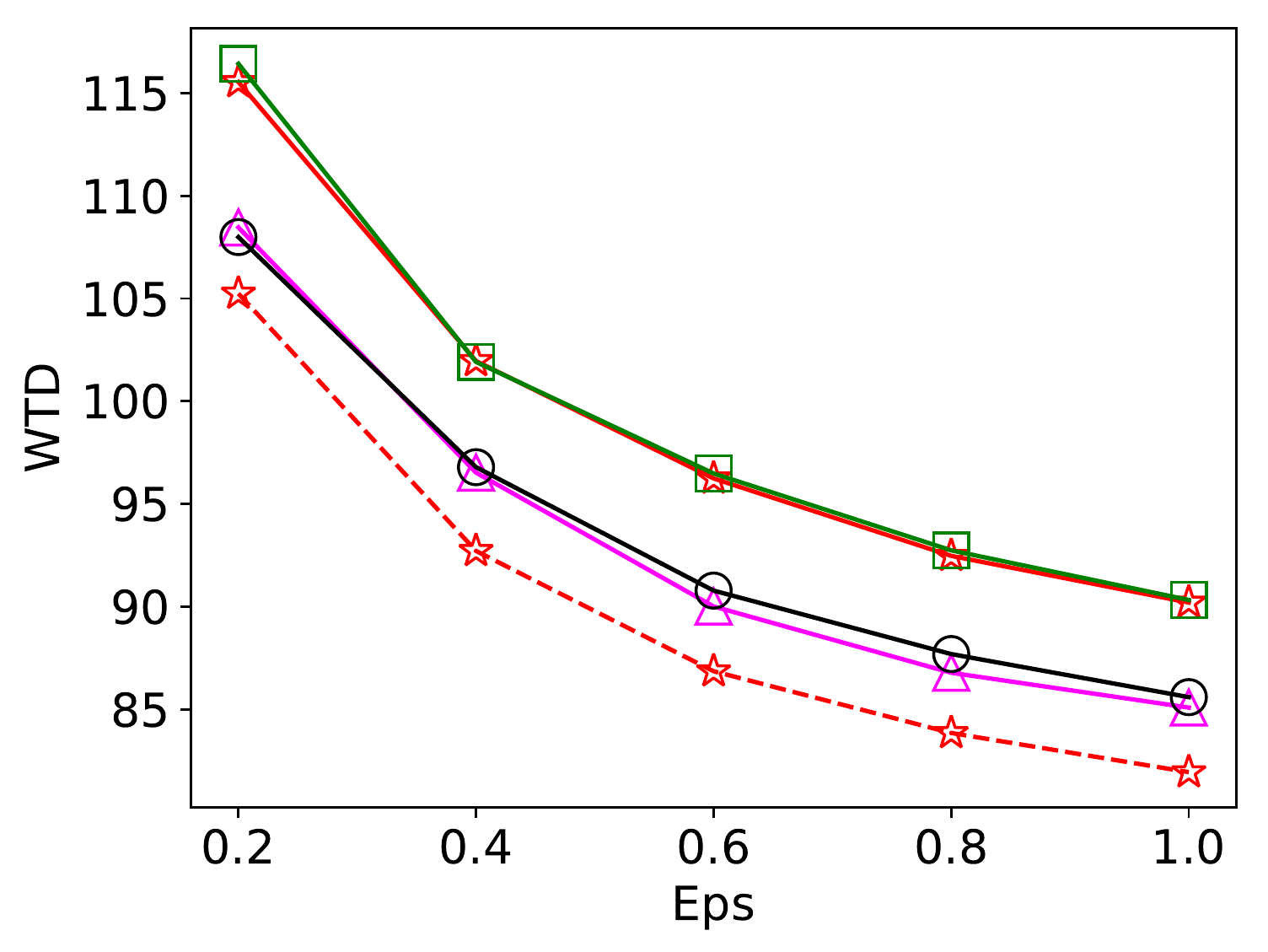}
	}
	\subfigure[HOP, Go.]{
		\includegraphics[height=0.99 in]{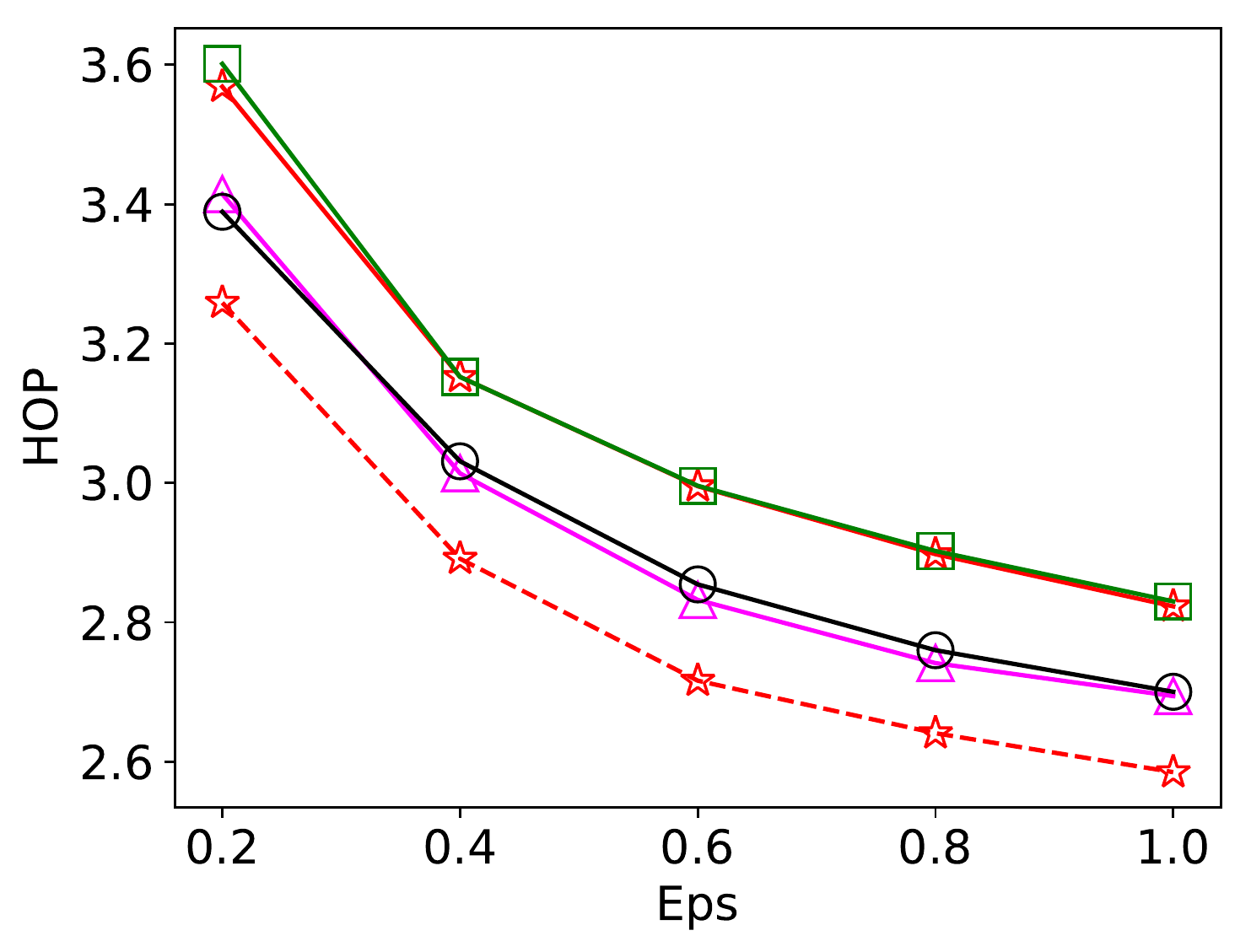}
	}
	\subfigure[ANW, Go.]{
		\includegraphics[height=0.99 in]{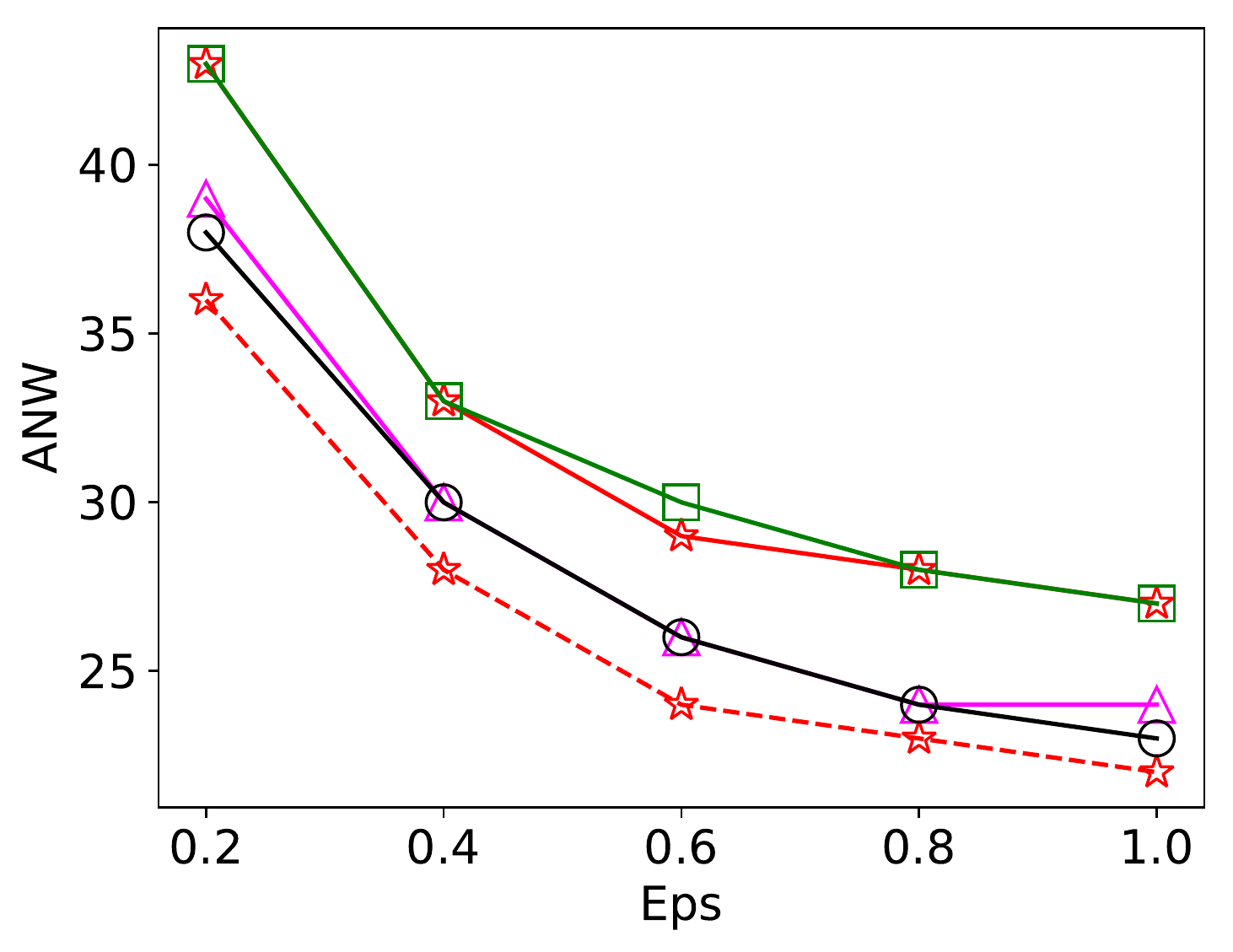}
	}
	\subfigure[CELL, Go.]{
		\includegraphics[height=0.99 in]{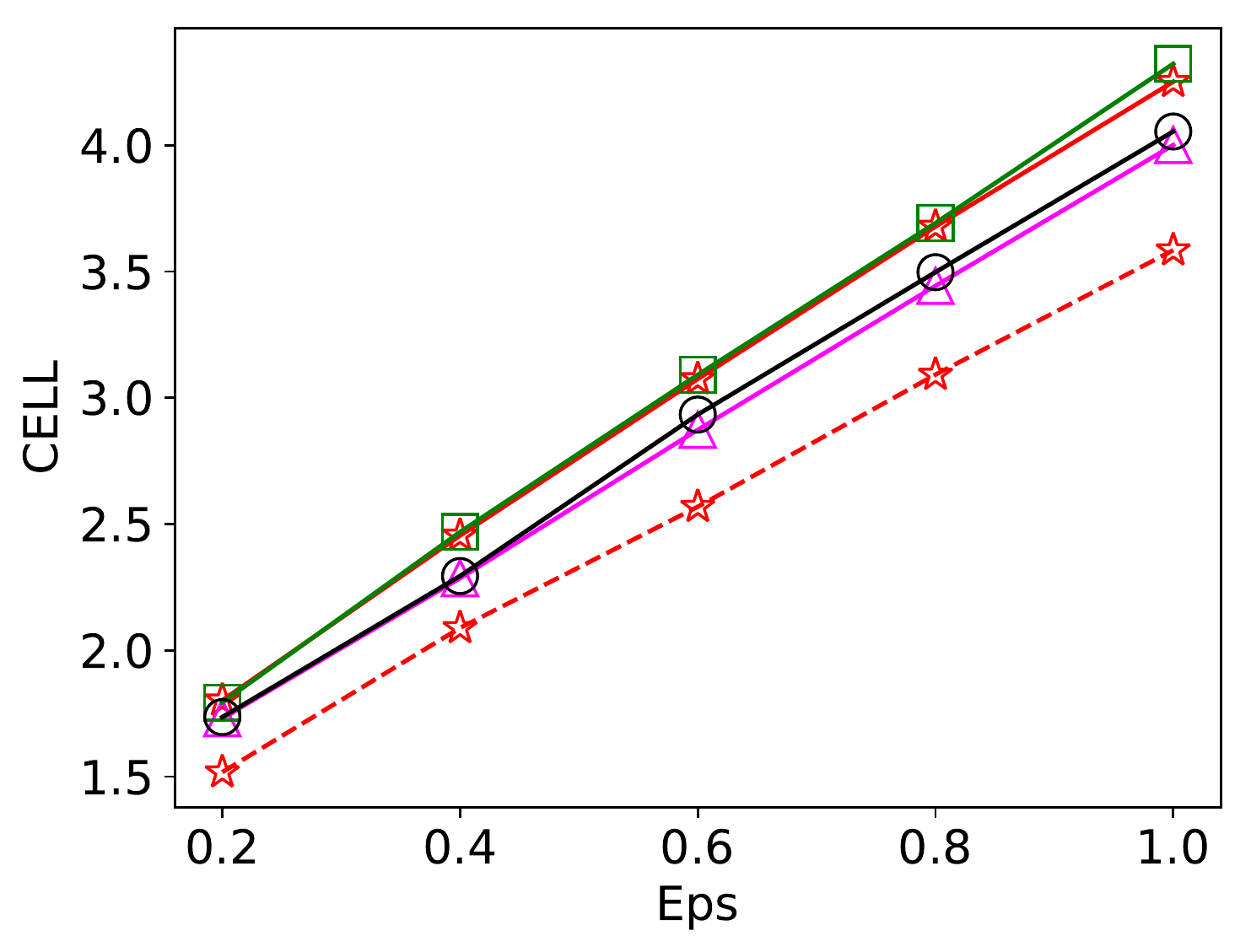}
	}
	\caption{Effect of applying ``Break'' strategy on scheme performance}
	\label{fig:T}
\end{figure*}
\subsection{Performance Metrics}\label{sec:goals}

Adding noise to protect worker locations in SC will inevitably reduce the validity of worker-task matching and efficiency of task assignment.

 To be specific, a notified region with a (noisy) positive count may contain no workers. Some workers may be notified of the task with long distance while nearer workers have no tasks. Moreover, some redundant messages for notification increase overhead.
%Due to the randomness of Laplace noise, the composition of $GR$ will show some random and irregular changes, and even cause the distance of workers to be far away.
In order to evaluate the performance of the framework, we concentrate on the following performance metrics proposed in \cite{TGF17}.

\begin{enumerate}[(1)]
\item {\bf Assignment Success Rate (ASR).}
%As the noise of privacy protection mechanism leads to the uncertainty of the task release area, the number of actual workers in the area may not enough or far away form task, affecting the success rate of task assignment.
As usual, ASR measures the ratio of tasks assigned successfully to the total number of task requests. The challenge for us is to ensure that ASR reaches the threshold $EU$ in the average sense of many task assignments.

\item {\bf Worker Travel Distance (WTD).}
%According to the framework setting, SC-server doesn't know the actual worker-task distance,
Long distance will inevitably affect the efficiency of task execution. The goal is to keep the actual worker-task distance as small as possible on average.

\item {\bf Average Number of Notified Workers (ANW).}
ANW affects both the communication overhead of the $GR$ and the computation overhead of the matching algorithm. Its goal is to inform as few workers as possible without compromise on the ASR.

\item {\bf Average Hop Count Required for Geocast (HOP).} In practice, task notifications to workers in $GR$ are sent by hop-by-hop wireless communication. HOP means the hop count required to disseminate the task request to all workers in given region. We approximate HOP as diameter of the $GR$ divided by the diameter of the communication range ($100$ meters for WiFi).

\item {\bf Digital Compactness Measurement (DCM).} Based on the assumption that the communication cost is proportional to the minimum bounding circle that covers $GR$, the ratio of the $GR$ area to the area of the smallest circumscribing circle, denoted by DCM, is adopted to measure the compactness of $GR$. Generally, the high compactness of the $GR$ is helpful in reducing communication costs. The challenge is to make DCM as close to 1 as possible.

\end{enumerate}

\section{Selection Strategies and Privacy Analysis}\label{sec:skill}

In this section, we introduce three practical cell selection strategies involved in the R-HT scheme in detail and give privacy analysis of the whole system framework.

\subsection{Cell Selection by Quality Scoring}\label{sec:scoring}
The goal of our scheme is firstly to reduce communication cost and improve the success rate of task assignment when constructing $GR$ for a single task, which requires suitable strategy of cell selection. The utility function involving worker count and worker-task distance is usually used in previous schemes, such as G-GR. It selects a neighboring cell with the maximum utility in each step, which would makes the number of cells in $GR$ as small as possible. Indeed, in the two-level grids structure, the difference of worker counts between cells is often small. For the neighboring level-2 cells, cell-task distances are close to each other, but the gaps on area are often very large, even dozens of times, while the utility function ignores the influence of the area. This will inevitably increase communication overhead. For this reason, we consider to harness quality scoring function derived from the exponential mechanism which comprehensively takes all of the three factors into account. We compare the effects of utility function and quality scoring on Yelp (Ye.) dataset, see Fig. \ref{fig:score}. Both functions $f_s$ and $f_d$ are assigned as linear mappings to $[1,10]$ based on series of experiments. For more detailed experimental settings, see Section \ref{sec:dataset}.

From the perspective of the quality scoring function versus the utility function, experimental results on real-time data show that the G-GS (using scoring) performs a little better than G-GR (using utility), and G-GS improves $3.1\%$, $1.1\%$, $1.2\%$ and $2.9\%$ on WTD, HOP, ANW and DCM, respectively. This has no increase on ASR, while both of the new schemes with historical data learning, G-HS and G-HU, have $4.0\%$ increase on ASR. The G-HS (using  scoring) maintains the above advantages on the above four indexes, since G-GS and G-HS prefer small cells with high worker density. The above analysis reflects fully the advantages of quality scoring function strategy.

From the perspective of using historical data, it can be seen that G-HS and G-HU (with historical data) have higher ASR than $EU$, while most of results for G-GS and G-GU failed to meet $EU$. Indeed, some negative noises in level-2 cells reduce worker count to be less than 0 and the cells' utility is usually set zero. In the $GR$ construction stage, the cells with a high noisy count are preferred which makes that more cells with positive noise are selected into $GR$ in probability and more positive virtual counts generate. Then the noisy utility of $GR$ is generally higher than the real value. In other words, when the noisy $U$ reaches $EU$, sometimes the real $U$ may not. As for using historical data learning, the zero allocation of privacy budget in the level-2 partition makes the (counting) noise scale smaller, the deviation between noisy and real $U$ is smaller so that the (real) ASR increases naturally. By the comparison of G-HS\_Ad (adjusting $EU$ of G-HS point by point to make its ASR value close to that of G-GS) to G-GS, we find out that in the average sense the indexes, WTD, HOP, ANW and DCM using historical data are $3.7\%$, $6.4\%$, $3.9\%$ and $3.7\%$ better than those using real data, respectively.

\subsection{Local Maximum Geocast Reigon}\label{sec:LGR}
In order to improve the compactness of $GR$, we set a local maximum geocast radius $r_{\rm loc}$ by adaptive search when selecting cells. We consider all cells in \emph{Local Maximum Geocast Reigon (LGR)} as a whole, calculate the average distance by weighting with the absolute value of noisy counts, and then estimate the utility of the area. The initial value of $r_{\rm loc}$ is the average distance from the task to the four corners of the cell covering the task, and it increases with the fixed step that equals half the width of the smallest cell in whole domain, until the approximate utility reaches $EU$. The algorithm for finding $LGR$ is shown in Algorithm \ref{alg:R}.

\begin{algorithm}[h]
	\caption{Finding $LGR$}
	\label{alg:R}
	\begin{algorithmic}[1]
		\REQUIRE
		PSD, task $t$
		\ENSURE
		local radius $r_{\rm loc}$\\
		\STATE Find the level-2 cell $c_t$ that covers $t$
		\STATE Comput the distance $r_0$ between $t$ and $c_t$
		\STATE Comfirm the half width $D_{\rm{min}}$ of the minimal cell
		\STATE Compute ${c_t}$\,'s utility $U_t$
		\STATE Init $U=U_t$,~$r_{\rm loc}=r_0$
		\STATE If $U\geq EU$ , return $r_{\rm loc}$\label{state:R_while}
		\STATE $r_{\rm loc}=r_{\rm loc}+D_{\rm{min}}$
		\STATE Add all cells in the $r_{\rm loc}$ area to the set $\widetilde{R}=\left\{c_1,c_2,...,c_n\right\}$
		\STATE $\widetilde{N}=\left(N_1,N_2,...,N_n\right)$ includes the noisy worker counts of cells in $\widetilde{R}$
		\STATE Compute the average distance of workers to $t$ in $r_{\rm loc}$ area, $\bar{d}=\frac{\sum_{i=1}^{n}\left( {\rm distance}(c_i,t)\times\mid N_i \mid \right)}{\sum_{i=1}^{n} \mid N_i \mid}$
		\STATE $N_{\rm{sum}}=\sum_{i=1}^{n}N_i$
		\STATE The average $\overline{AR}$ in $LGR$ is $(1-\bar{d} / MTD)\times MAR$
		\STATE If $N_{\rm{sum}}>0$ then $U=1-(1-\overline{AR})^{N_{\rm{sum}}}$
		\STATE Goto Line \ref{state:R_while}
	\end{algorithmic}
\end{algorithm}

We consider the effect of the $LGR$ trick by performing experiments on NYTaxi (Ta.) dataset, and the results are shown in Fig. \ref{fig:R}.

From the comparisons of R-GS (with LGR trick) to G-GS, and R-HS (with LGR trick) to G-HS, respectively, we observe that applying $LGR$ does not obviously weaken the ASR. However, it performs effectively on other metrics, especially improves $13.9\%$ and $19.3\%$ (WTD), $3.5\%$ and $5.1\%$ (HOP), $5.0\%$ and $6.8\%$ (ANW), $12.3\%$ and $19.1\%$ (DCM), respectively, with the privacy budget of 1.0. After adjusting the $EU$ to make the ASR of the four schemes approximately the same, the advantage of the R-HS is more prominent on the metrics. Compared with R-GS, R-HS improves $5.0\%$, $5.6\%$, $6.8\%$ and $3.0\%$, respectively. In Section \ref{sec:evaluation}, a series of experiments will demonstrate the effects of our budget allocation strategy with partition simulation by historical data learning from multiple perspectives.

\subsection{Break for Nonpositive Neighbor Case}\label{sec:break}

As is mentioned in Section \ref{sec:scoring}, adding noise generates some negative cells (where noisy count of workers is negative) which have no contributions on utility. For this, we set a ''Break'' strategy. Except for the initial cell that covers task $t$, when the alternative neighboring cells are all non-positive, the current $GR$ are returned directly. We focus on the ''Break'' strategy with experiments on the Gowalla (Go.) dataset, and the results are given in Fig. \ref{fig:T}.

It can be seen from the CELL index (number of cells in $GR$)  that with application of  ''Break'' strategy the CELL is reduced by $1.0\%$ without causing a significant decrease in ASR. This indicates that the real utility contribution of so-called negative cell is relatively small in the average sense. Besides, the WTD and HOP are reduced by about $0.4\%$ and $0.3\%$, respectively.

\begin{figure}[htb]
	\centering
	\includegraphics[scale=0.7]{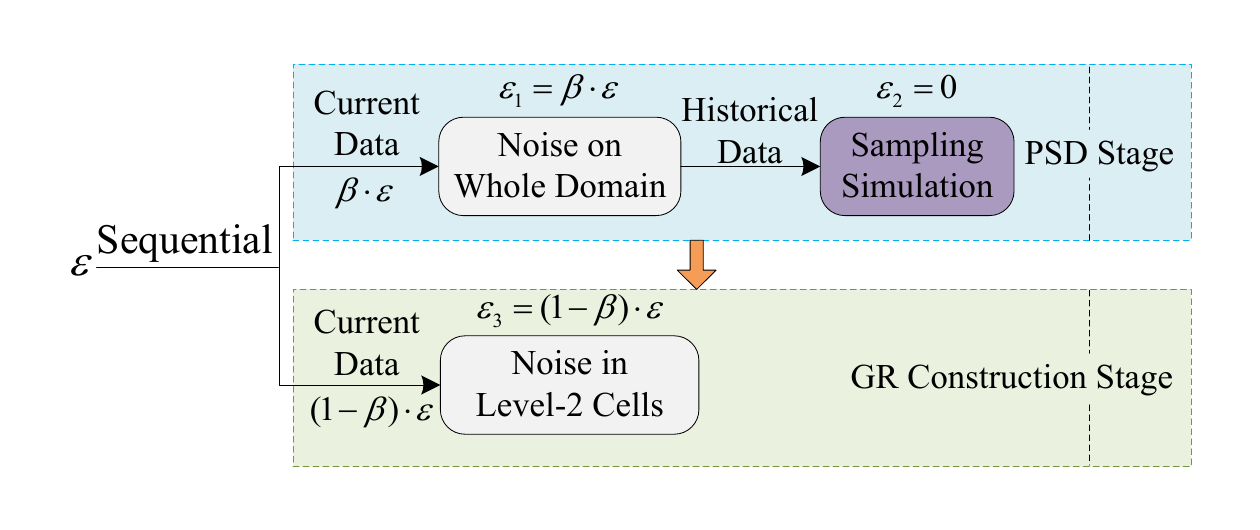}
	\caption{Privacy budget allocation of R-HT scheme}
	\label{fig:budget_allocation}
\end{figure}

\subsection{Privacy Analysis}\label{sec:privacy_analysis}

In this section, we focus on the analysis of privacy budget allocation and protection in the whole system framework. The budget allocation is detailed as in Fig. \ref{fig:budget_allocation}.

Firstly, due to the nature of unbounded DP, we have to allocate a small fraction of privacy budget to noisy count of the total locations in the domain for level-1 partition.

%the totally different original data adopted at two stages, PSD stage and $GR$ construction stage, one can allocate the given full privacy budget $\epsilon$ in parallel composition according to the parallel combination theorem (Theorem \ref{thm:par_comp}). That is, the privacy budget is assigned as $\epsilon$ in PSD stage and also $\epsilon$ in $GR$ construction stage. This mechanism of privacy budget improves greatly the quality of the whole framework.

Secondly, for level-2 partition we employ historical data learning to make sampling simulation of current distribution. Usually the prediction by learning of historical locations has correlation of real-time data no matter whether noises are added on original data. For this, we use the probability distribution determined by the predicted proportion on counts to perform random sampling (i.i.d.), which achieves a nice simulation of current distribution without privacy costs. Such a method helps us learn the overall distribution of locations in the statistical sense. Based on the sampled points (only used for local counting) generated independently, any adversary can not guess which (level-1) cell a specific worker is currently located in.
%The privacy budget $\epsilon_1 = \beta  \epsilon$ for Laplace noises added independently to the everyday (or each periodic) count of workers in the whole domain (for example, in the latest $20$ days or periods) are in parallel, and similarly the budget, $\epsilon_2 = (1-\beta)  \epsilon$, for the noise in each level-1 cell.
%That is, the sampling points in simulation has no relations to the real-time data.
Moreover, the level-2 partition costs no privacy budget.
\begin{figure*}[htbp]
	\centering
	\includegraphics[scale=0.5]{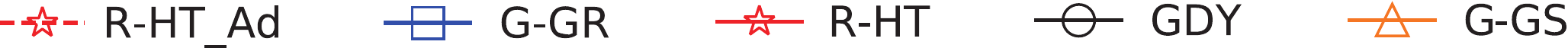}\\
	\subfigure[ASR, Ye.]{
		\includegraphics[height=0.97 in]{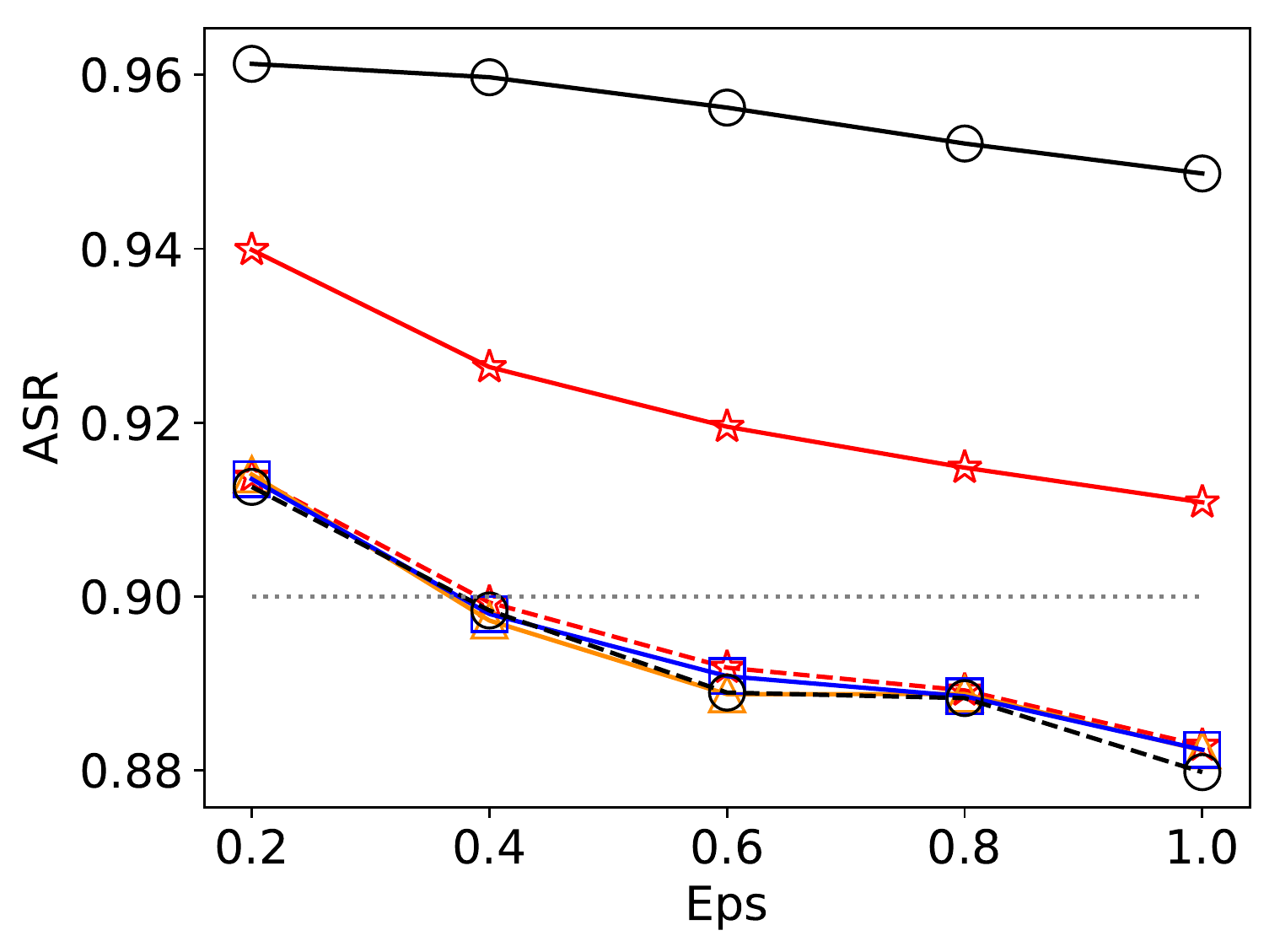}
	}
	\subfigure[WTD, Ye.]{
		\includegraphics[height=0.97 in]{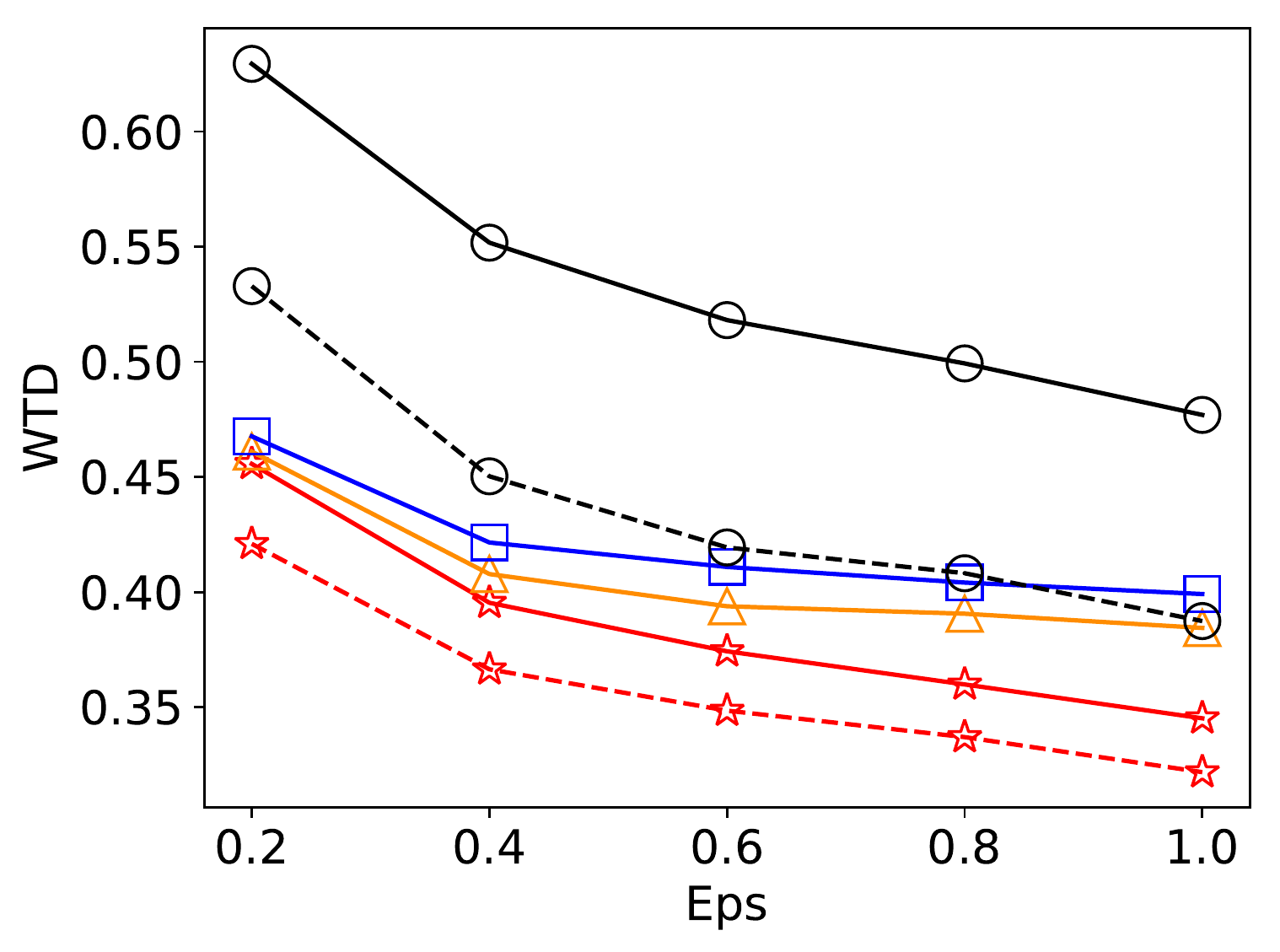}
	}
	\subfigure[HOP, Ye.]{
		\includegraphics[height=0.97 in]{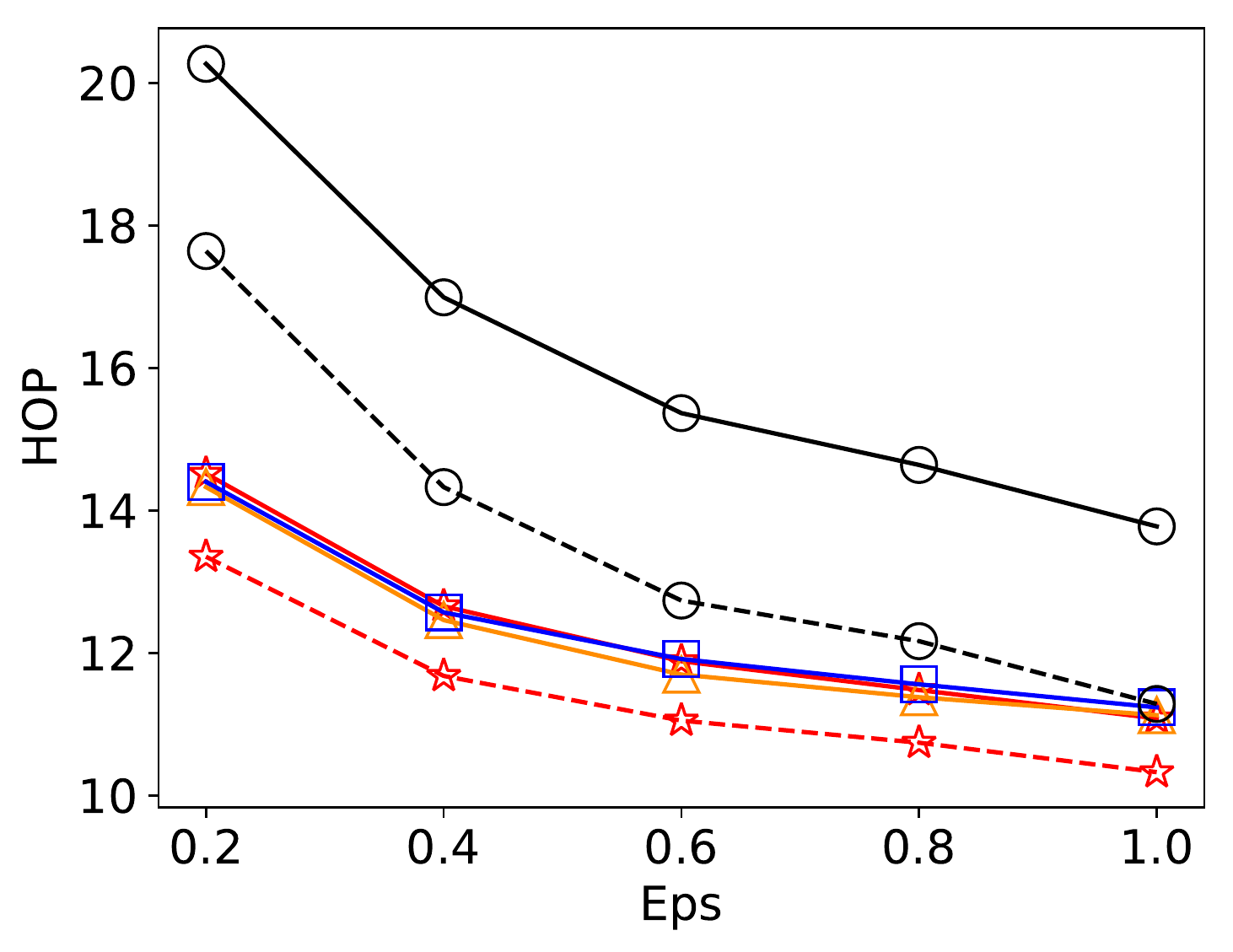}
	}
	\subfigure[ANW, Ye.]{
		\includegraphics[height=0.97 in]{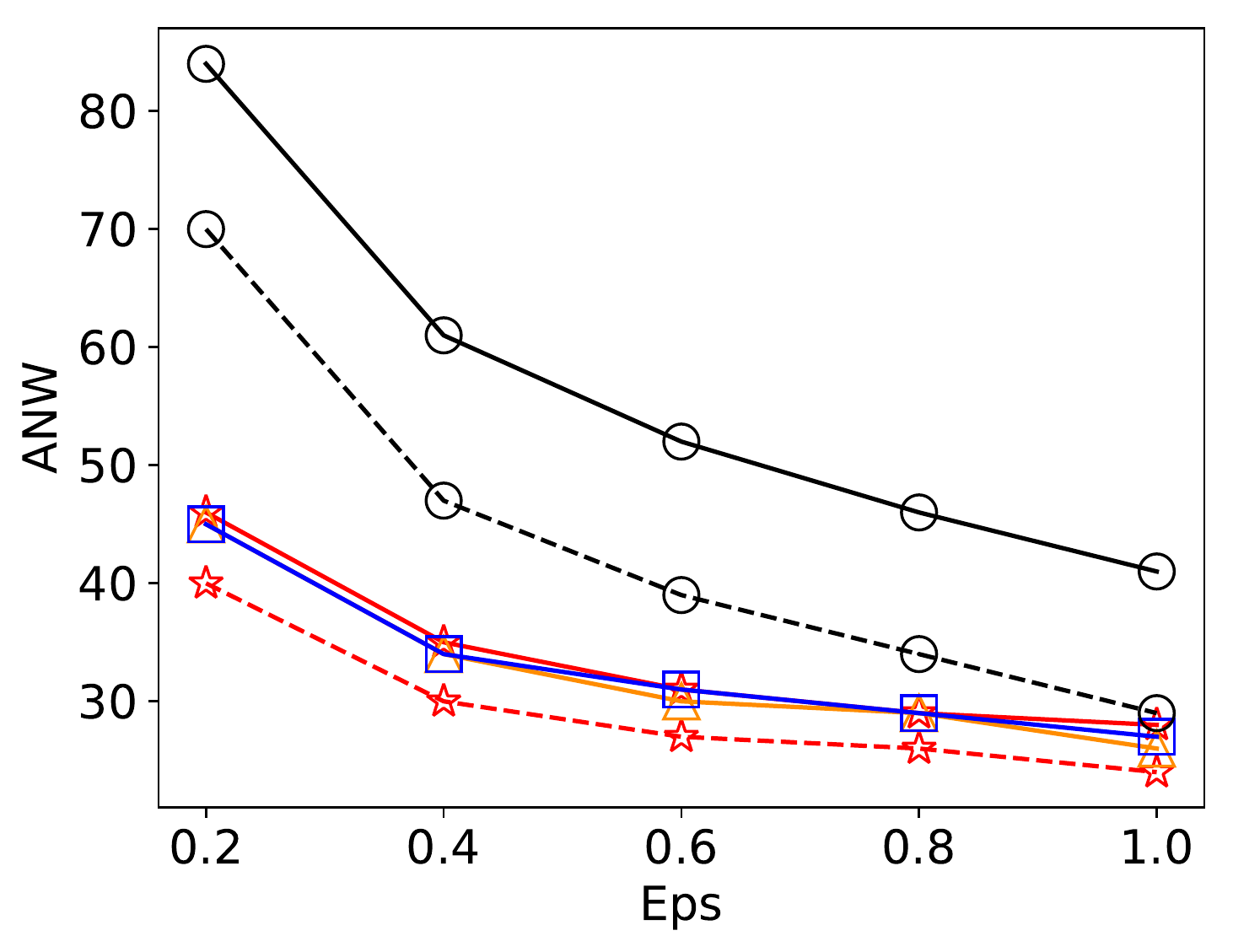}
	}
	\subfigure[DCM, Ye.]{
		\includegraphics[height=0.97 in]{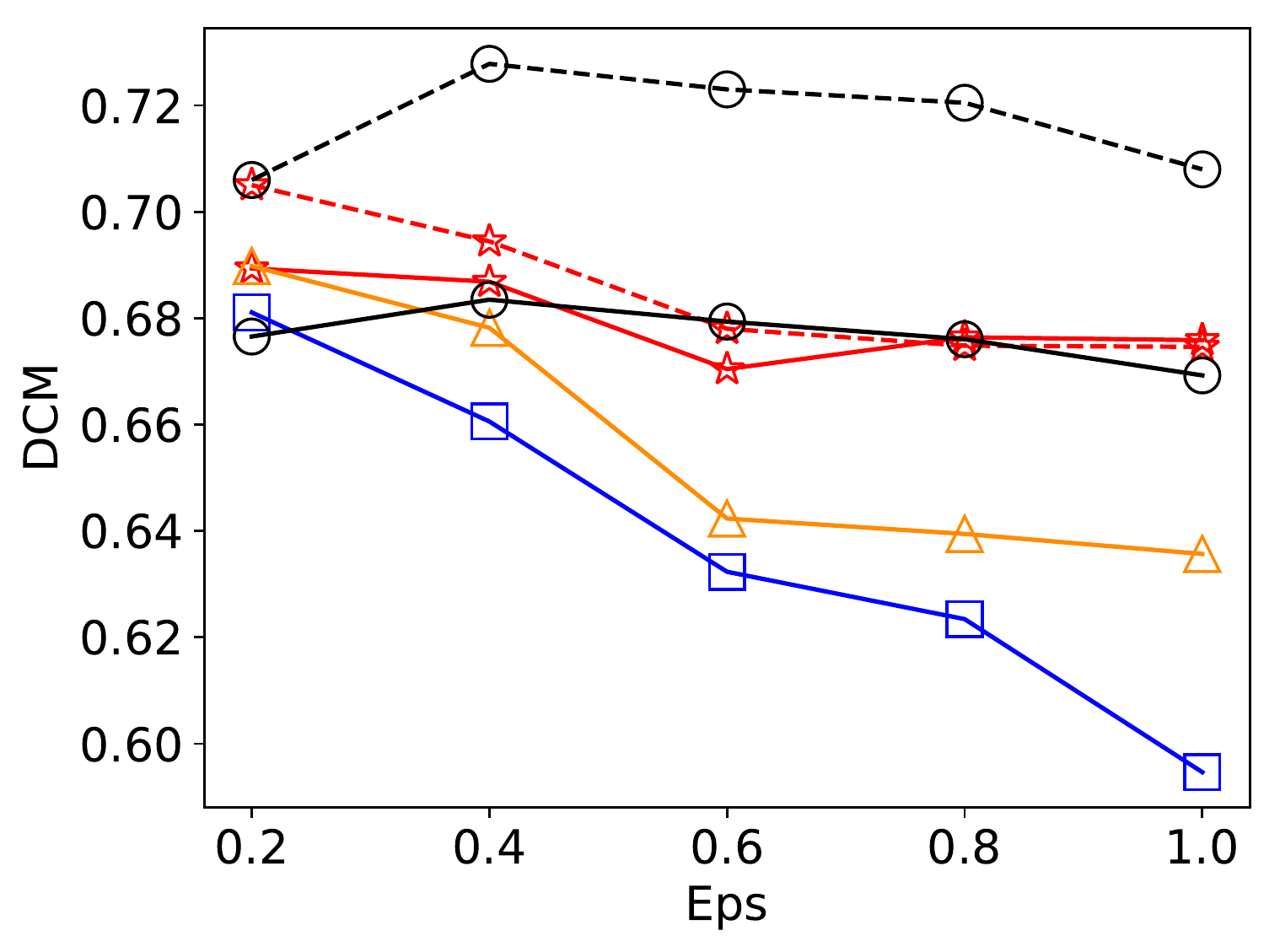}
	}
	\subfigure[ASR, Go.]{
		\includegraphics[height=0.97 in]{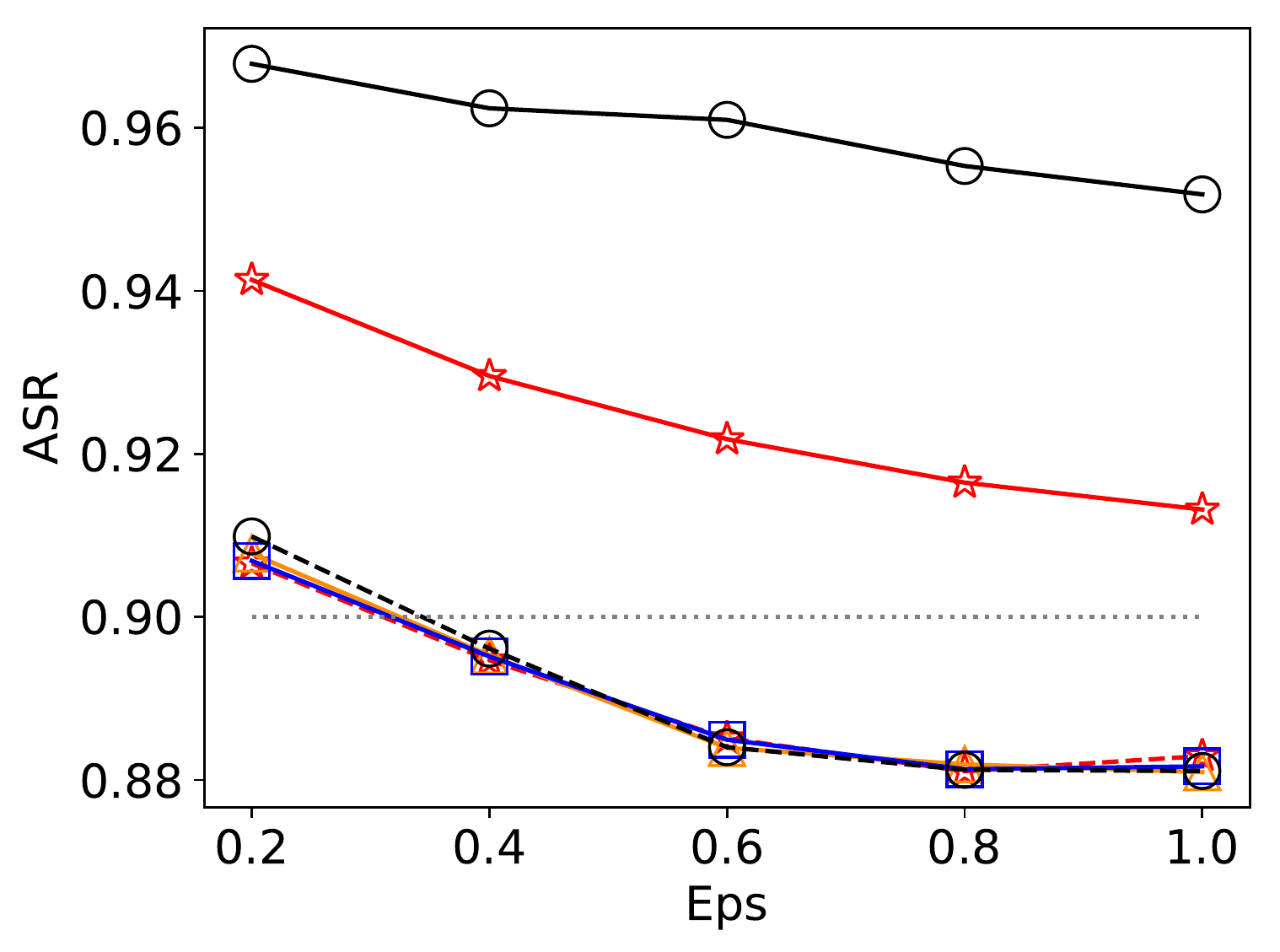}
	}
	\subfigure[WTD, Go.]{
		\includegraphics[height=0.97 in]{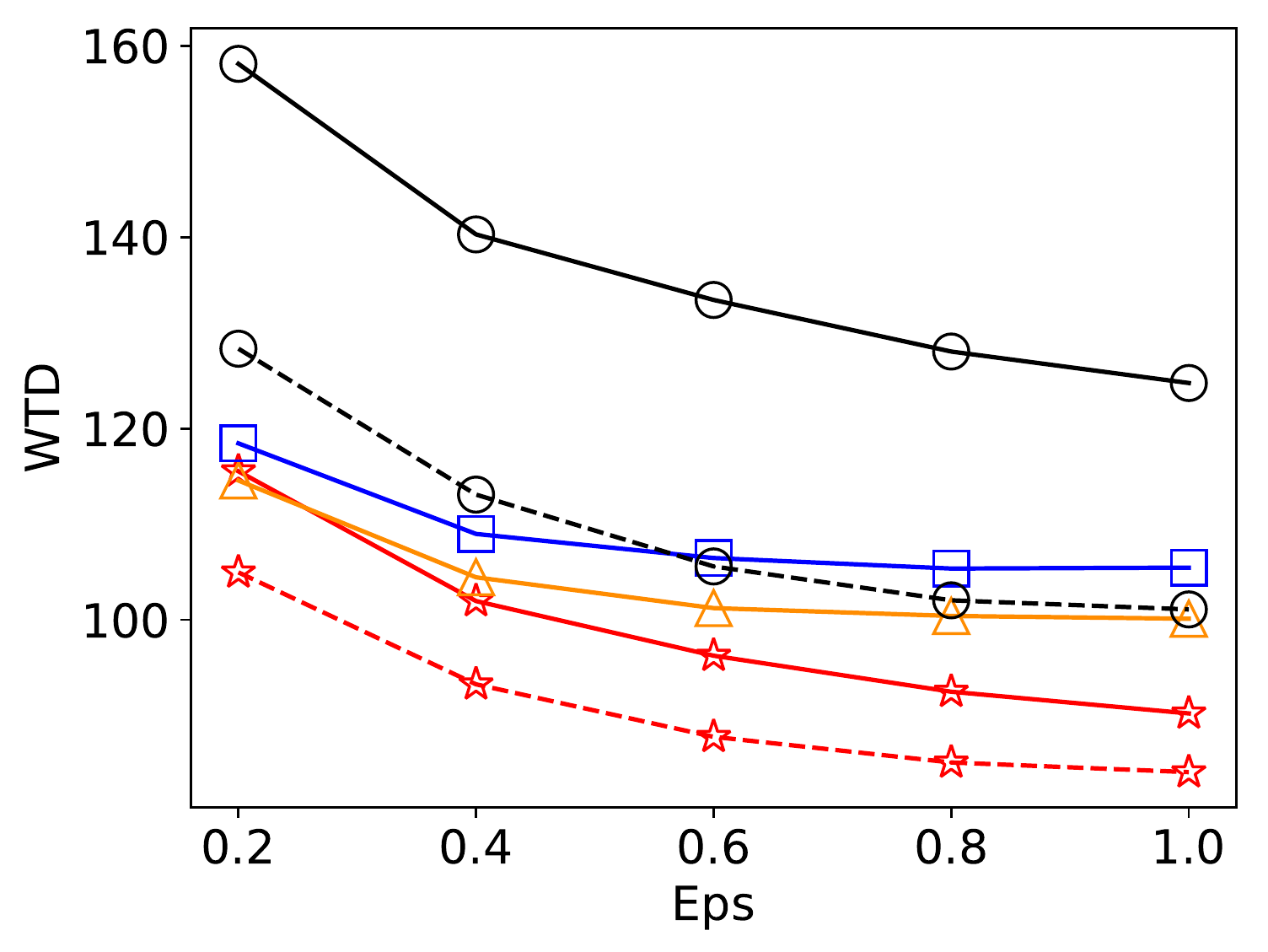}
	}
	\subfigure[HOP, Go.]{
		\includegraphics[height=0.97 in]{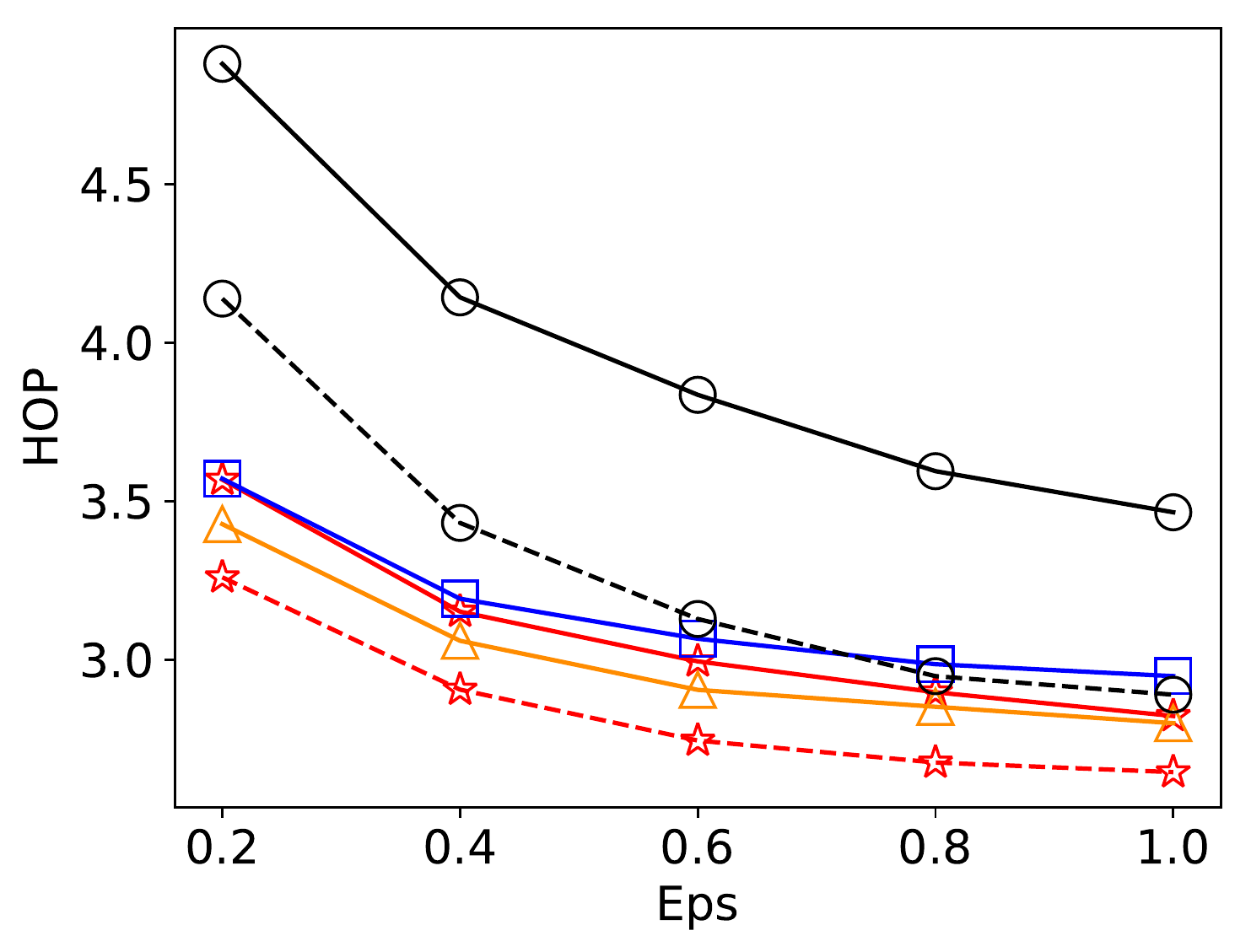}
	}
	\subfigure[ANW, Go.]{
		\includegraphics[height=0.97 in]{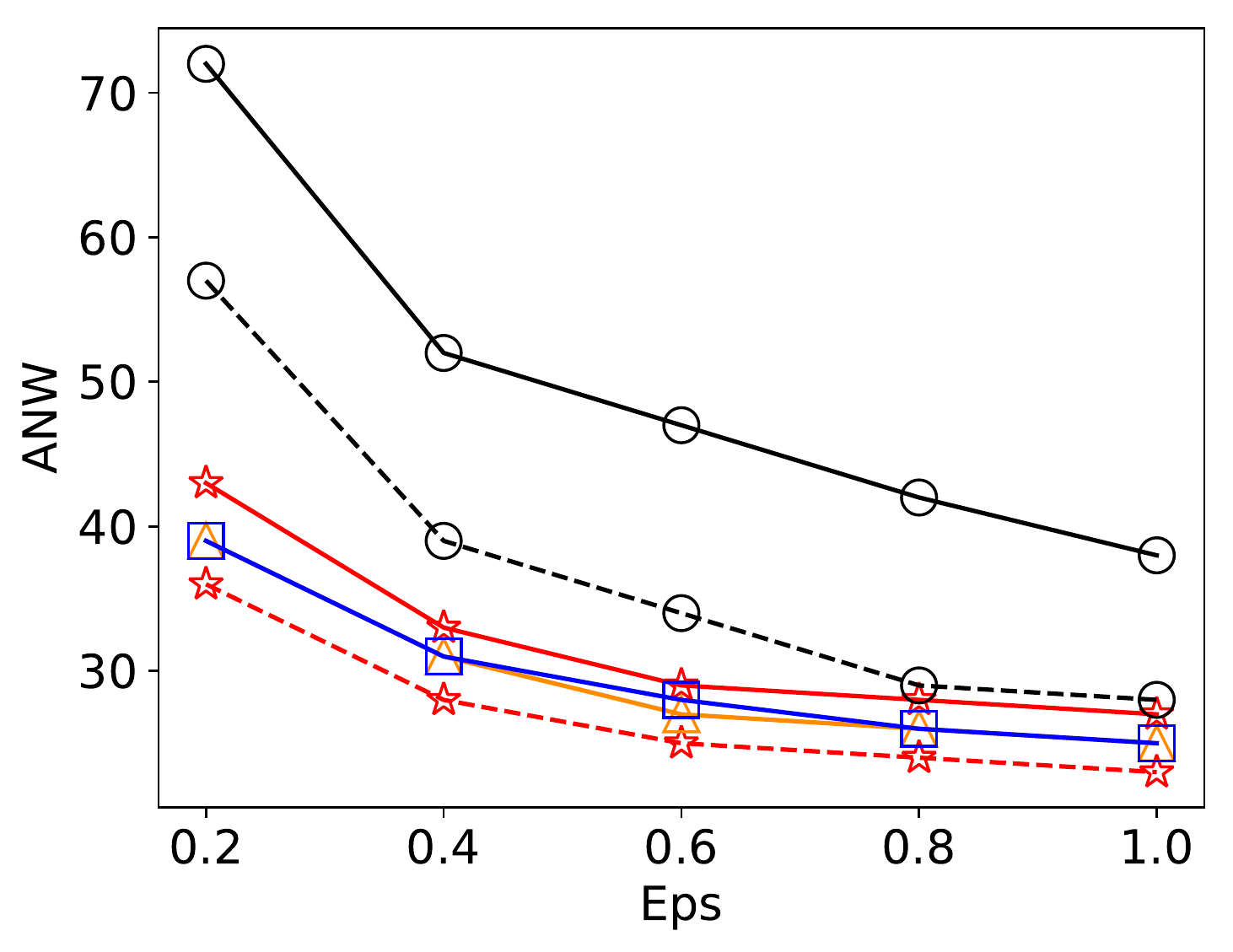}
	}
	\subfigure[DCM, Go.]{
		\includegraphics[height=0.97 in]{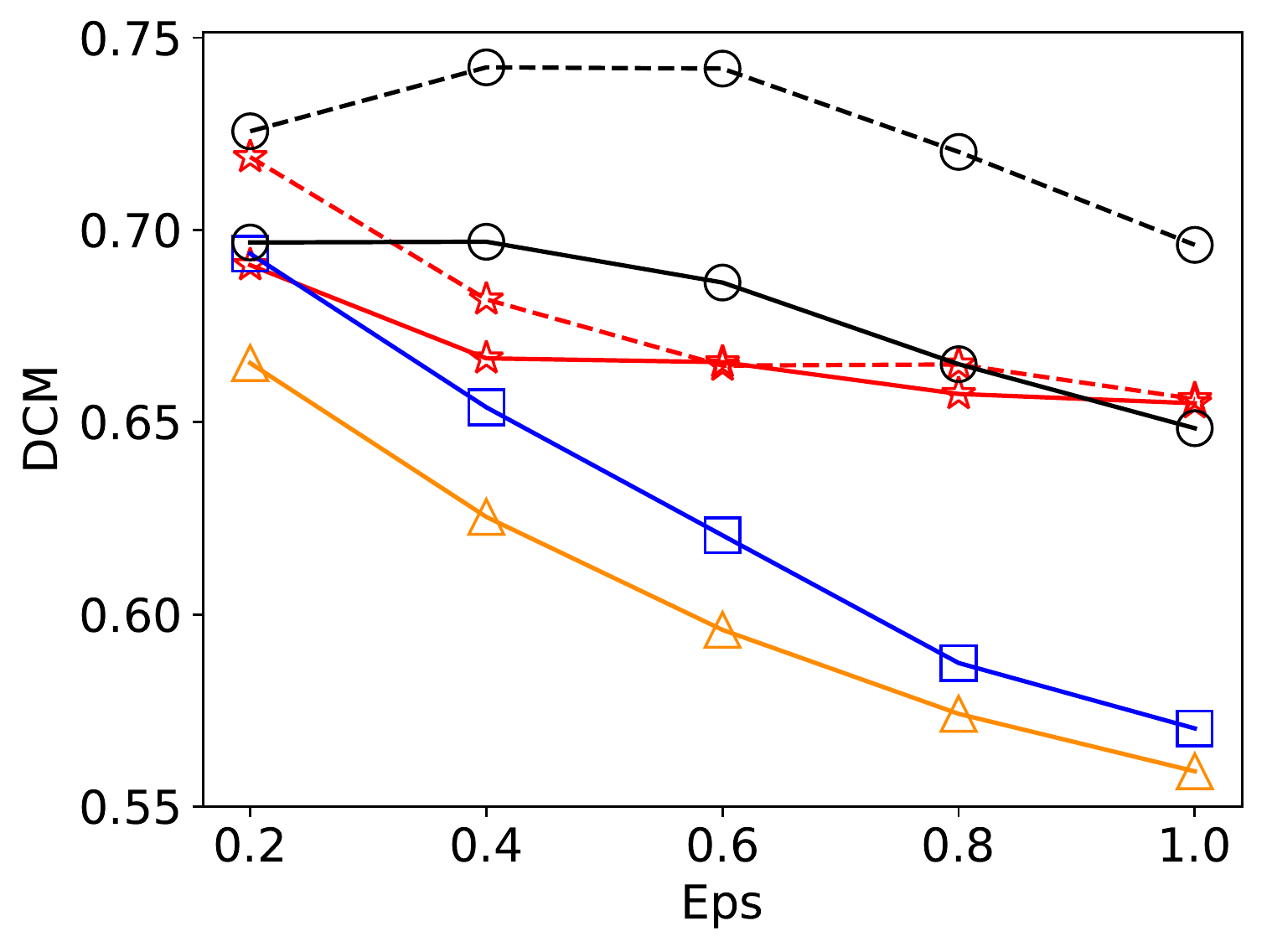}
	}
	\subfigure[ASR, Ta.]{
		\includegraphics[height=0.97 in]{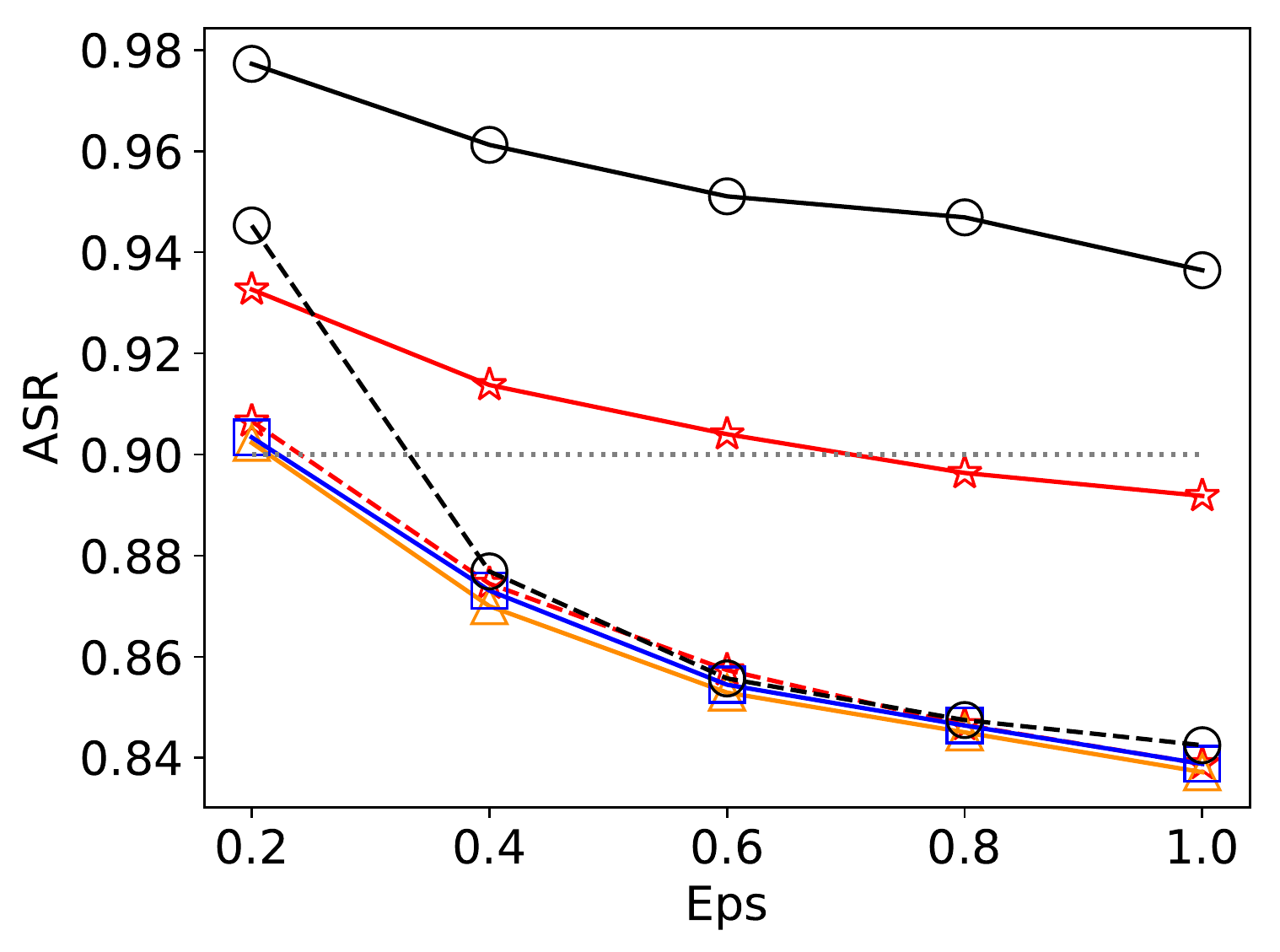}
	}
	\subfigure[WTD, Ta.]{
		\includegraphics[height=0.97 in]{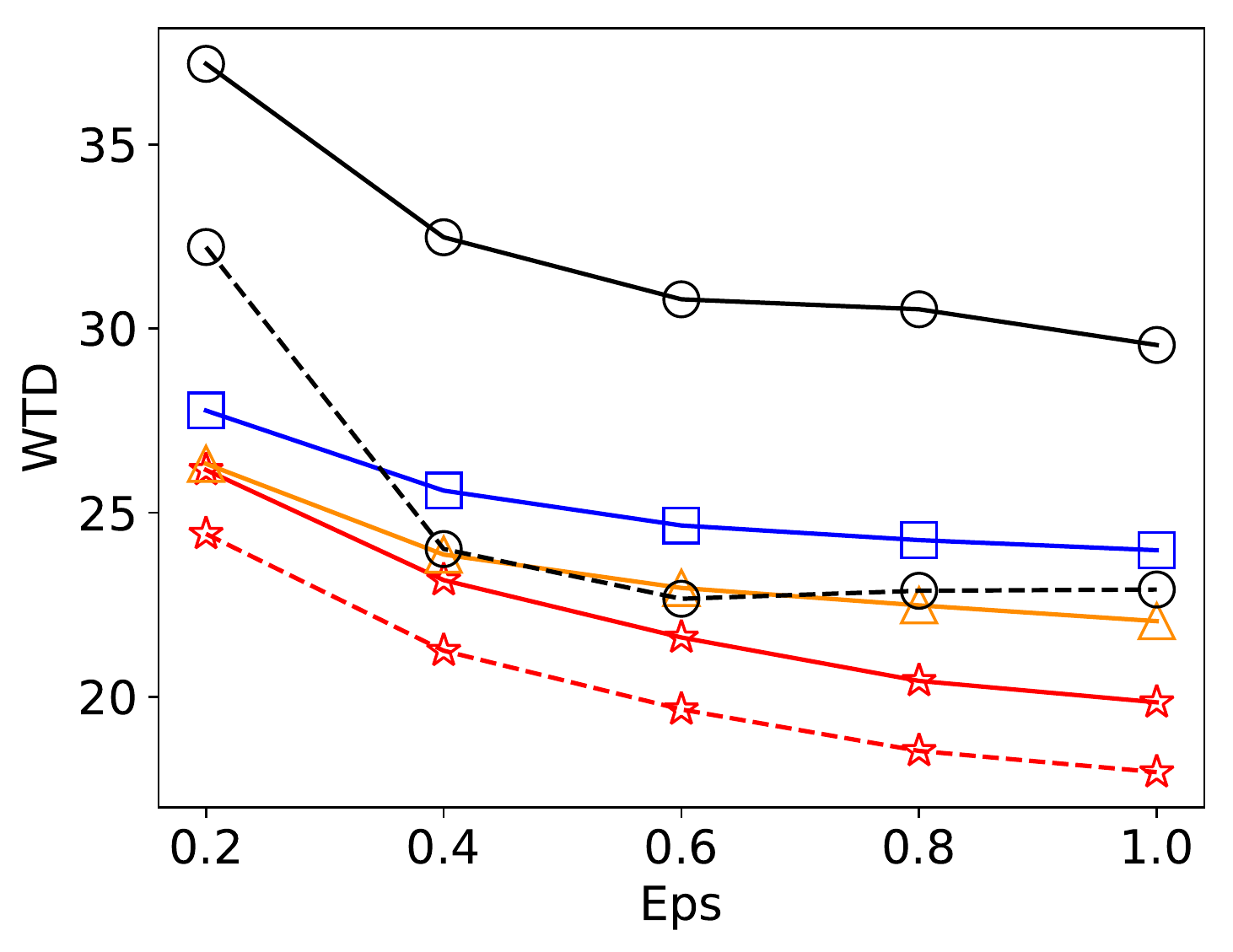}
	}
	\subfigure[HOP, Ta.]{
		\includegraphics[height=0.97 in]{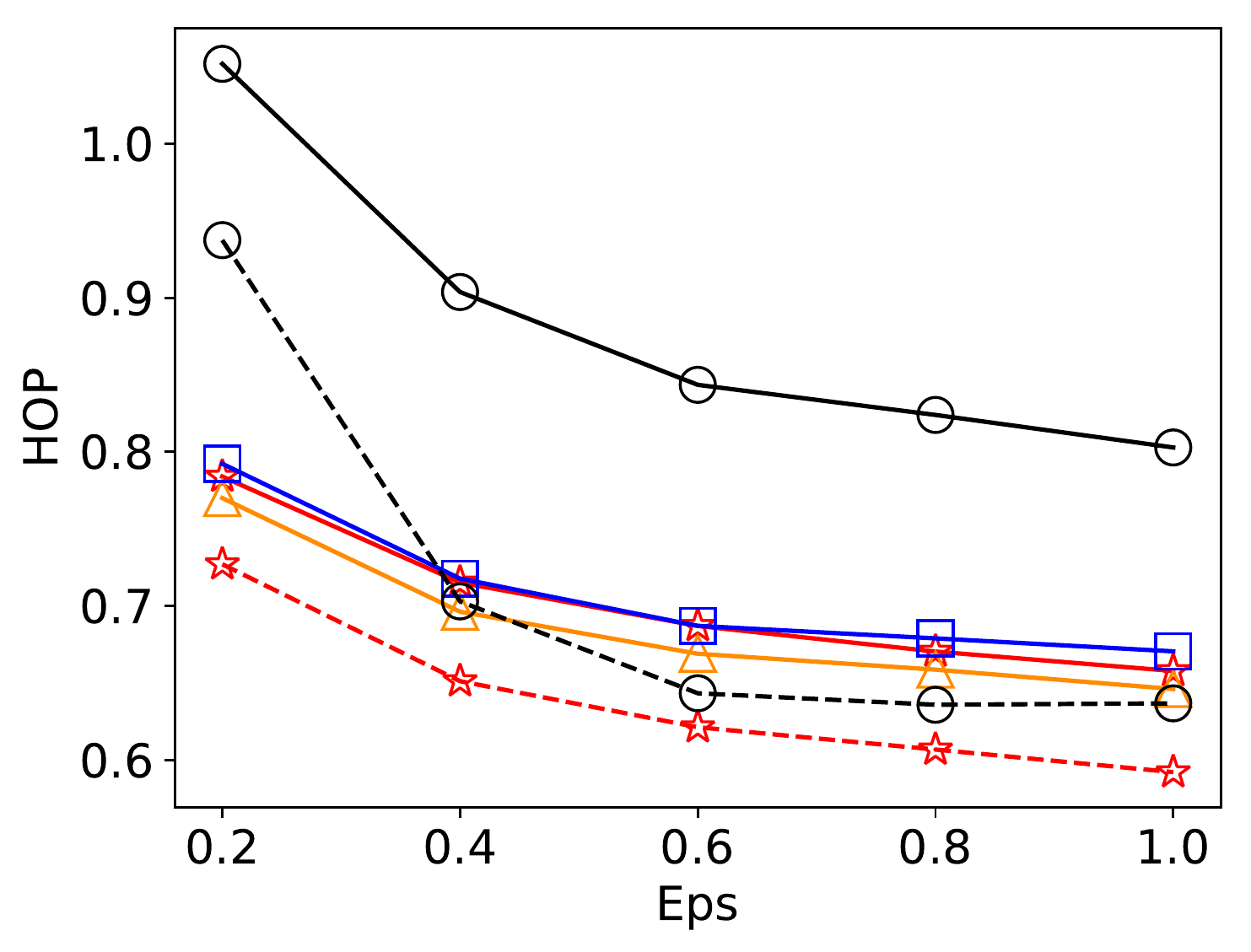}
	}
	\subfigure[ANW, Ta.]{
		\includegraphics[height=0.97 in]{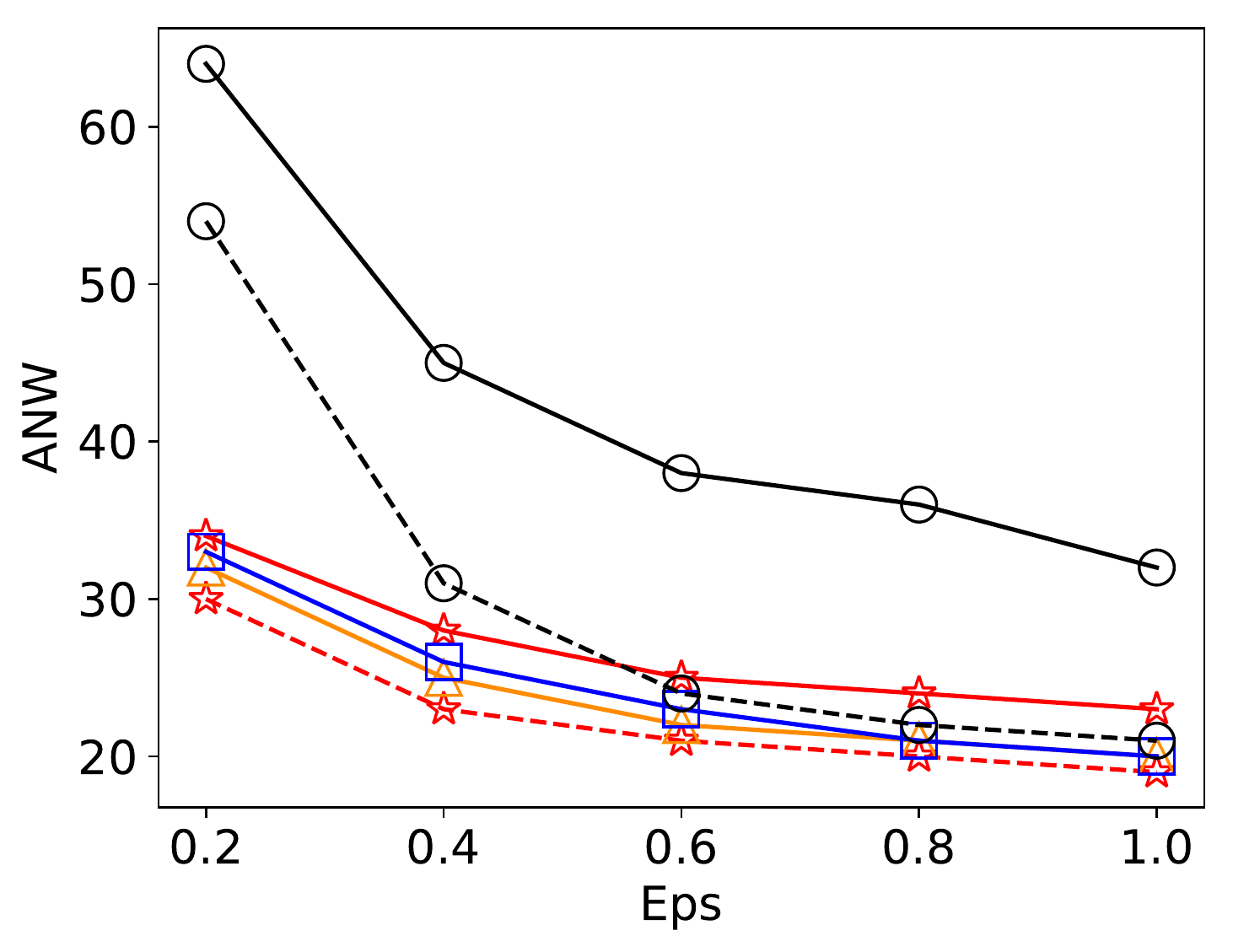}
	}
	\subfigure[DCM, Ta.]{
		\includegraphics[height=0.97 in]{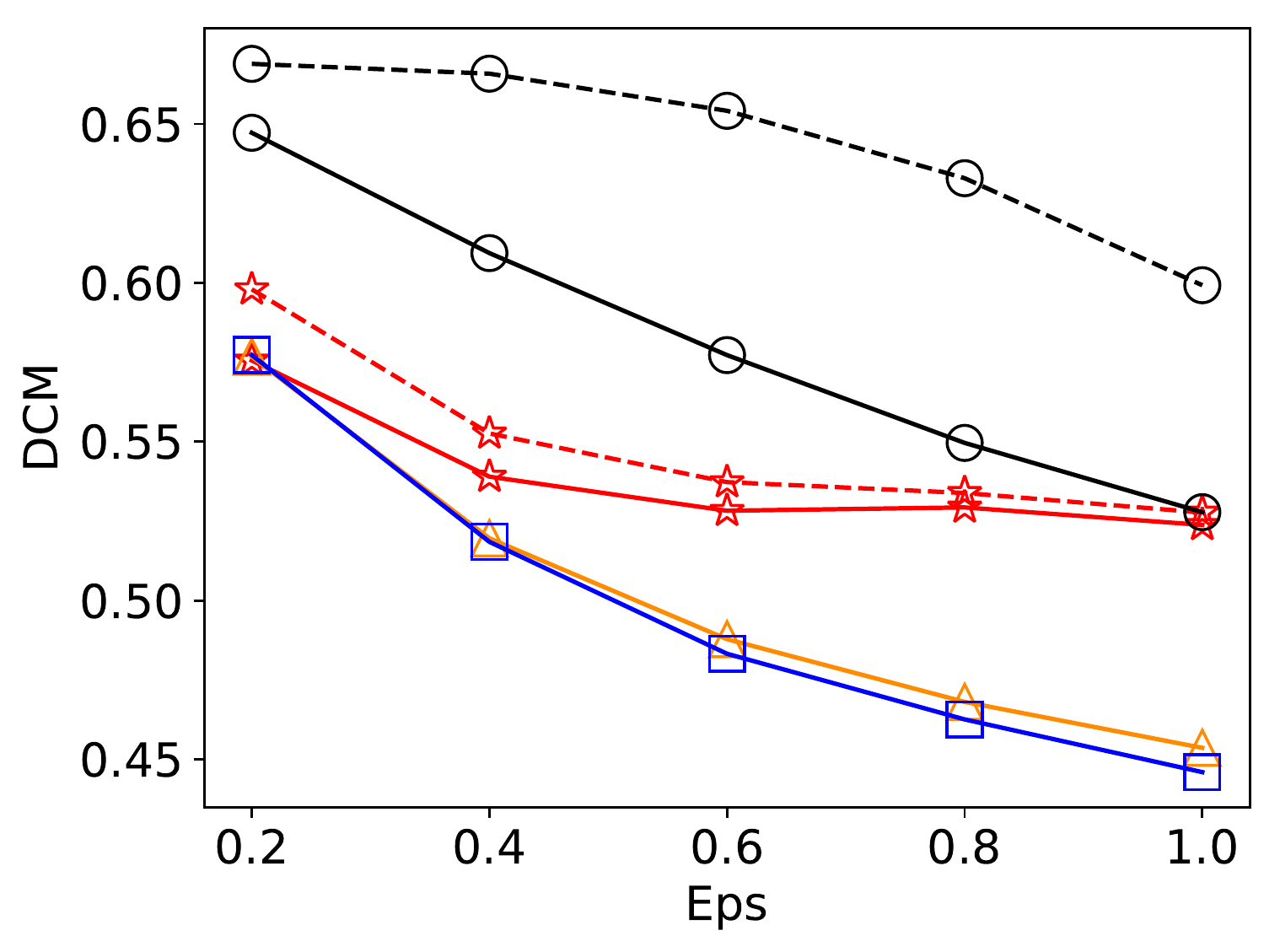}
	}
	\caption{Performance of geocast schemes with varying privacy budget $\epsilon$}
	\label{fig:Eps}
\end{figure*}

Thirdly, in the $GR$ construction stage, real-time data is imported into the partitioned grid, and Laplace noise of main privacy budget is added to each level-2 cell in parallel.

Finally, the involved learning on historical data, ``Break'' strategy and $LGR$ operations suit the characteristics of post-processing (Theorem \ref{thm:pos-pro}), and the whole system scheme satisfies $\epsilon$-DP protection.

\section{Performance Evaluation}\label{sec:evaluation}
In this section, we mainly present an experimental evaluation of our R-HT scheme. In particular, we compare R-HT with previous schemes, like G-GR.
\subsection{Experimental Methodology}\label{sec:dataset}
\textbf{Experimental comparison.} Historical data learning technique has been proved to provide relatively high predictive accuracy \cite{JHC18,LCZ18}. In order to verify the system quality of the proposed R-HT scheme, we carry out a series of experiments to compare it with the GDY \cite{QYL13}, G-GR \cite{TGF17} and G-GS. Indeed, the G-GS scheme is modified from G-GR, where the utility rule is replaced by quality scoring for cell selections.

These privacy frameworks are achieved based on the notion of unbounded DP. Then the sensitivity of workers' count in each grid (or region) is $1$, that is, while the maximum increase or decrease is 1 for the total number of workers, such a change may occur in any cell \cite{LLS16}. Compared with unbounded DP, the sensitivity in bounded DP case is $2$, and the increase of noise scale would significantly affect the success rate of task assignment.

\textbf{Selection of Datasets.} We use three real datasets: NYTaxi (NYC's Taxi Trip Data), Gowalla and Yelp.

NYTaxi is a New York taxi location dataset. We extract taxi pickup positions in New York City during 21 days, May 1 to 21, 2013. The first 20 days are used as historical data, and 27,165 positions on May 21 as real-time data. Each pickup position is modeled as a worker position in \emph{spatial crowdsourcing (SC)}. Since most of the taxi pickup positions are distributed on the city's main roads, in order to better simulate the actual positions of workers, we randomly and uniformly blur each position into a circle centered at the current position and with the radius (\emph{Blur Radius, BR}) of 80 meters.

Gowalla is a social network check-in dataset. We extract the check-in locations for 42 days from September 5 to October 16, 2010. Due to the large geographic span of data, we reduce the geographical distance by a ratio of $1:280$. Every two days are regarded as a period, and the last period includes 6,736 location points for real-time data. Each restaurant location is modeled as a worker position in SC, and the BR is 250 meters.

Yelp corresponds to some data of the Phoenix area of Arizona. We take location data from March 2014 to August 2017 and assign 2 months as a period. The last period with 17,730 locations was used as real-time data. We also model the location of each restaurant as a worker position in SC, BR is 600 meters.

\newcommand{\tabincell}[2]{\begin{tabular}{@{}#1@{}}#2\end{tabular}}
\begin{table}[htbp]\renewcommand{\arraystretch}{1.5}
	\caption{Parameters of Datasets}\label{tlb:datasets}
	\centering
	\begin{tabular}{|c|c|c|c|c|}
		\hline
		Name&\tabincell{c}{Historical\\Locations}&\tabincell{c}{Real-time\\Workers}&MTD/m&Tasks\\
		\hline
		NYTaxi~(Ta.)&841080&27165&300&\multirow{3}*{2000}\\
        \cline{1-4}
		Gowalla~(Go.)&133771&6736&1200&~\\
		\cline{1-4}
		Yelp~(Ye.)&363330&17730&3000&~\\
		\hline
	\end{tabular}
\end{table}

%We can set scenes for the above three datasets respectively. For example, NYTaxi can be regarded as a taxi ordering scene in the downtown area. The maximum distance for taxi drivers to take orders is 500 m. $MTD$ is 90\% of $MCD$ which is determined according to Eq. (\ref{eq:MCD}), calculated in 300 m. Considering Gowalla as a takeaway booking scenario, the maximum order distance is 2 km, and the $MTD$ is calculated to be 1.2 km. Yelp is considered to be a car repair scenario. The maximum order distance is 5 km, so $MTD$ is 3 km. The relevant parameters of the dataset are shown in Table \ref{tlb:datasets}.

We can set scenes for the above three datasets separately. NYTaxi dataset can be regarded as a taxi ordering scene in the downtown. The maximum distance for taxi drivers to perform orders is 500 m. $MTD$ is 300 m, usually equal to $90\%$ of $MCD$ which is determined by Eq. (\ref{eq:MCD}). For Gowalla, we can consider a takeaway booking scenario in which the maximum order distance is 2 km, and then the $MTD$ is 1.2 km. Yelp can be a model of an auto repair scenario. The maximum order distance is 5 km and $MTD$ is 3 km. Besides, 2,000 positions are randomly and uniformly selected from the worker positions as task points. The relevant parameters of the datasets are shown in Table \ref{tlb:datasets}.

In our experimental settings, privacy budget $\epsilon \in \{0.2,~0.4,~0.6,~0.8,~1.0 \}$, expected utility $EU \in \{0.6,~0.7,~0.8,~0.9\}$, and maximum task acceptance probability $MAR \in \{0.05,~0.1,~0.15,~0.2,~0.25\}$. The default values for $\epsilon$, $EU$ and $MAR$ are set to 0.5, 0.9 and 0.1, respectively. The functions $f_s$ and $f_d$ involved in quality scoring function are linear mappings to $[1,10]$ from the interval between minimum and maximum areas of cells, and between the half
length of diagonal line in the minimal cell and $MTD$, respectively.

The utility loss caused by DP can be seen more intuitively by performing a non-privacy scheme, which constructs $GR$ by selecting workers closest to the task one by one within the $MTD$ reigon. The HOP value of $GR$ is defined as the distance between the two farthest workers. As for GDY and G-GR, in order to avoid the influence of the randomness of noise on partitioning, we carry out the whole process 50 times to obtain 50 groups of results and get their average values. For R-HT, the simulation process of grids partition is independently executed 10 times, which results in 10 groups of partitioned grids. The real-time location counting noise in each level-2 cell is randomly added for 20 times separately, and then a total of 200 PSD snapshots are generated. For all schemes, 2000 single-task assignments are performed on each map, regardless of possible conflicts between task assignments, so as to obtain stable results of performance evaluation of each scheme under this scenario (single worker requested for single task). Next, we investigate the effects of varying privacy budget, $MAR$ and expected utility, respectively, and also evaluate $LGR$-based heuristics and the running time of the schemes for feasibility.
\begin{figure*}[htp]
	\centering
	\includegraphics[scale=0.5]{fig8910_menu.pdf}\\
	\subfigure[ASR, Ye.]{
		\includegraphics[height=0.97 in]{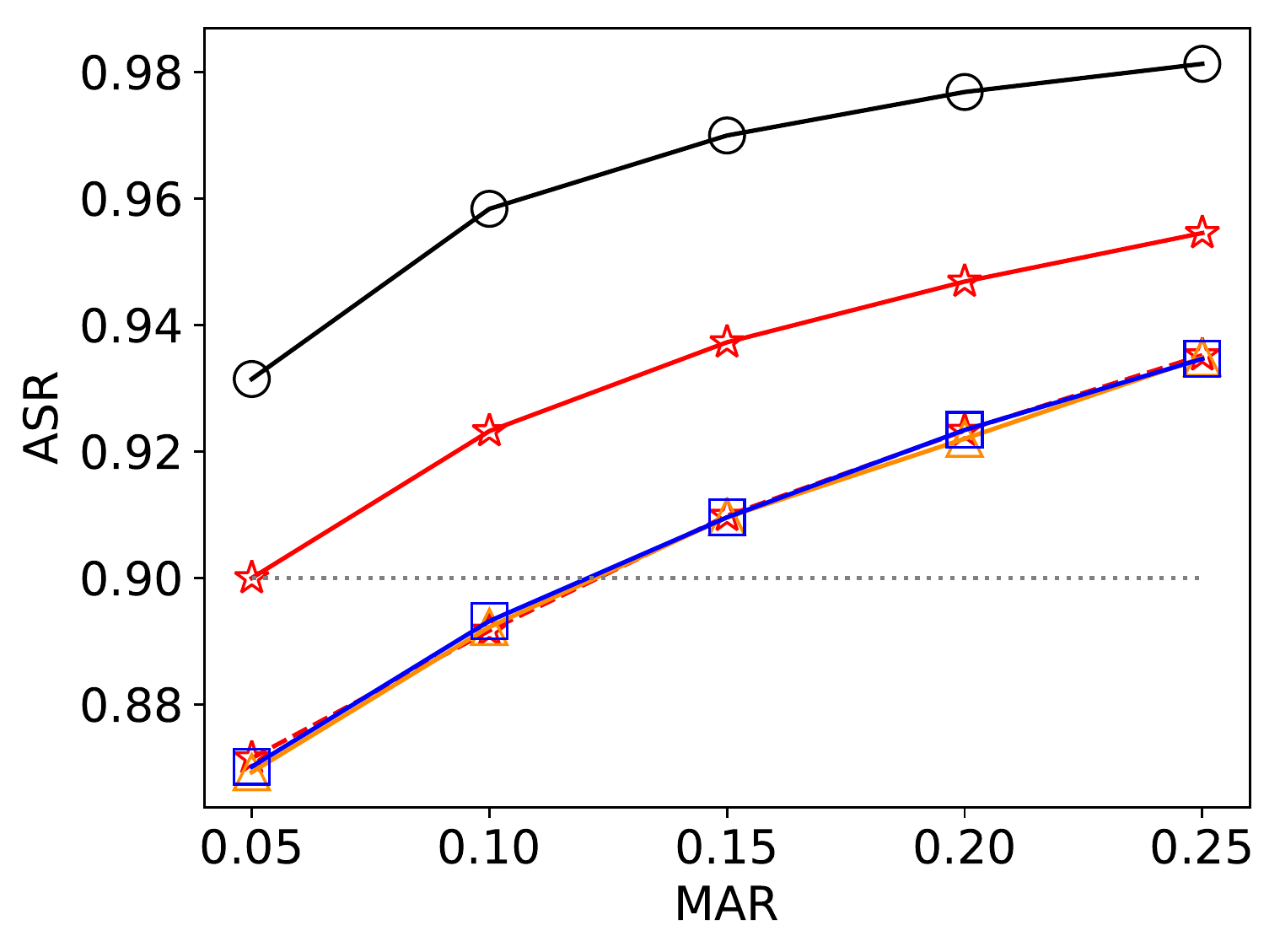}
	}
	\subfigure[WTD, Ye.]{
		\includegraphics[height=0.97 in]{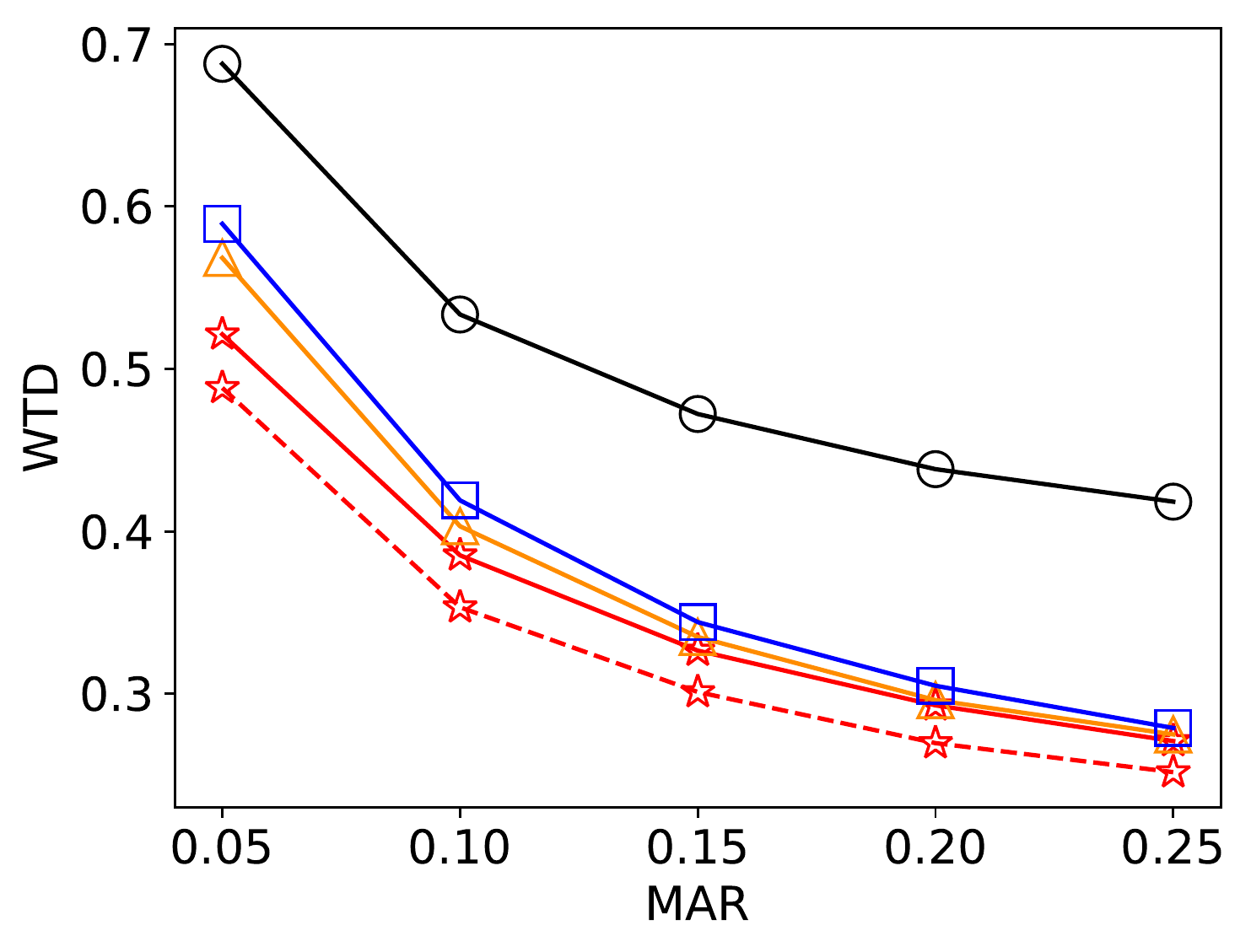}
	}
	\subfigure[HOP, Ye.]{
		\includegraphics[height=0.97 in]{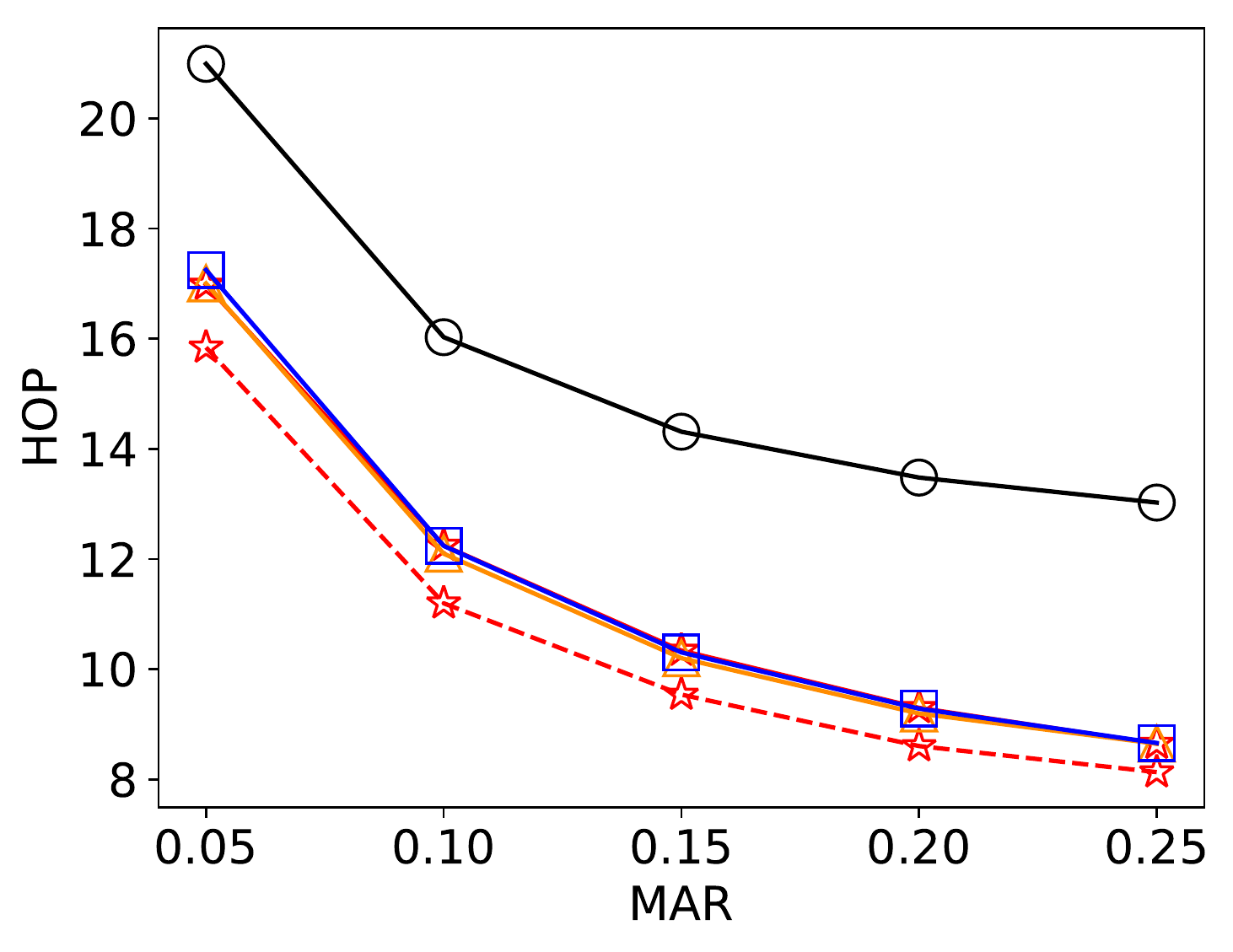}
	}
	\subfigure[ANW, Ye.]{
		\includegraphics[height=0.97 in]{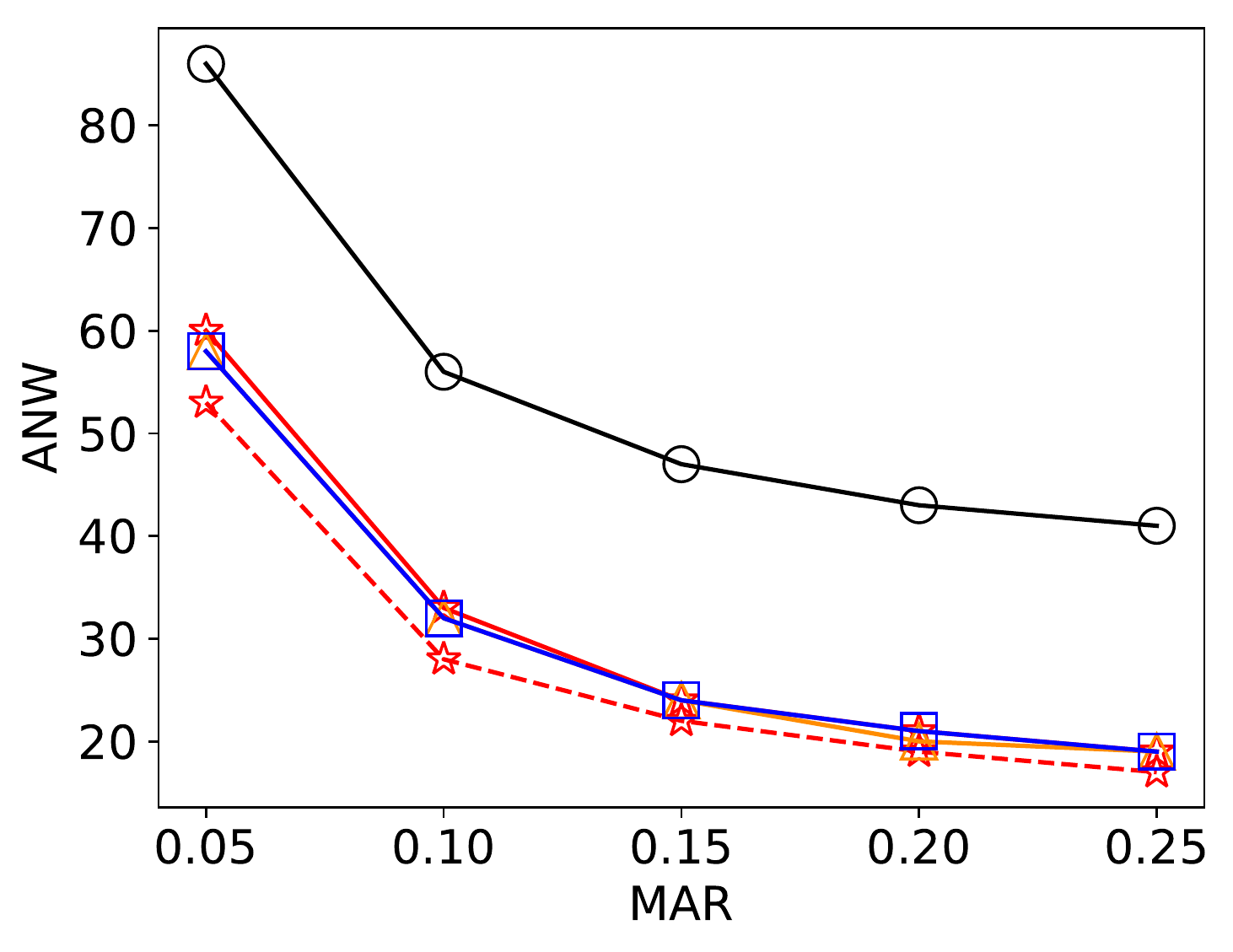}
	}
	\subfigure[DCM, Ye.]{
		\includegraphics[height=0.97 in]{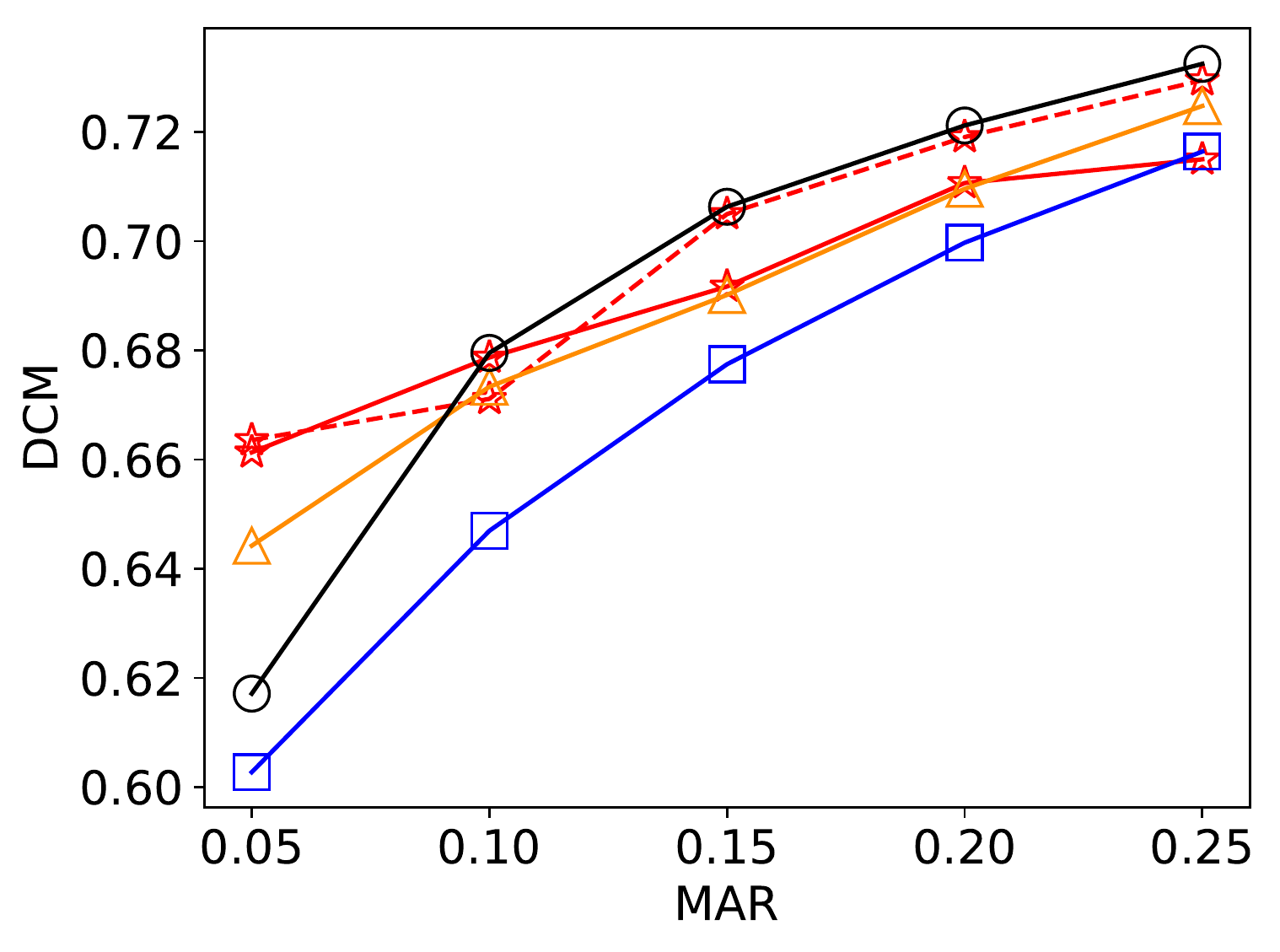}
	}
	\subfigure[ASR, Go.]{
		\includegraphics[height=0.97 in]{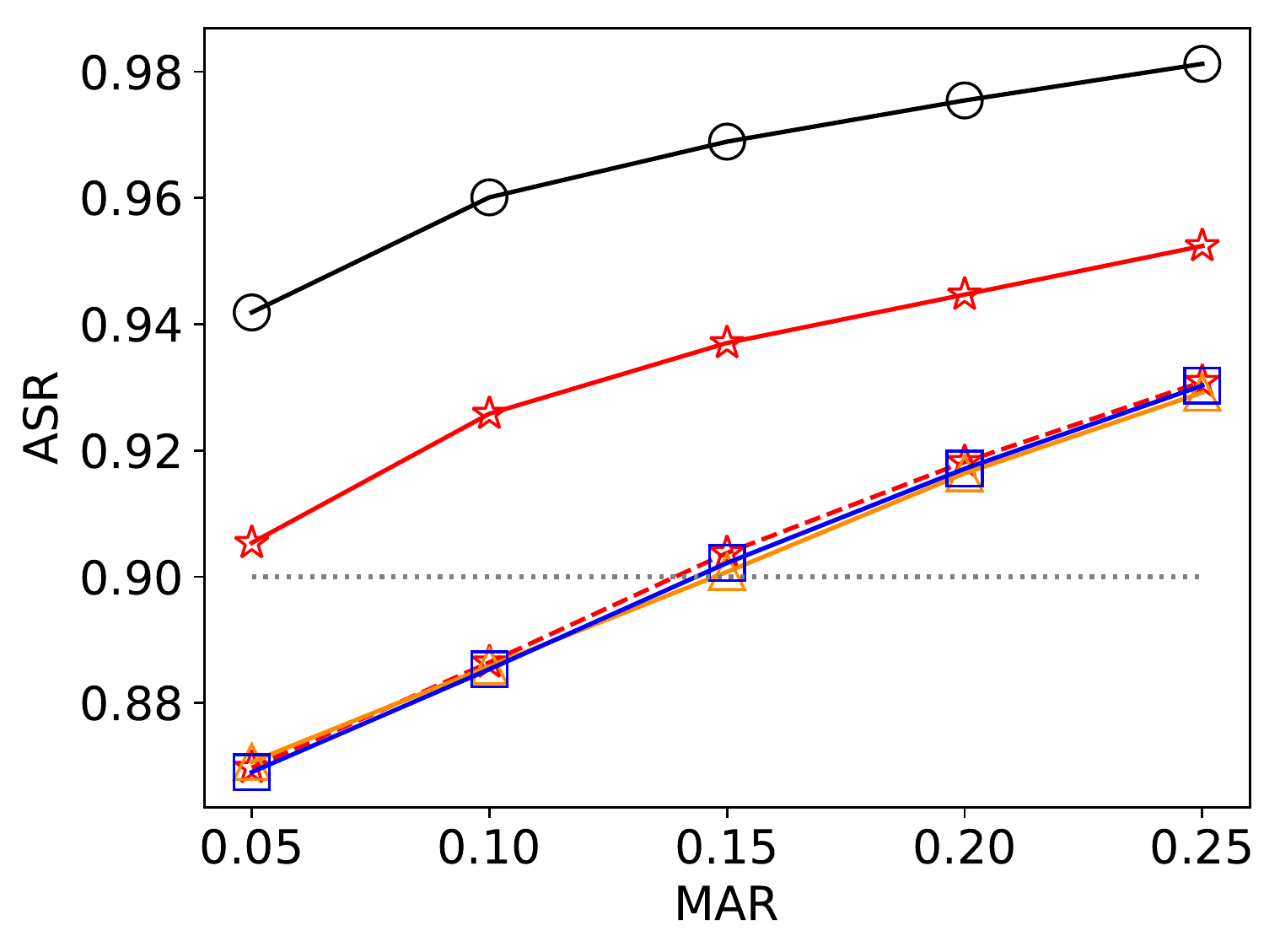}
	}
	\subfigure[WTD, Go.]{
		\includegraphics[height=0.97 in]{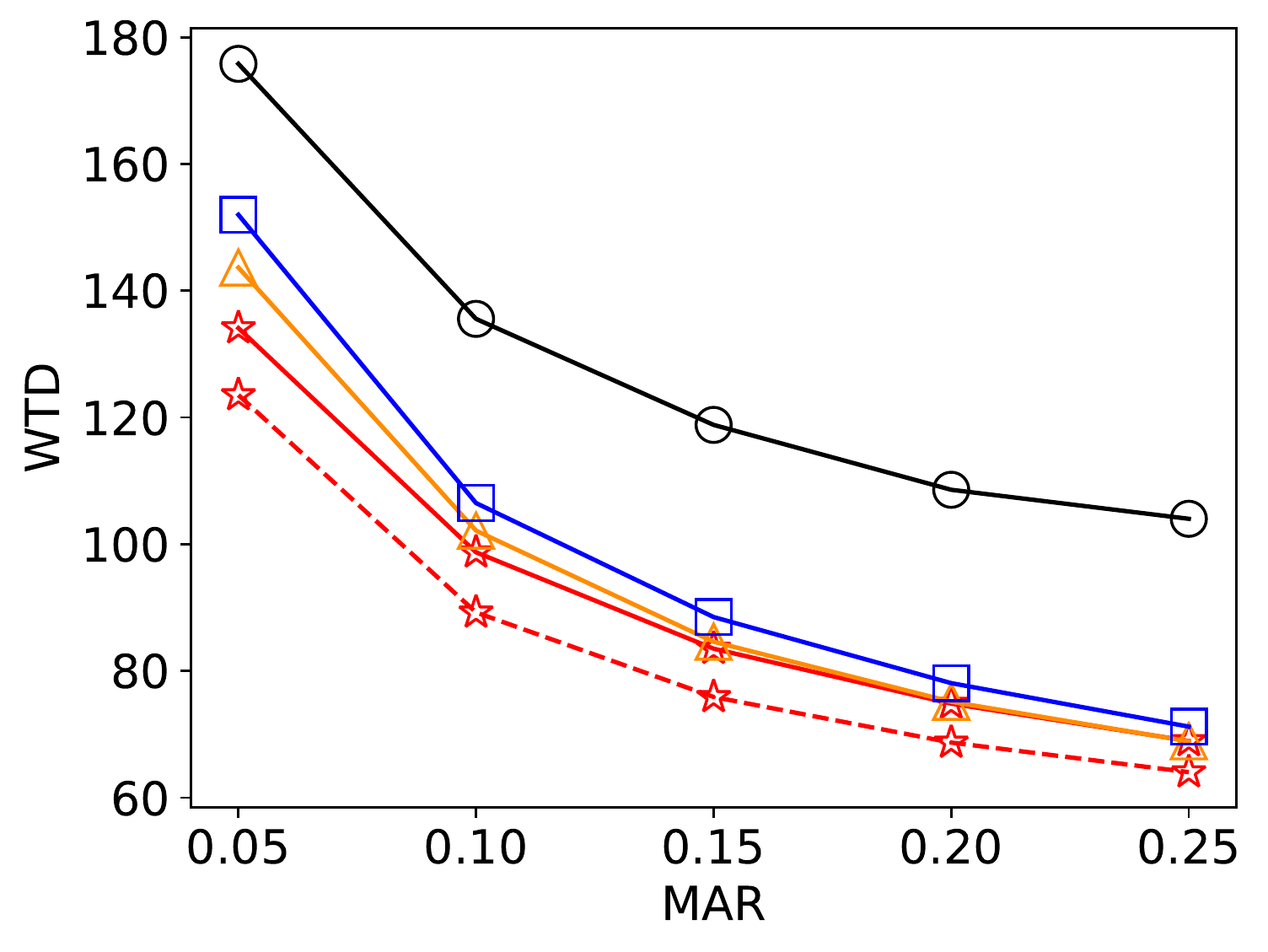}
	}
	\subfigure[HOP, Go.]{
		\includegraphics[height=0.97 in]{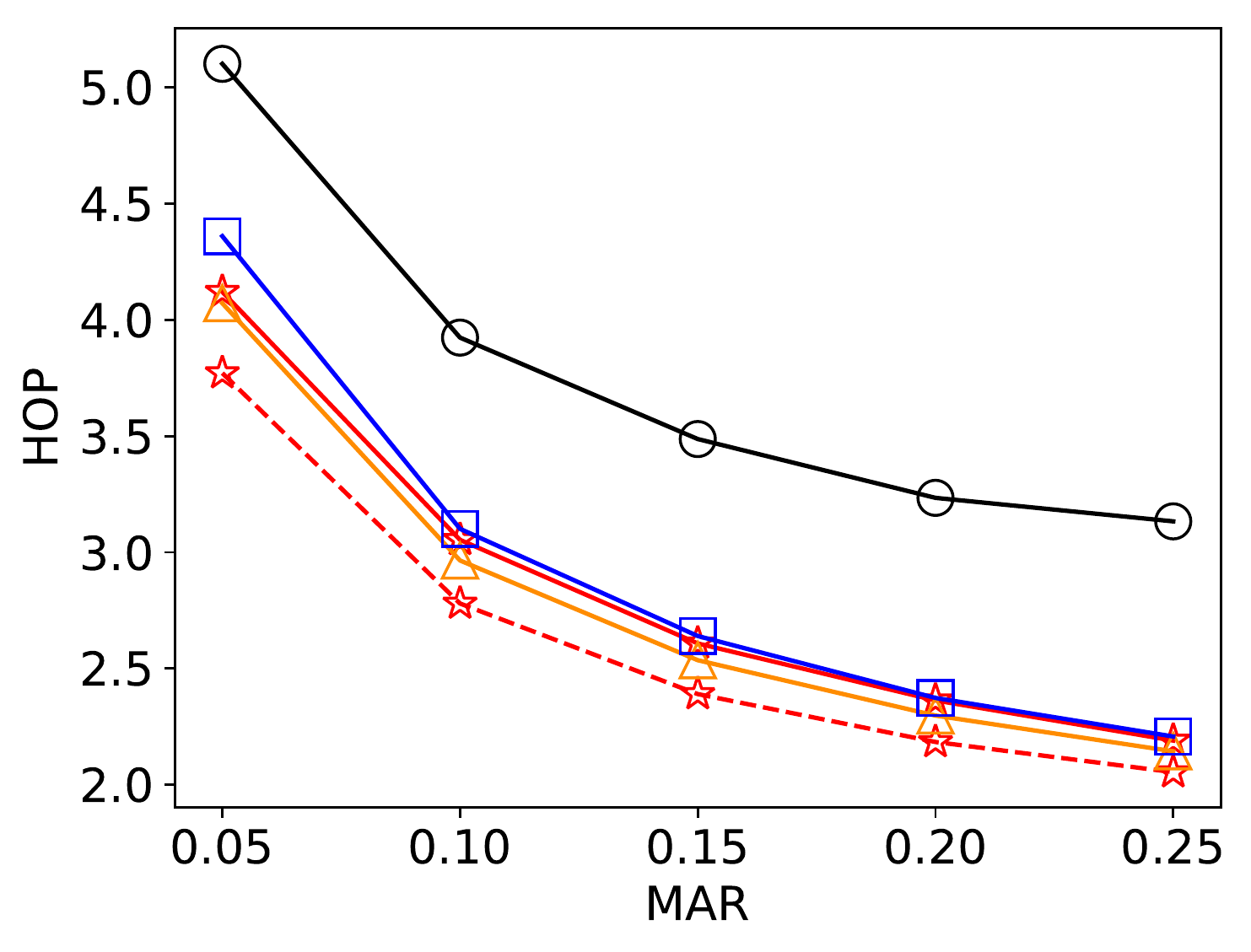}
	}
	\subfigure[ANW, Go.]{
		\includegraphics[height=0.97 in]{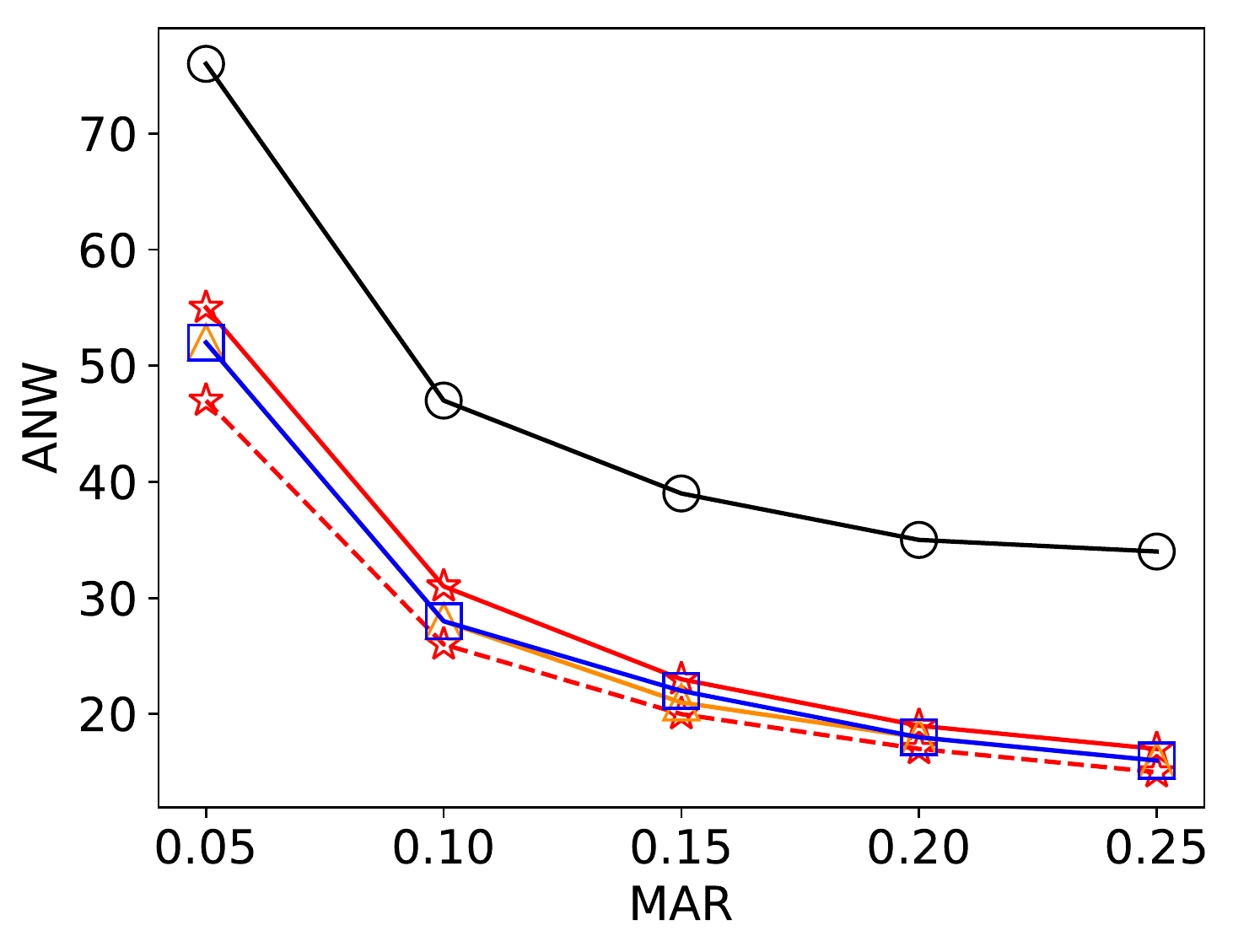}
	}
	\subfigure[DCM, Go.]{
		\includegraphics[height=0.97 in]{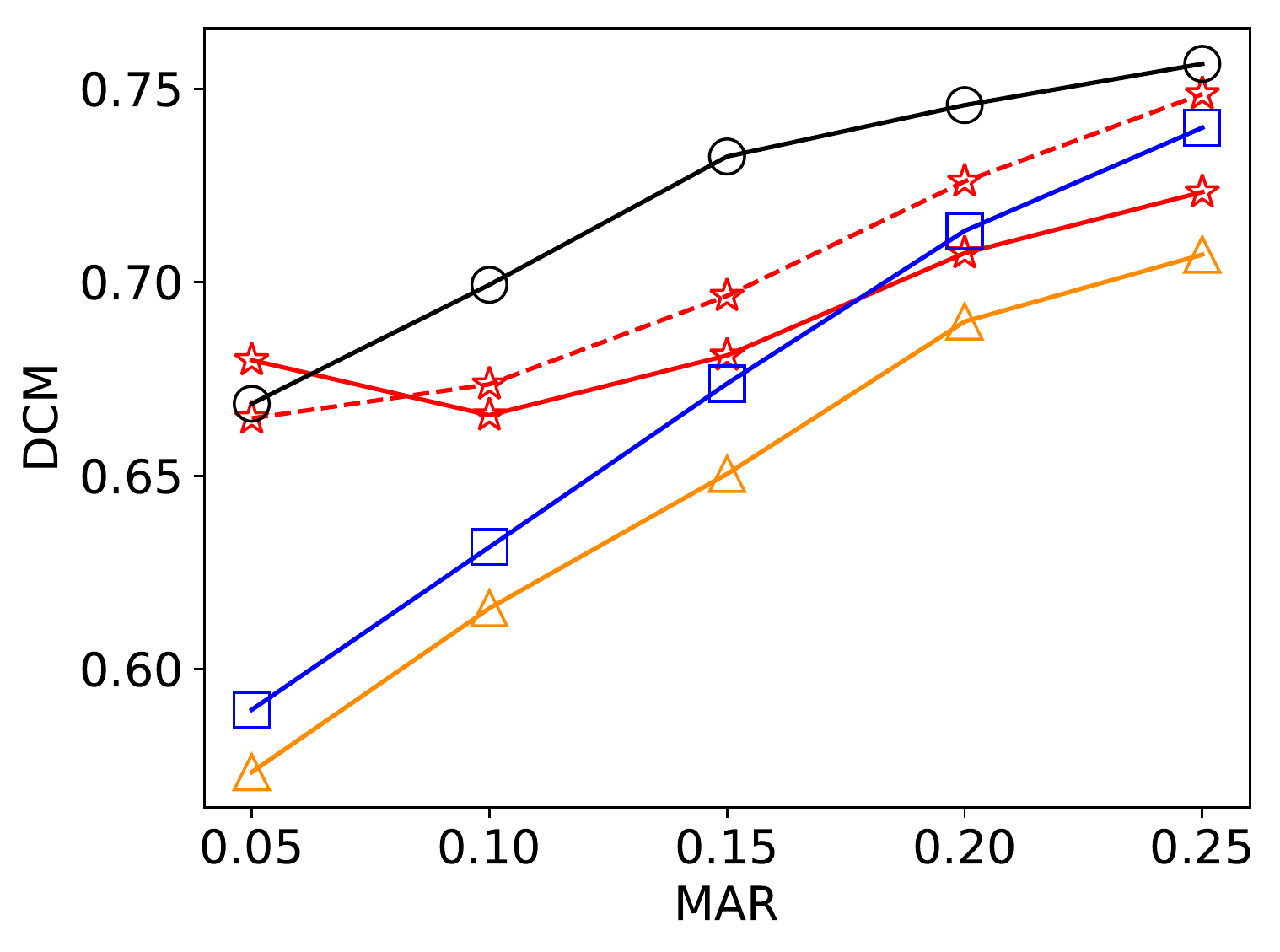}
	}
	\subfigure[ASR, Ta.]{
		\includegraphics[height=0.97 in]{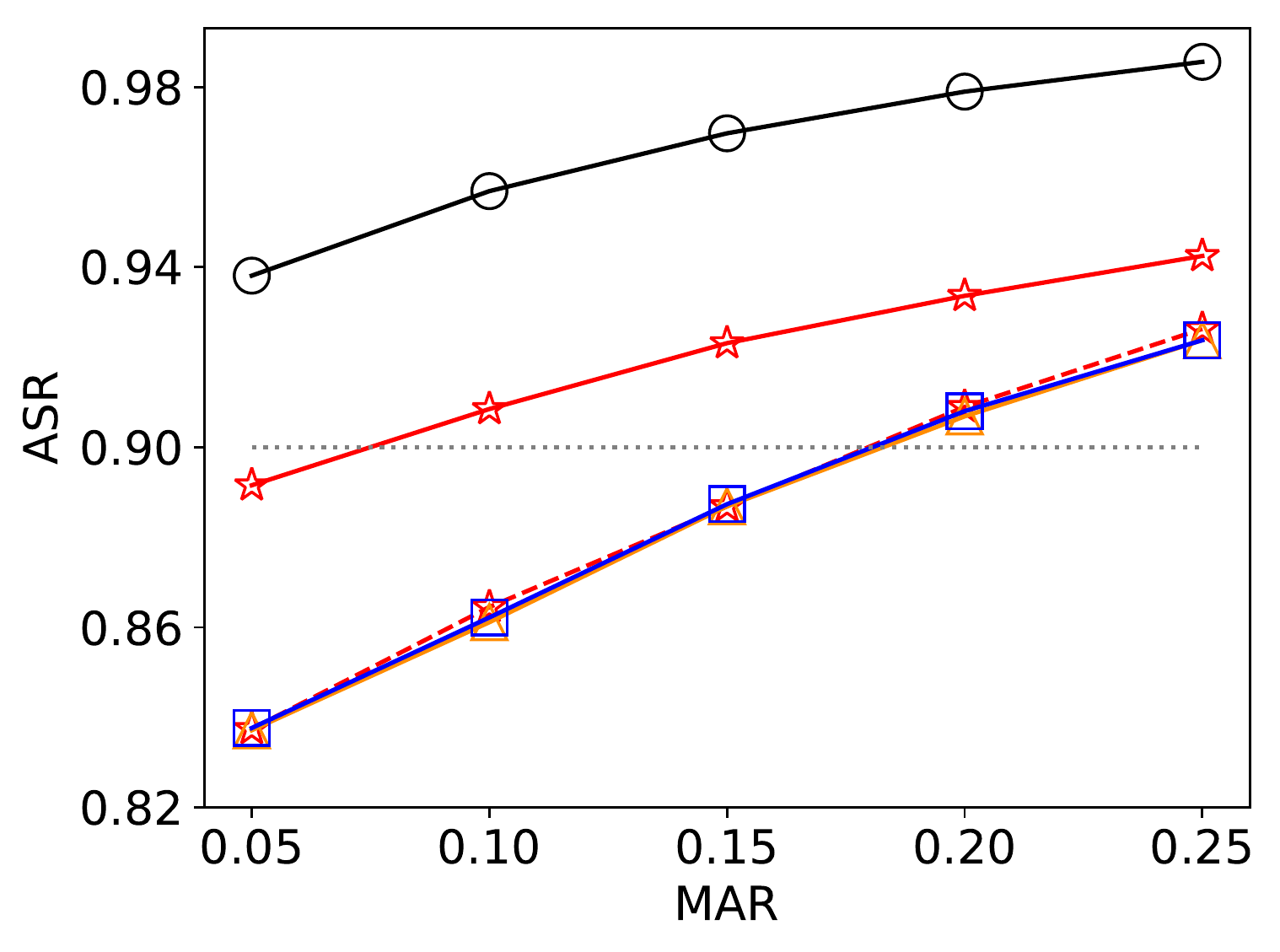}
	}
	\subfigure[WTD, Ta.]{
		\includegraphics[height=0.97 in]{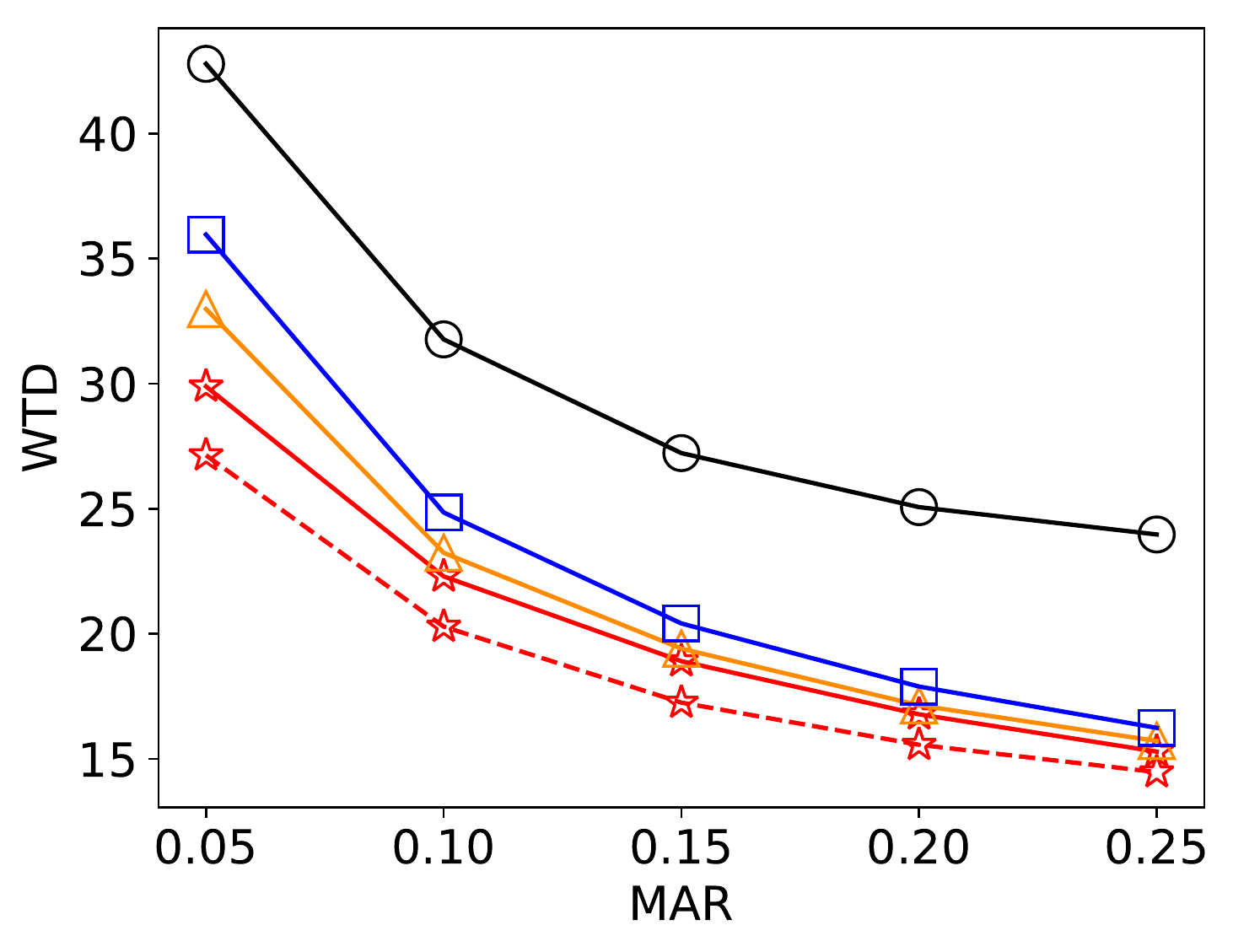}
	}
	\subfigure[HOP, Ta.]{
		\includegraphics[height=0.97 in]{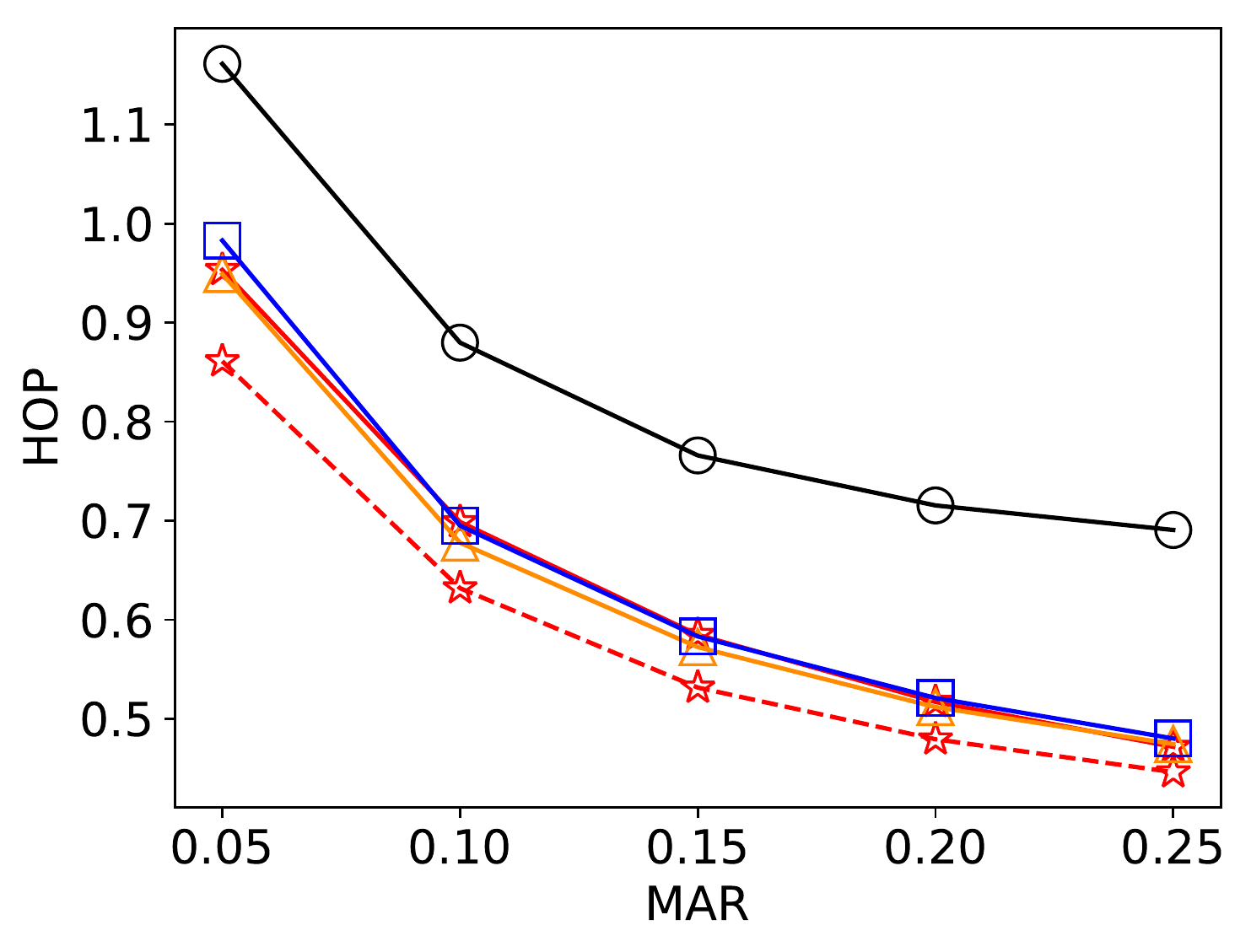}
	}
	\subfigure[ANW, Ta.]{
		\includegraphics[height=0.97 in]{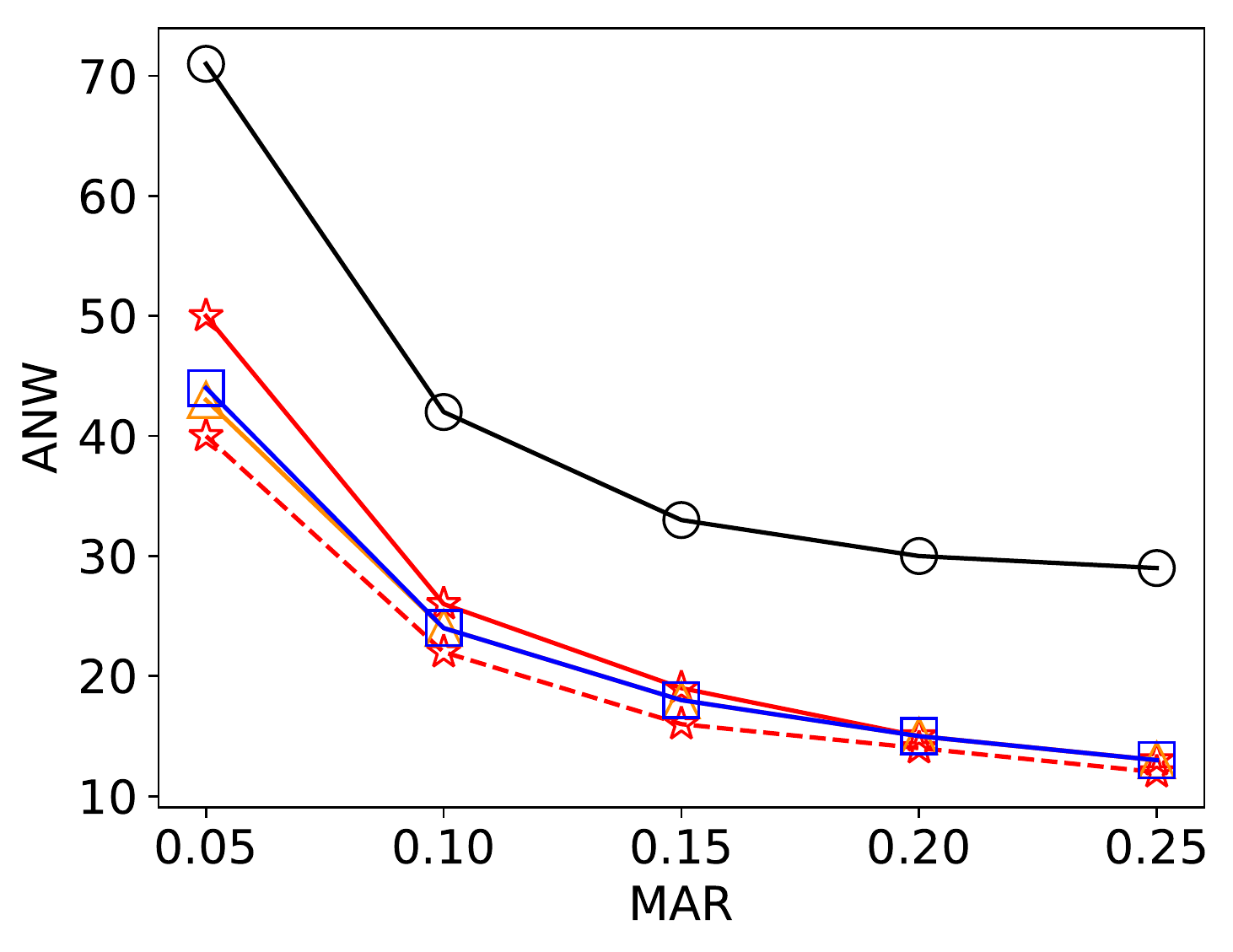}
	}
	\subfigure[DCM, Ta.]{
		\includegraphics[height=0.97 in]{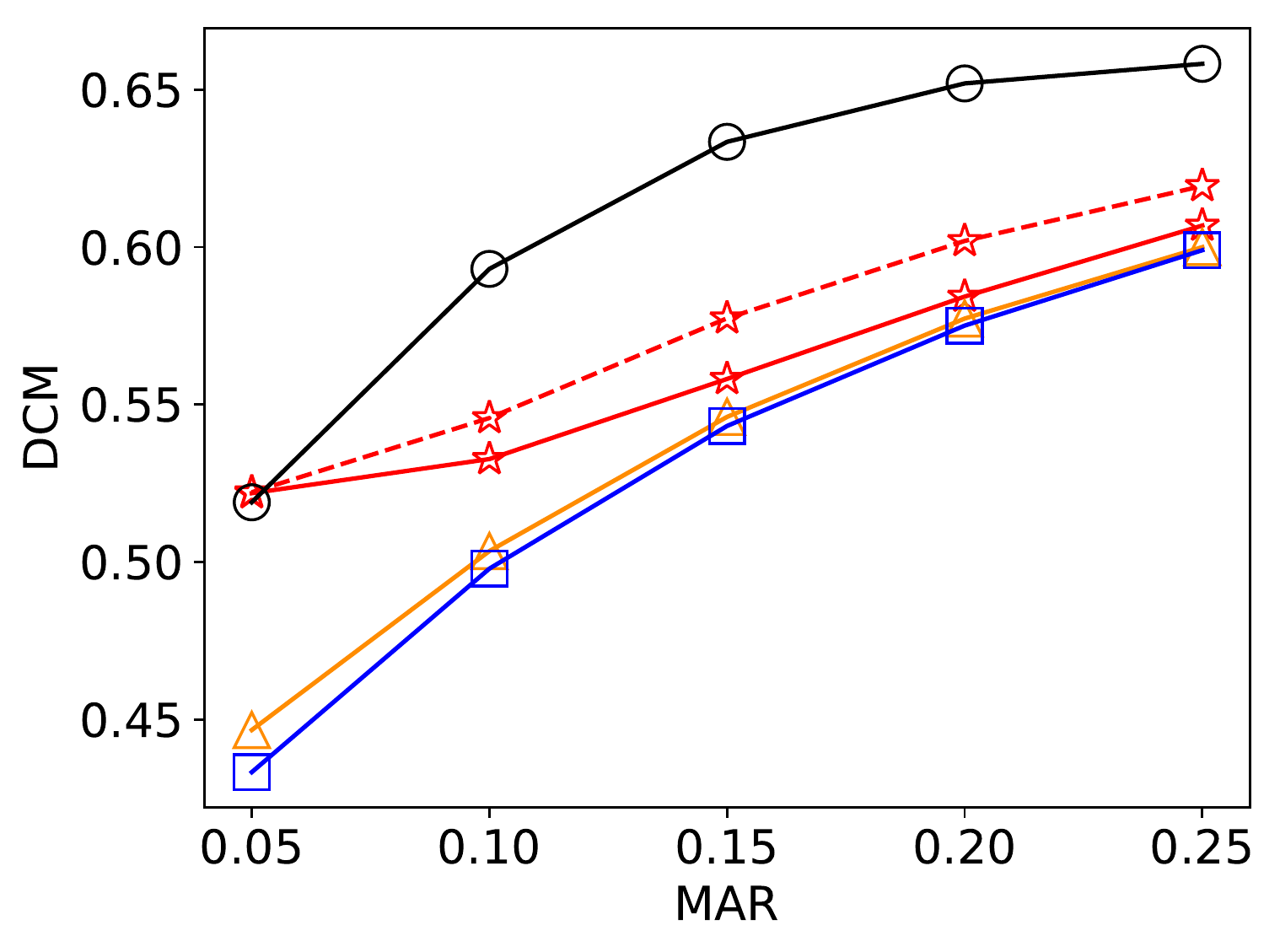}
	}
	\caption{Performance of geocast schemes with varying acceptance rate $MAR$}
	\label{fig:MAR}
\end{figure*}
\subsection{Task Allocation Evaluation}
In this section, we evaluate the performance of R-HT and some closely related schemes on three datasets by varying parameters, including privacy budget $\epsilon$, maximal acceptance rate $MAR$, and expected utility $EU$. Performance is measured by the five main metrics presented in Section \ref{sec:goals}.

\subsubsection{Effect of Varying Privacy Budget}\label{sec:epsilon}
We firstly investigate the impact of varying the privacy budget while keeping the other parameters at their defaults, see Fig. \ref{fig:Eps}.
%With the increase of privacy budget, the ASR of each scheme decreases gradually, which is mainly related to the partition granularity. When the privacy budget is low, in order to resist the noise, the granularity increases caused more larger cells, so there is a significant oversupply when adding the last cell to $GR$.
With the default $EU$=0.9, almost all ASR values of R-HT reach the threshold, while G-GR fails in most cases. As for WTD, HOP and DCM, R-HT is $9.7\%$, $1.0\%$ and $7.2\%$ better than G-GR on average, and has a maximum increase of $11.9\%$, $2.1\%$ and $8.4\%$, respectively. This demonstrates that R-HT ensures basically that the task acceptance probability reaches $EU$, and it behaves better than G-GR in other metrics. The reason is that the use of historical data significantly reduces the noise added to the grid partition and real-data publication. Smaller noise makes the adaptively generated grids more suitable and yields that cell selections are inclined to be of higher density and closer distance.
 Besides, the ASR of R-HT in NYTaxi is relatively low. In fact, the location distribution is very concentrative on roads, which yields many smaller cells locally. $MTD$ is set small, which makes $AR$ decreasing sharply with the increase of distance. Then the larger count of cells needed for $GR$ results in more virtual locations, and the real ASR will be lower when noisy ASR reaches $EU$.

By adjusting $EU$ to R-HT and GDY to make the ASR of all schemes almost the same, the advantage of R-HT\_Ad (R-HT with adjusted $EU$) is more obvious. Compared with G-GR, the average decrease of HOP for R-HT\_Ad and G-GS is $9.0\%$ and $2.9\%$, respectively.
%Especially in NYTaxi, the decrease of HOP in R-HT\_Ad is 10.2\%.
%This is because using scoring function involved area and distance, both R-HT\_Ad and G-GS prefer to choose cells with high distribution density or closer distance.
R-HT\_Ad behaves perfect on WTD, with an average reduction of $16.8\%$, while it reduced the notified workers (ANW) by $9.4\%$ compared with G-GR. In addition, we observe that for the change of DCM, GDY makes the best performance. Indeed, GDY has the coarsest partitioning granularity, which results in the smallest number of selected cells, even the cell covering task is enough for $GR$. The DCM of R-HT\_Ad is $8.7\%$ higher than that of G-GR because of applying $LGR$ strategy.

To sum up, within the wide range of privacy budget, R-HT can basically meet $EU$, effectively reduce the actual system cost for task assignments, and also improve the compactness of $GR$ for effectively saving communication costs, which improves system operation efficiency comprehensively.

\subsubsection{Effect of Varying MAR }\label{sec:mar}
We observe the metrics for R-HT by changing $MAR$ and make some comparisons, see Fig. \ref{fig:MAR}. The experimental results show that the increase of $MAR$ reduces the HOP and WTD soon, because it increases AR of each worker and fewer workers nearby are needed for achieving the requirements of $EU$. Further, with the increase of $MAR$,  the decrease of cell count in $GR$ is more obvious than the impact of changing $EU$. Except for a few points, the ASR of R-HT meets $EU$, while the ASR of G-GR fails on nearly a half of points. Compared with the G-GR, the average improvement of R-HT\_Ad on the WTD, HOP, ANW and DCM are $15.7\%$, $9.2\%$,  $8.9\%$ and $5.8\%$, respectively.

\subsubsection{Effect of Varying $EU$ }\label{sec:eu}
We observe the influence of changing $EU$ on the metrics of R-HT and other schemes. Obviously, when $EU$ increases, the $GR$ area increases, and so do both WTD and HOP. The results show that for varied $EU$, each ASR of R-HT meets the threshold and the WTD and HOP are always smaller than those of G-GR, respectively. With the increase of $EU$, the gaps are wider due to more cells necessarily selected. When $EU$ equals 0.9, the gaps between R-HT\_Ad and G-GR on WTD, HOP, ANW and DCM are up to $12.6\%$, $7.0\%$, $7.4\%$ and $4.0\%$, respectively.

\begin{figure*}[htp]
	\centering
	\includegraphics[scale=0.5]{fig8910_menu.pdf}\\
	\subfigure[ASR, Ye.]{
		\includegraphics[height=0.97 in]{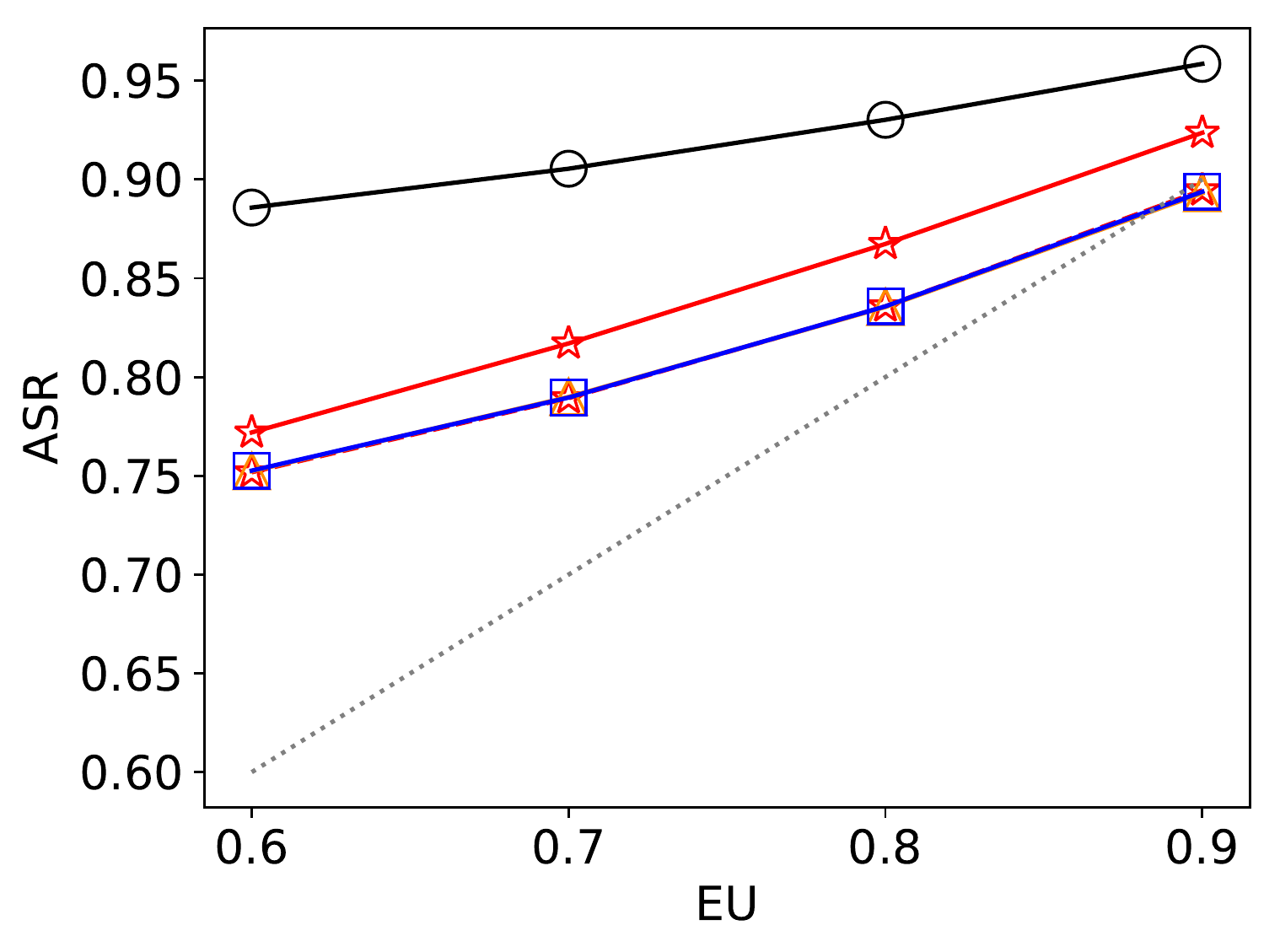}
	}
	\subfigure[WTD, Ye.]{
		\includegraphics[height=0.97 in]{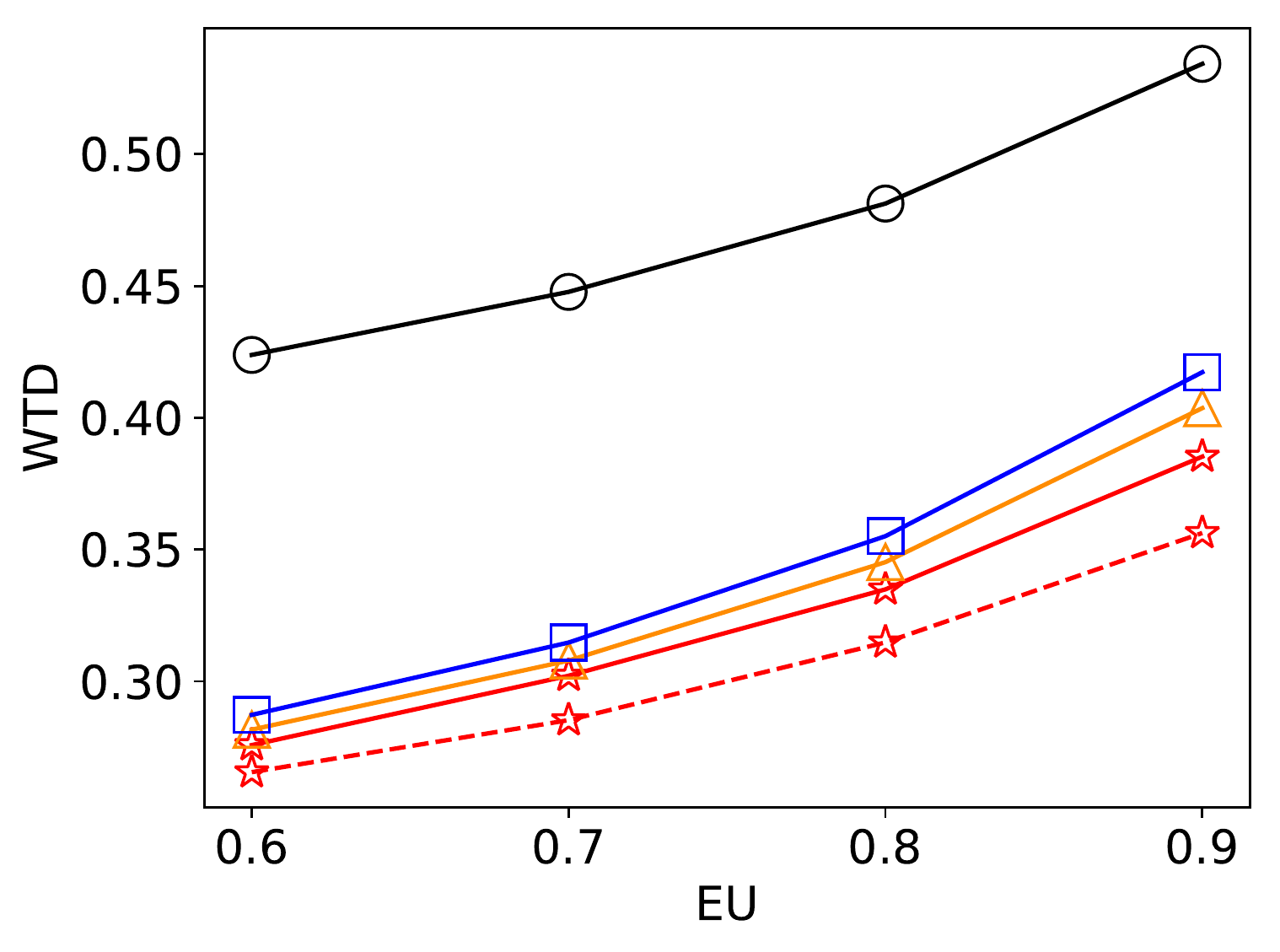}
	}
	\subfigure[HOP, Ye.]{
		\includegraphics[height=0.97 in]{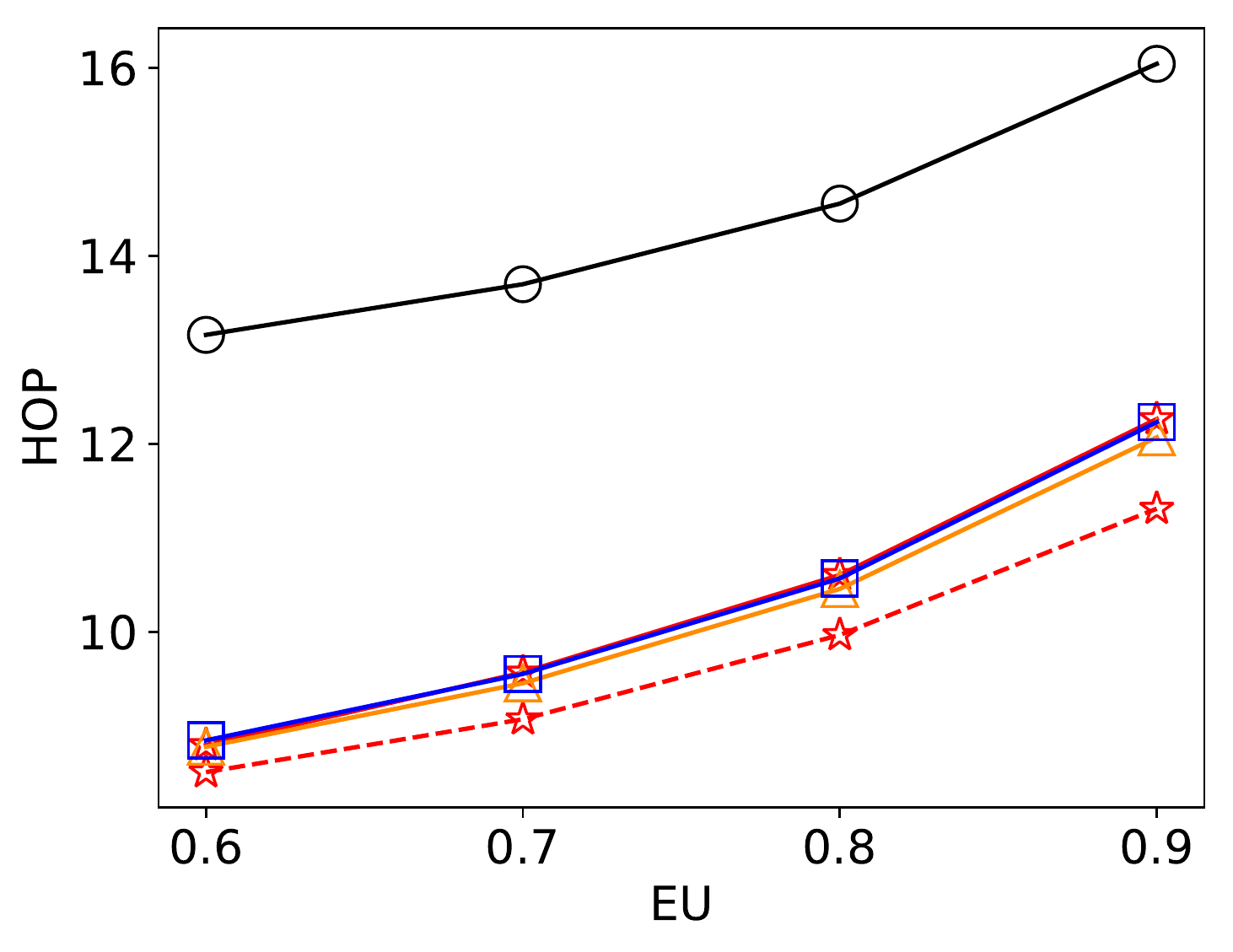}
	}
	\subfigure[ANW, Ye.]{
		\includegraphics[height=0.97 in]{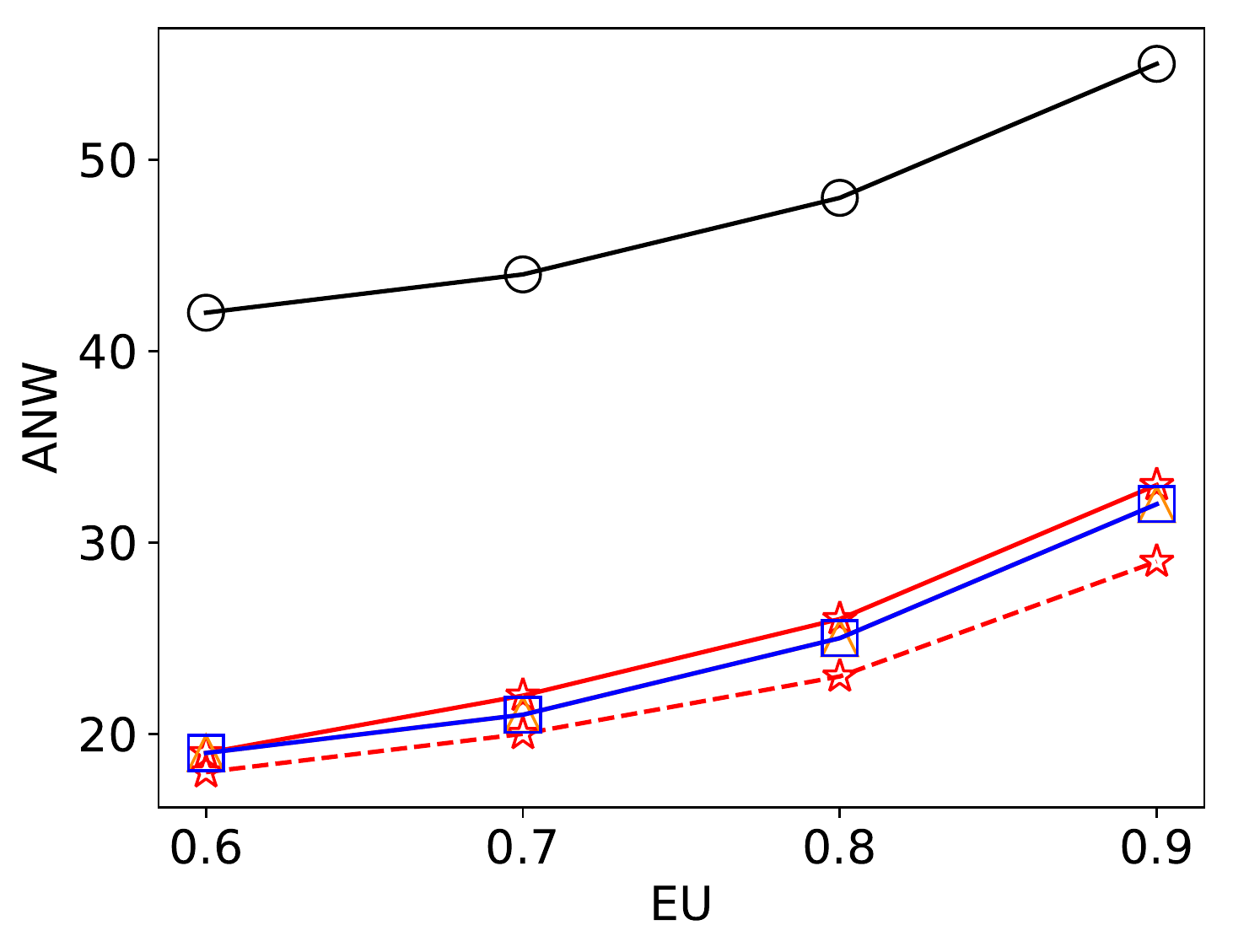}
	}
	\subfigure[DCM, Ye.]{
		\includegraphics[height=0.97 in]{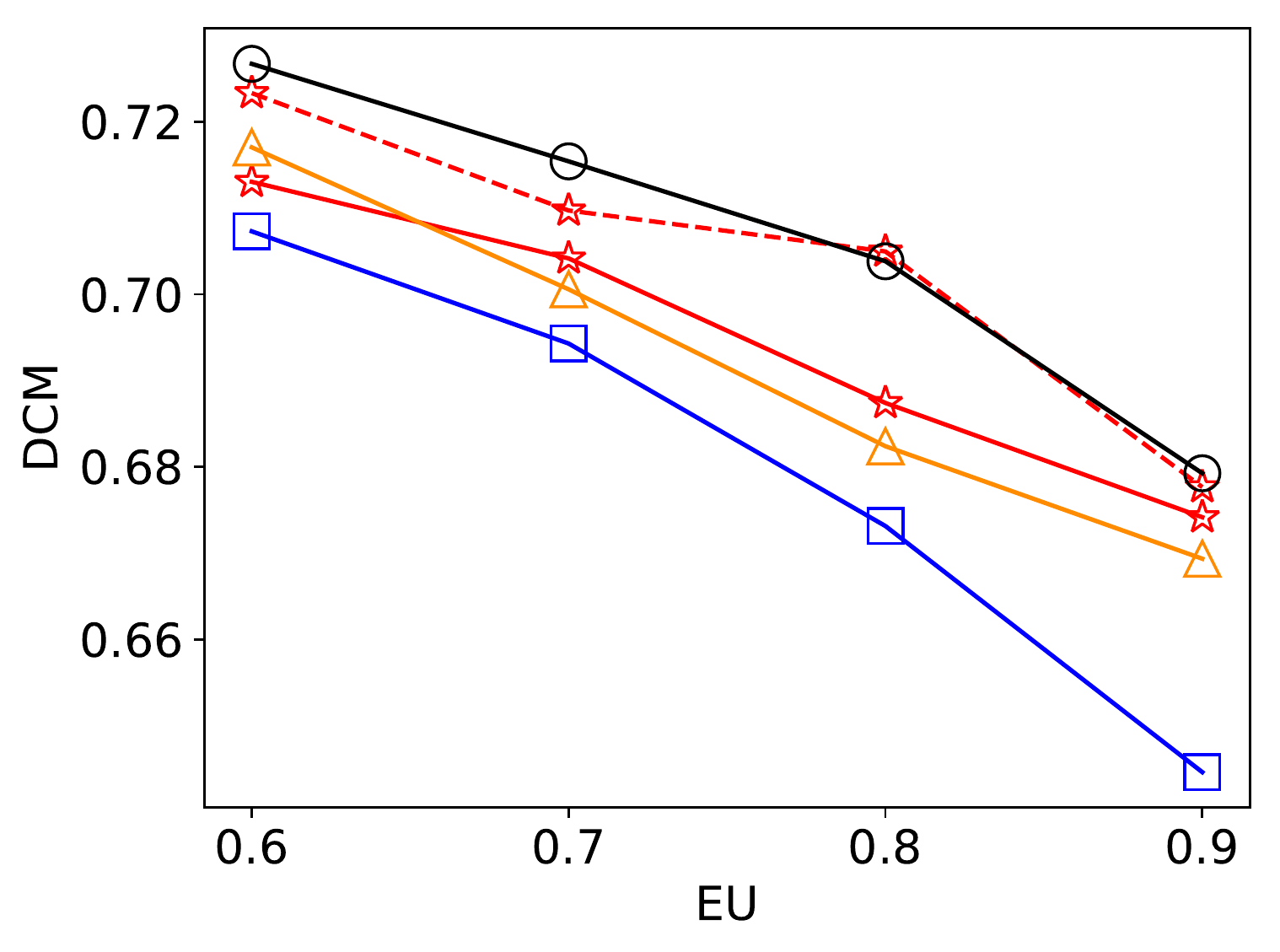}
	}
	\subfigure[ASR, Go.]{
		\includegraphics[height=0.97 in]{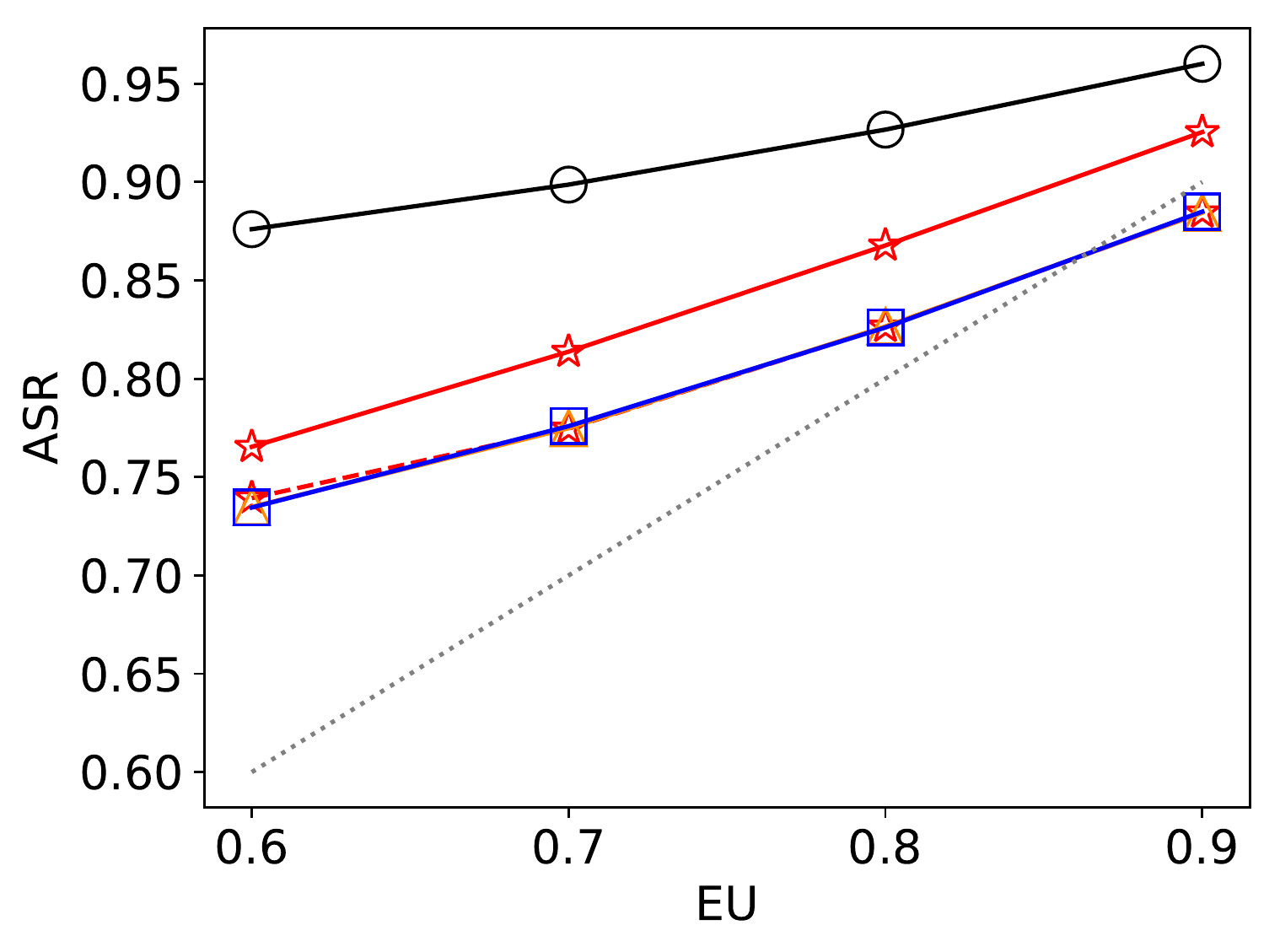}
	}
	\subfigure[WTD, Go.]{
		\includegraphics[height=0.97 in]{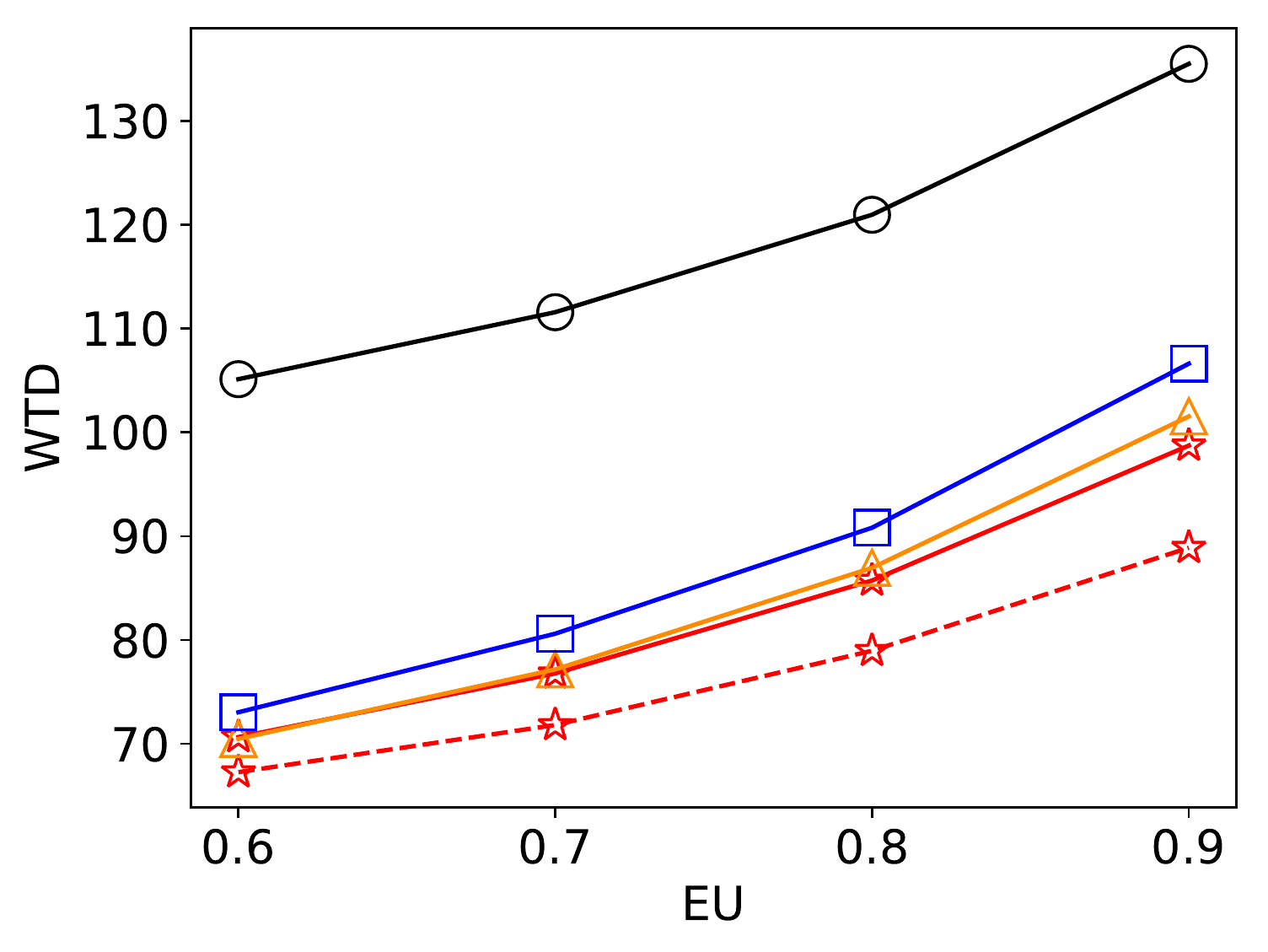}
	}
	\subfigure[HOP, Go.]{
		\includegraphics[height=0.97 in]{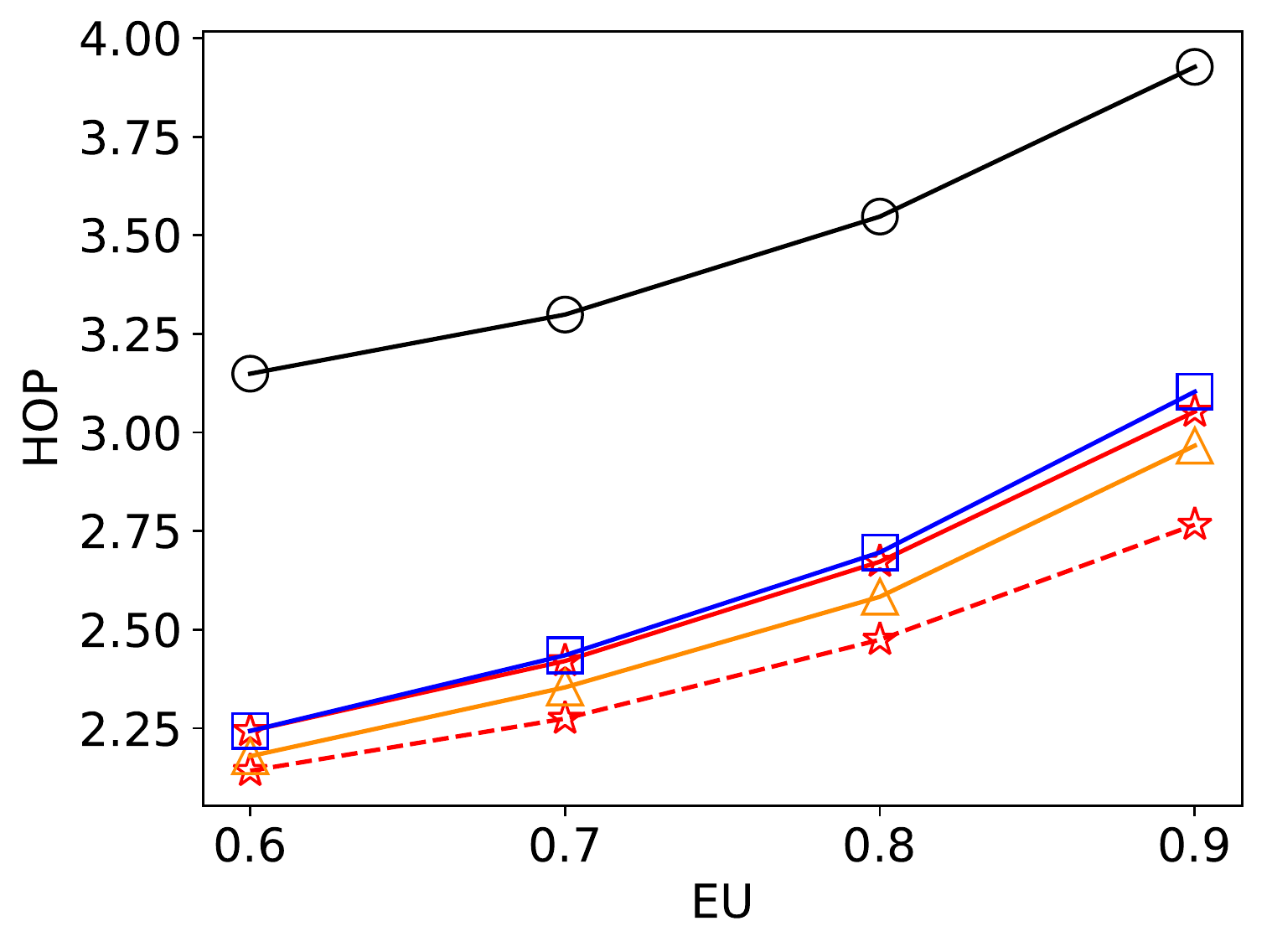}
	}
	\subfigure[ANW, Go.]{
		\includegraphics[height=0.97 in]{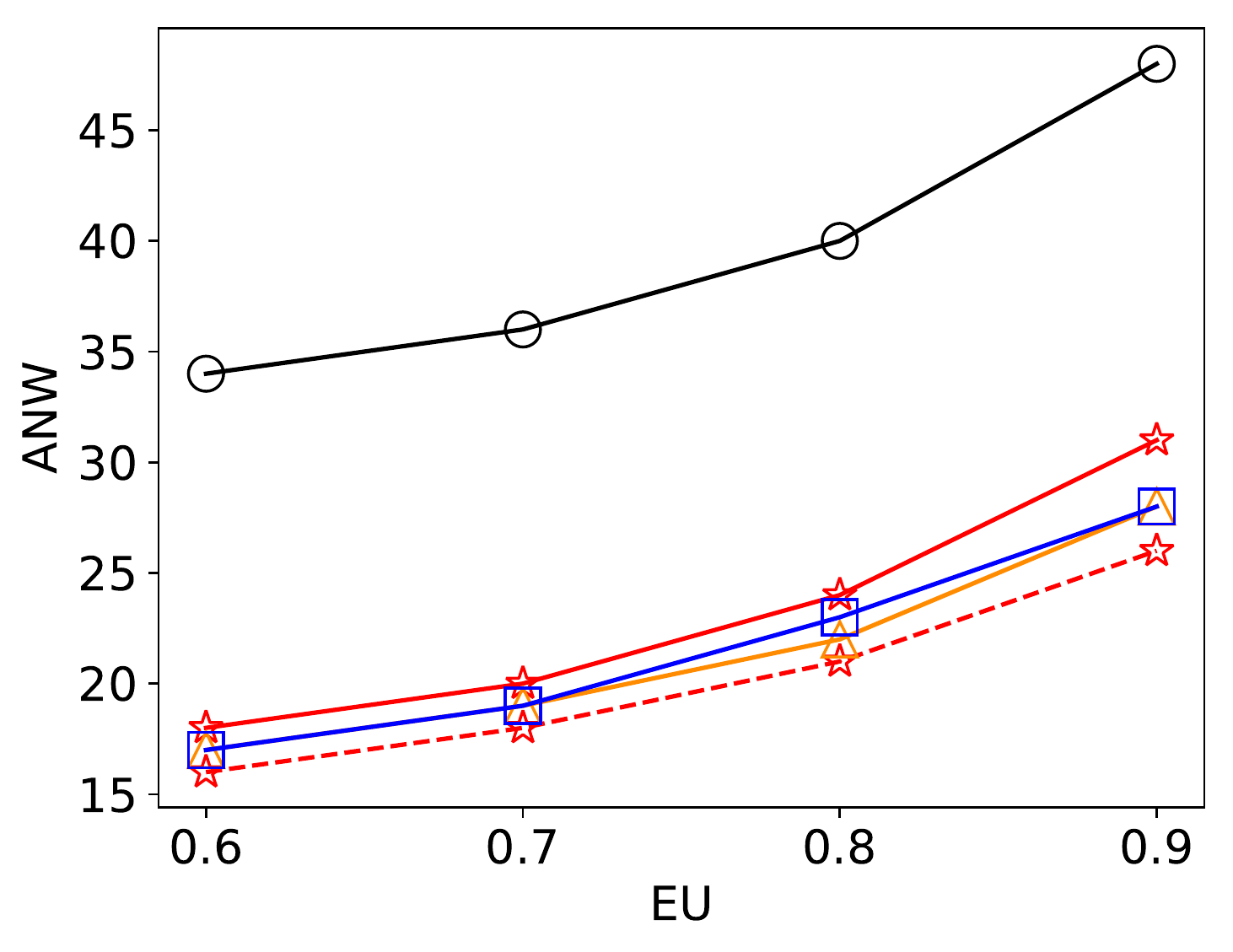}
	}
	\subfigure[DCM, Go.]{
		\includegraphics[height=0.97 in]{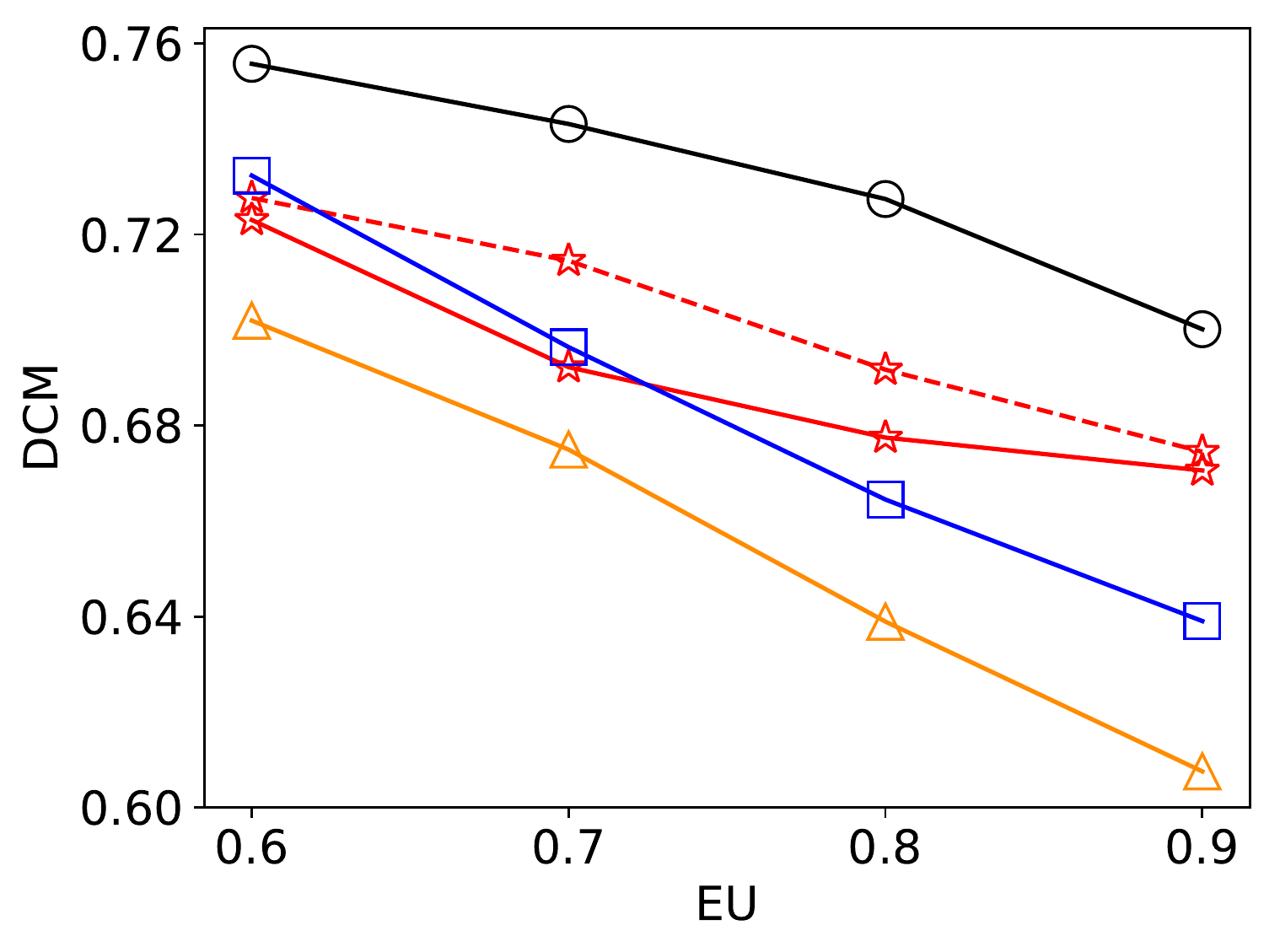}
	}
	\subfigure[ASR, Ta.]{
		\includegraphics[height=0.97 in]{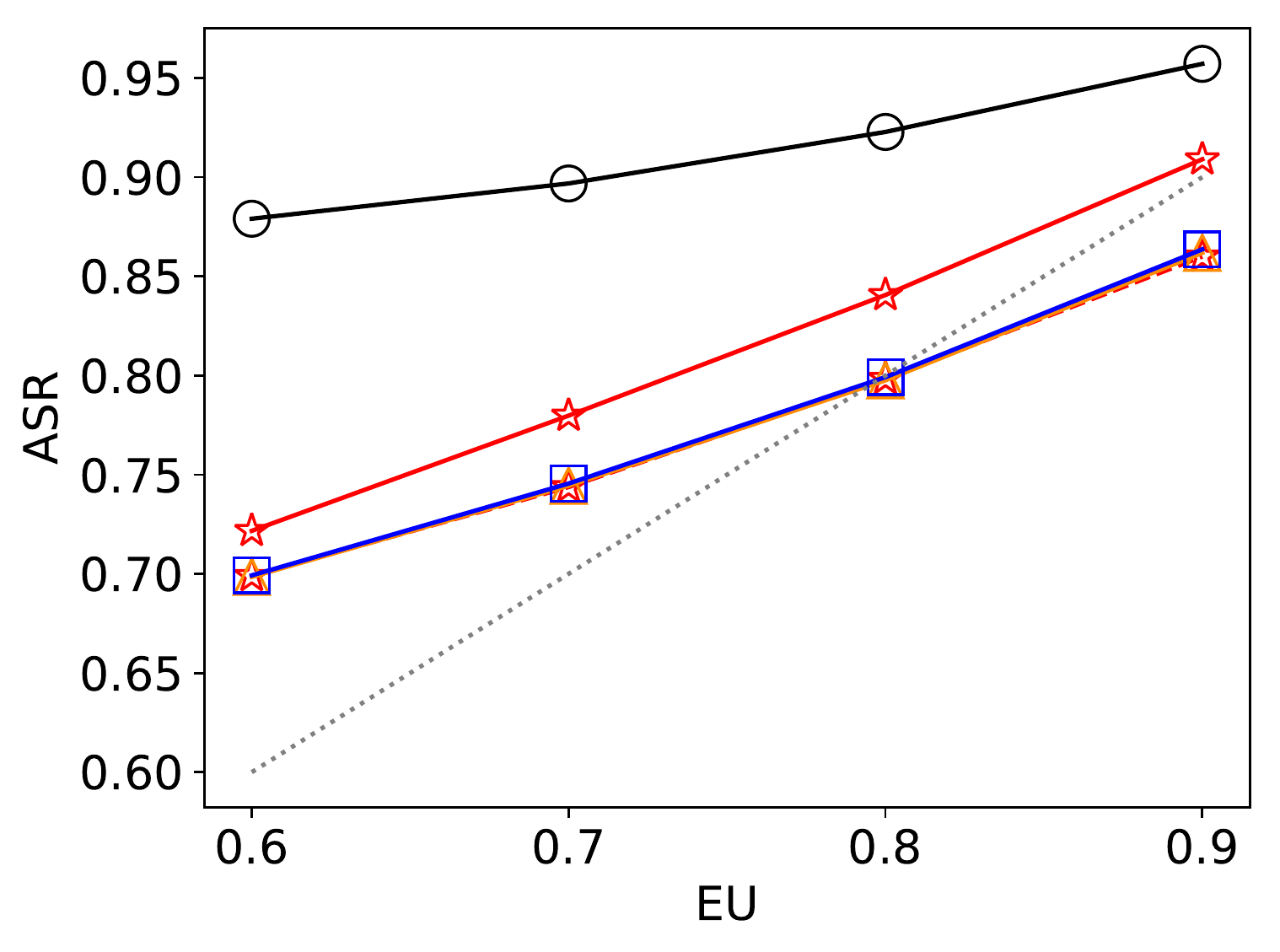}
	}
	\subfigure[WTD, Ta.]{
		\includegraphics[height=0.97 in]{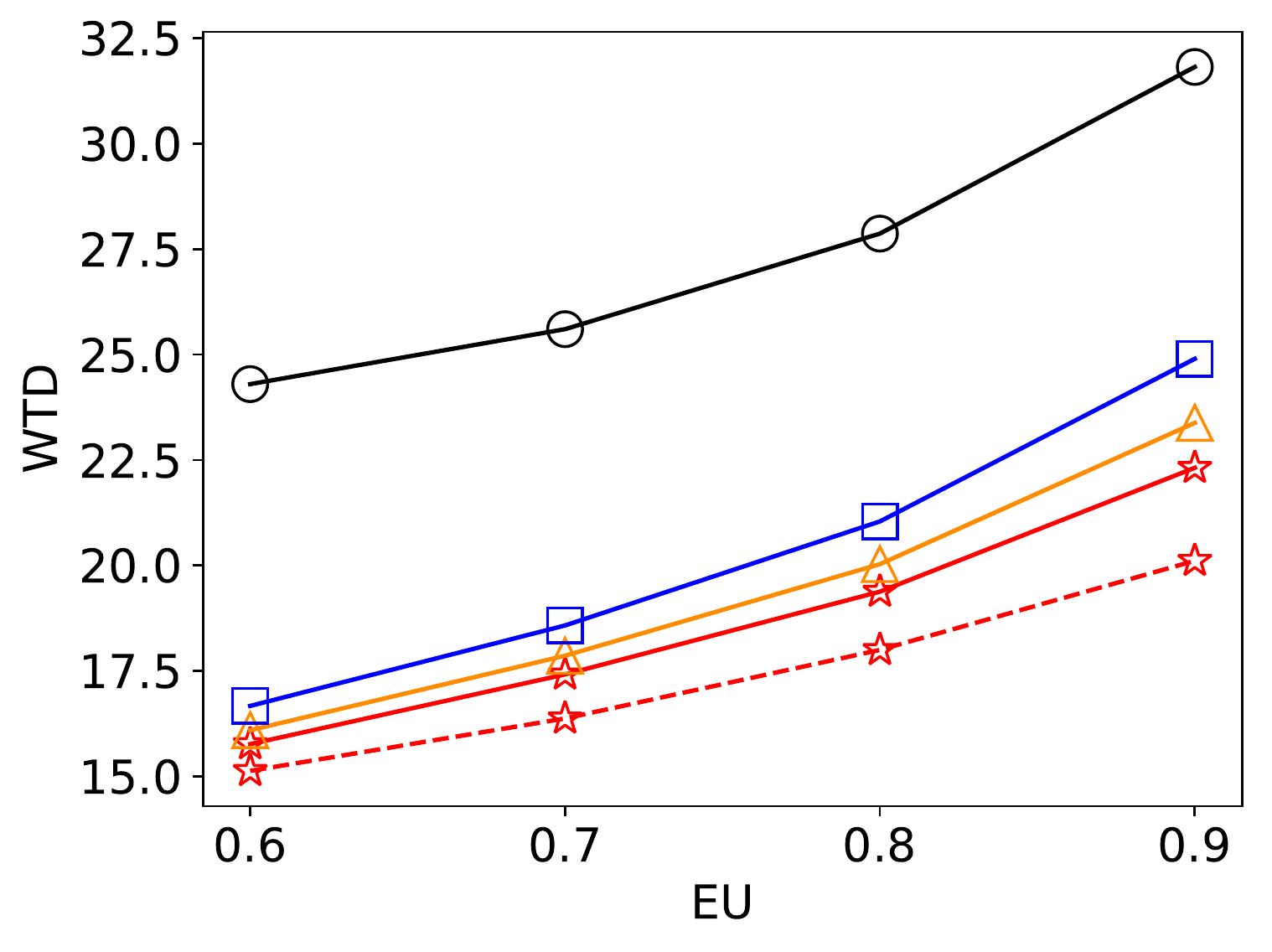}
	}
	\subfigure[HOP, Ta.]{
		\includegraphics[height=0.97 in]{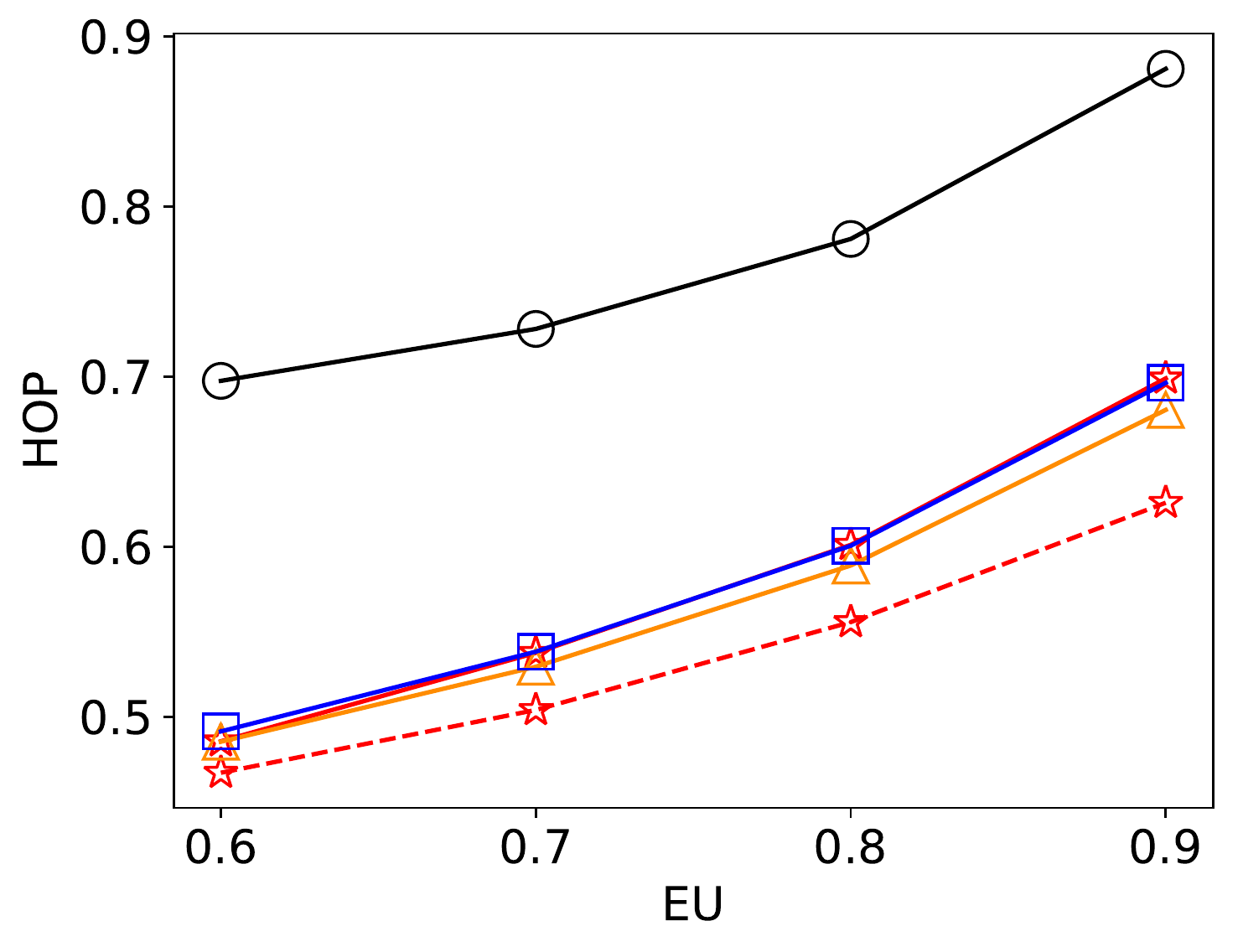}
	}
	\subfigure[ANW, Ta.]{
		\includegraphics[height=0.97 in]{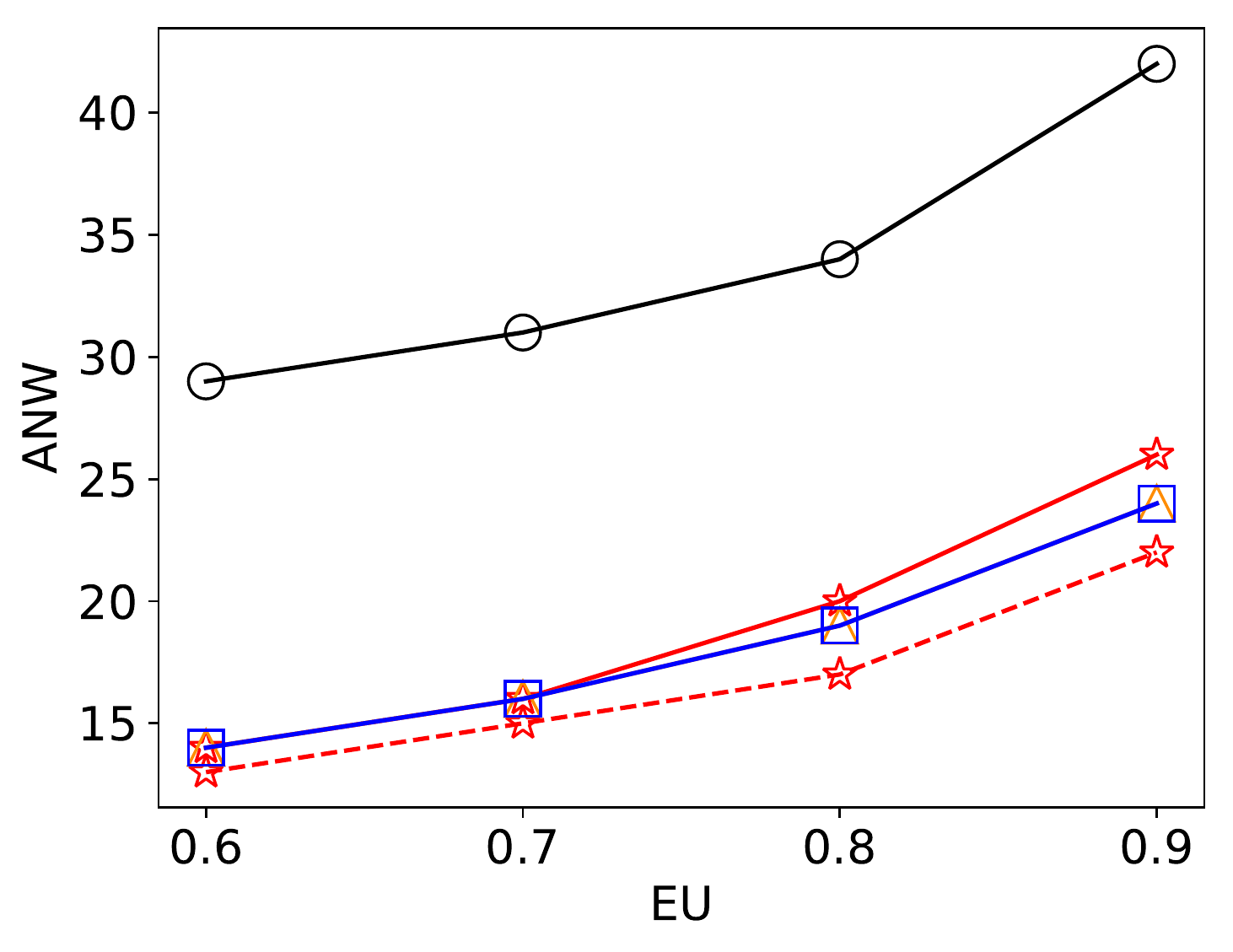}
	}
	\subfigure[DCM, Ta.]{
		\includegraphics[height=0.97 in]{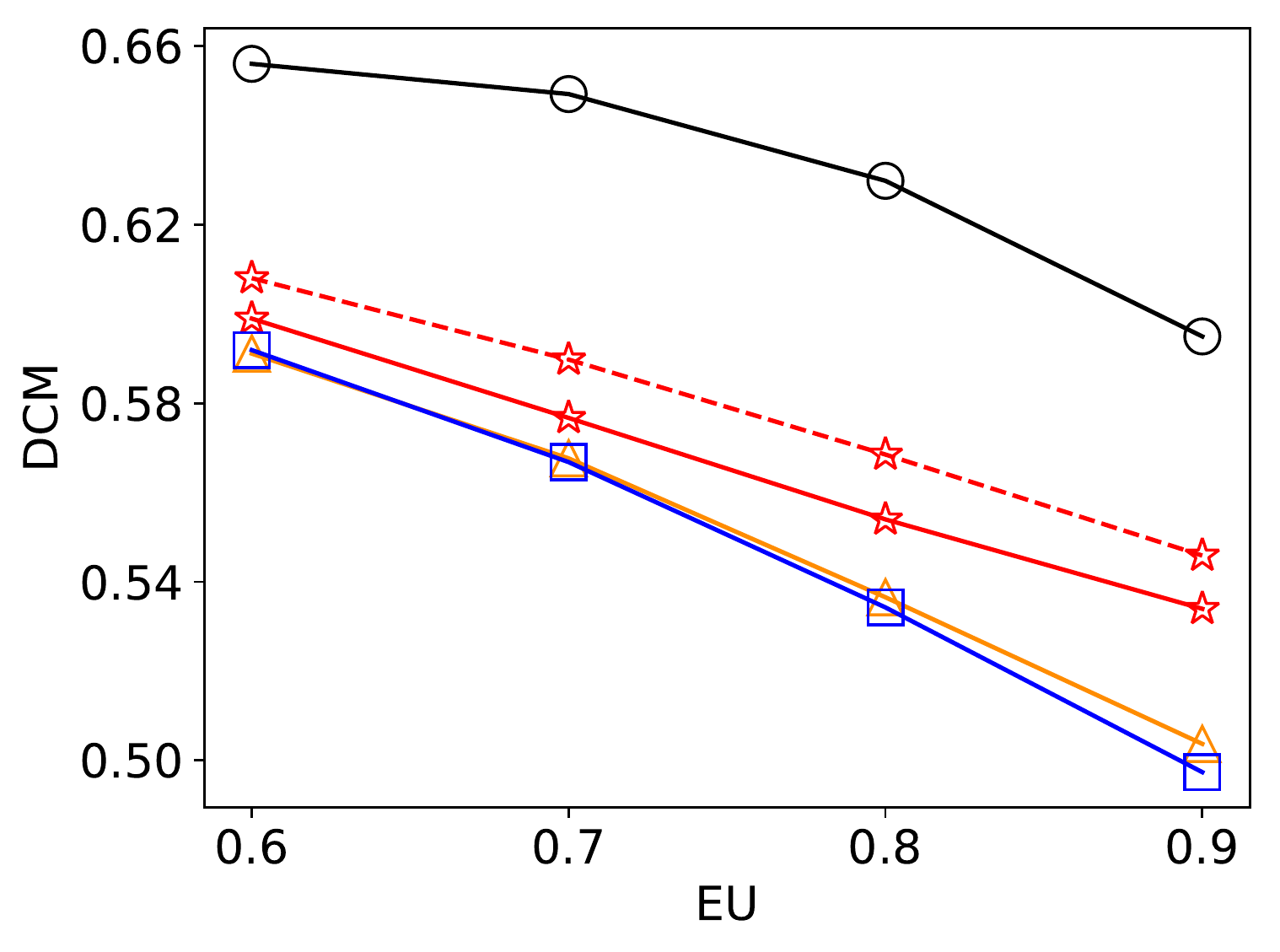}
	}
	\caption{Performance of geocast schemes with varying expected utility $EU$}
	\label{fig:EU}
\end{figure*}
In conclusion for Sections \ref{sec:epsilon} to \ref{sec:eu}, according to the ASR results above, the R-HT basically reaches $EU$ under various parameter settings. In most cases, the G-GR fails to reach $EU$, as is mentioned in Section 7.2.1 of \cite{TGF17}. Therefore, G-GR does not address the challenge of achieving a high success rate for the task assignment. To be specific, a total of 42 groups of ASR values are compared in different settings of parameters $\epsilon$, $EU$ and $MAR$ on three datasets. Among them, 19 points of G-GR reach $EU$, accounting for $45.2\%$, while 38 points of R-HT accounting for $92.9\%$, twice as much as G-GR. We find out that the compliance rates of ASR on Yelp, Gowalla and NYTaxi, G-GR is $50.0\%$, $50.0\%$ and $35.7\%$, while R-HT $100.0\%$, $100.0\%$ and $78.6\%$, respectively. Both of the two schemes have some gaps on compliance rate in NYTaxi due to the extremely concentrative distribution of locations as is mentioned above.
\begin{figure*}[htp]
	\centering
	\includegraphics[scale=0.5]{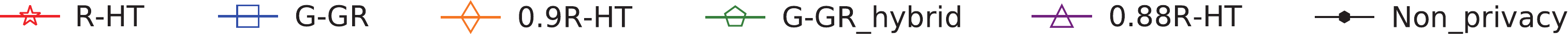}\\
	\subfigure[ASR, Go.]{
		\includegraphics[height=0.97 in]{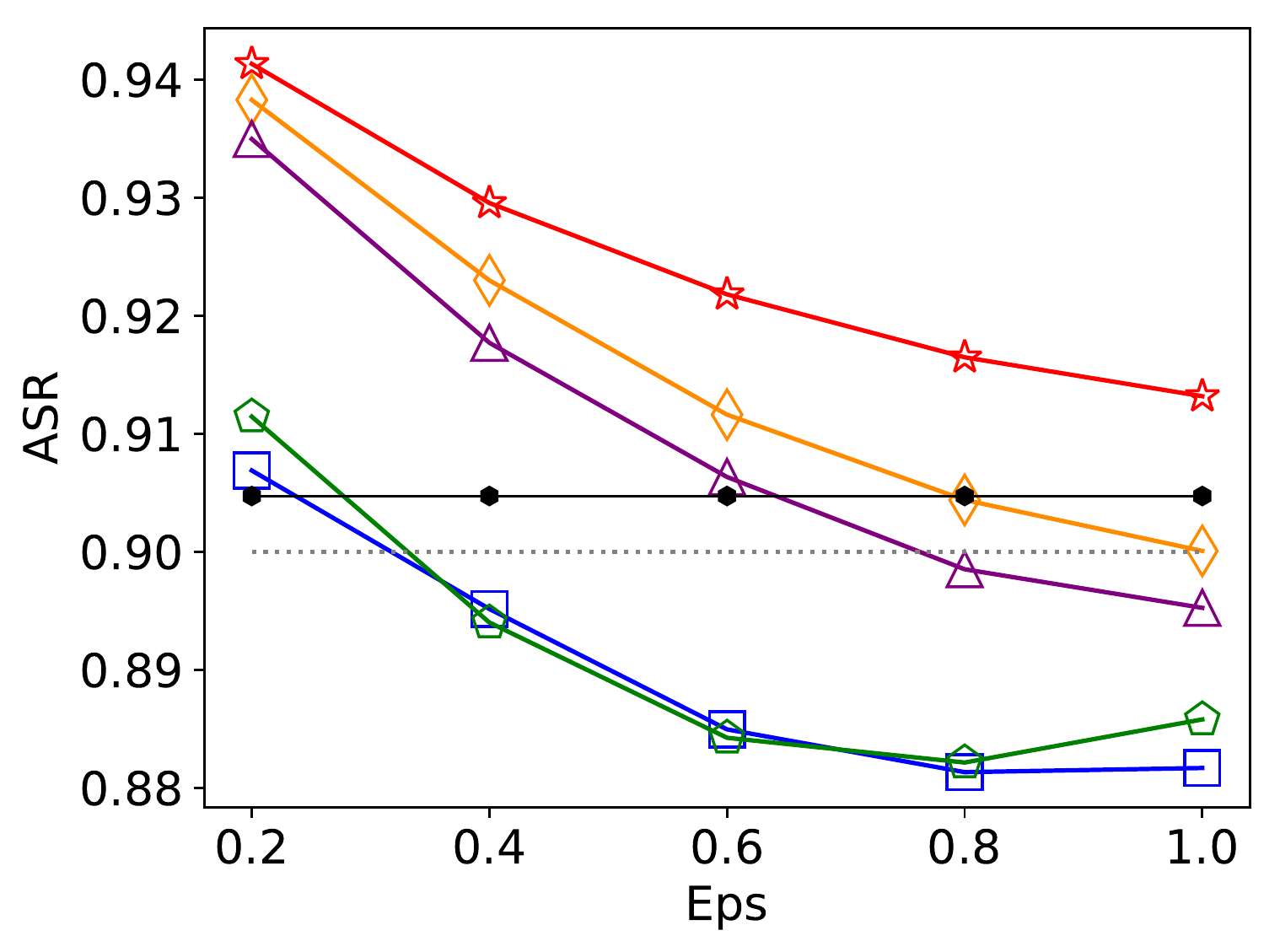}
	}
	\subfigure[WTD, Go.]{
		\includegraphics[height=0.97 in]{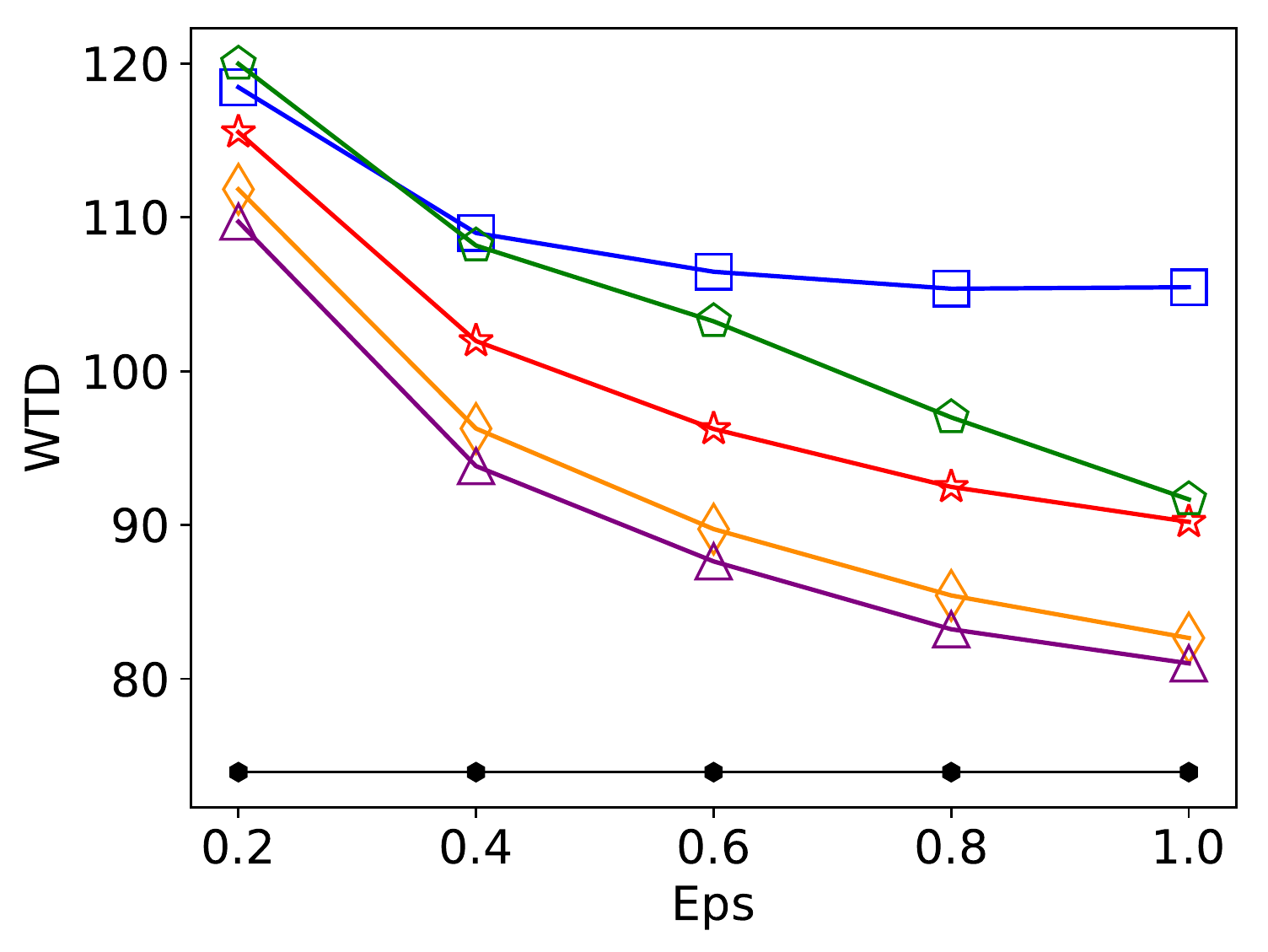}
	}
	\subfigure[HOP, Go.]{
		\includegraphics[height=0.97 in]{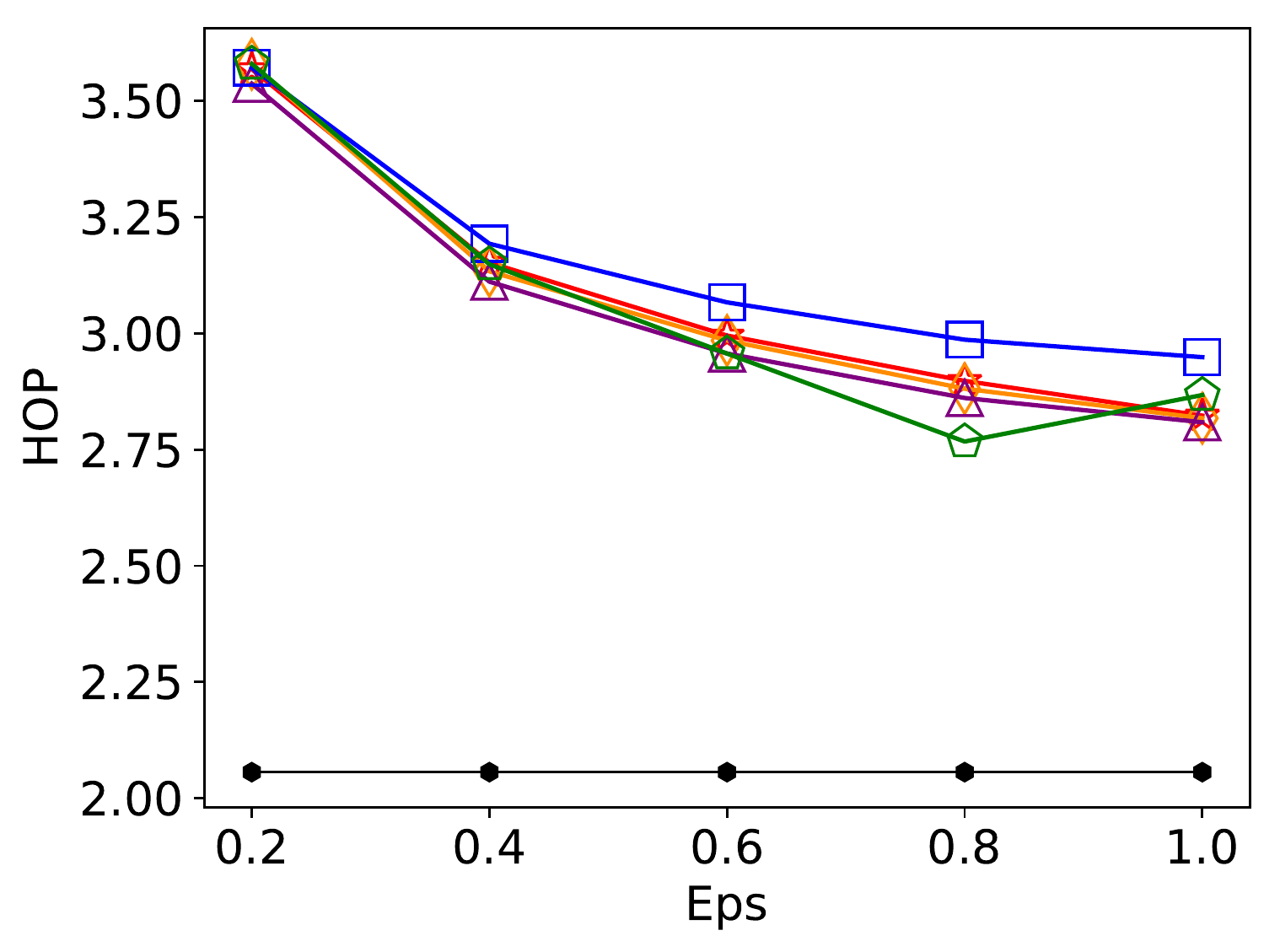}
	}
	\subfigure[ANW, Go.]{
		\includegraphics[height=0.97 in]{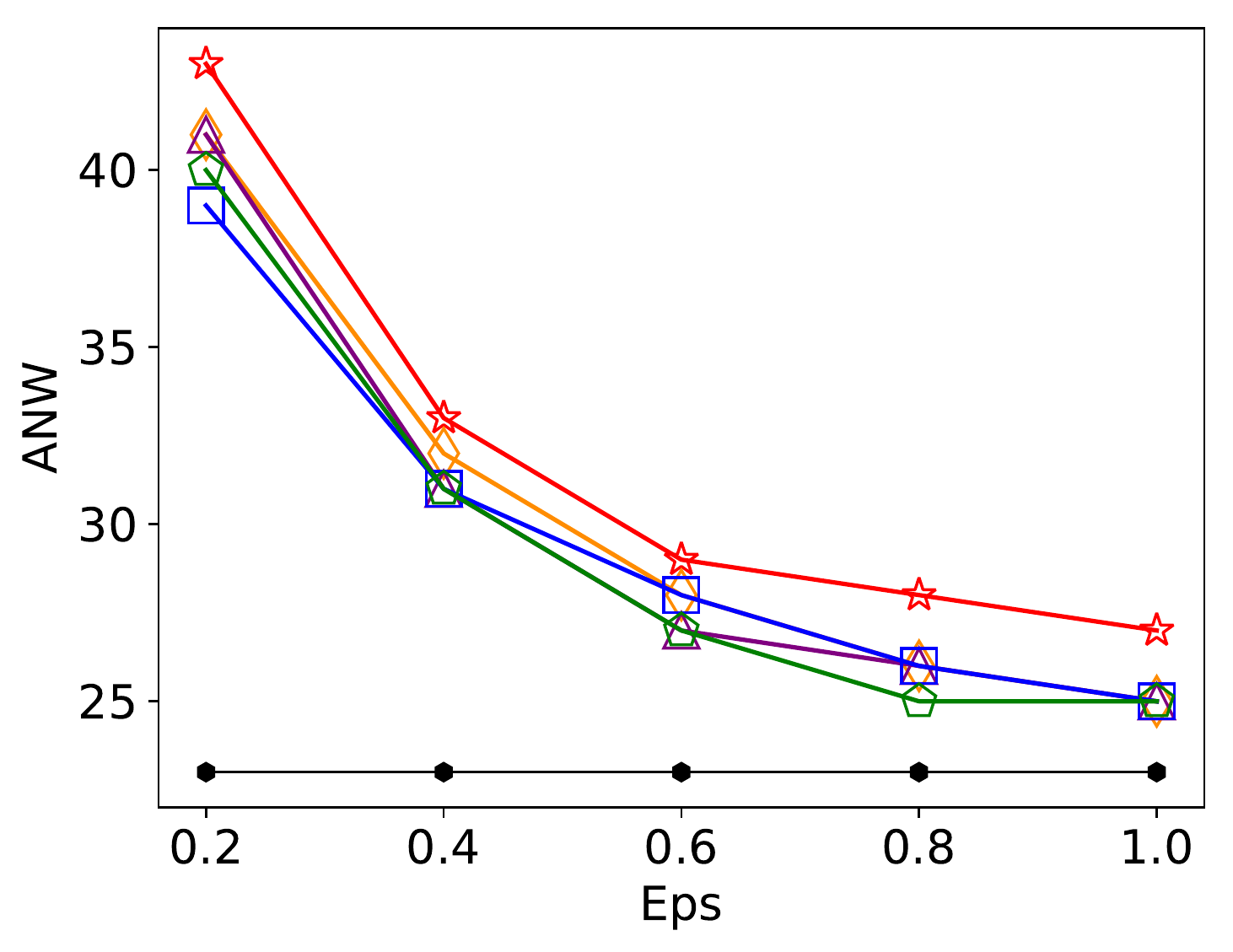}
	}
	\subfigure[DCM, Go.]{
		\includegraphics[height=0.97 in]{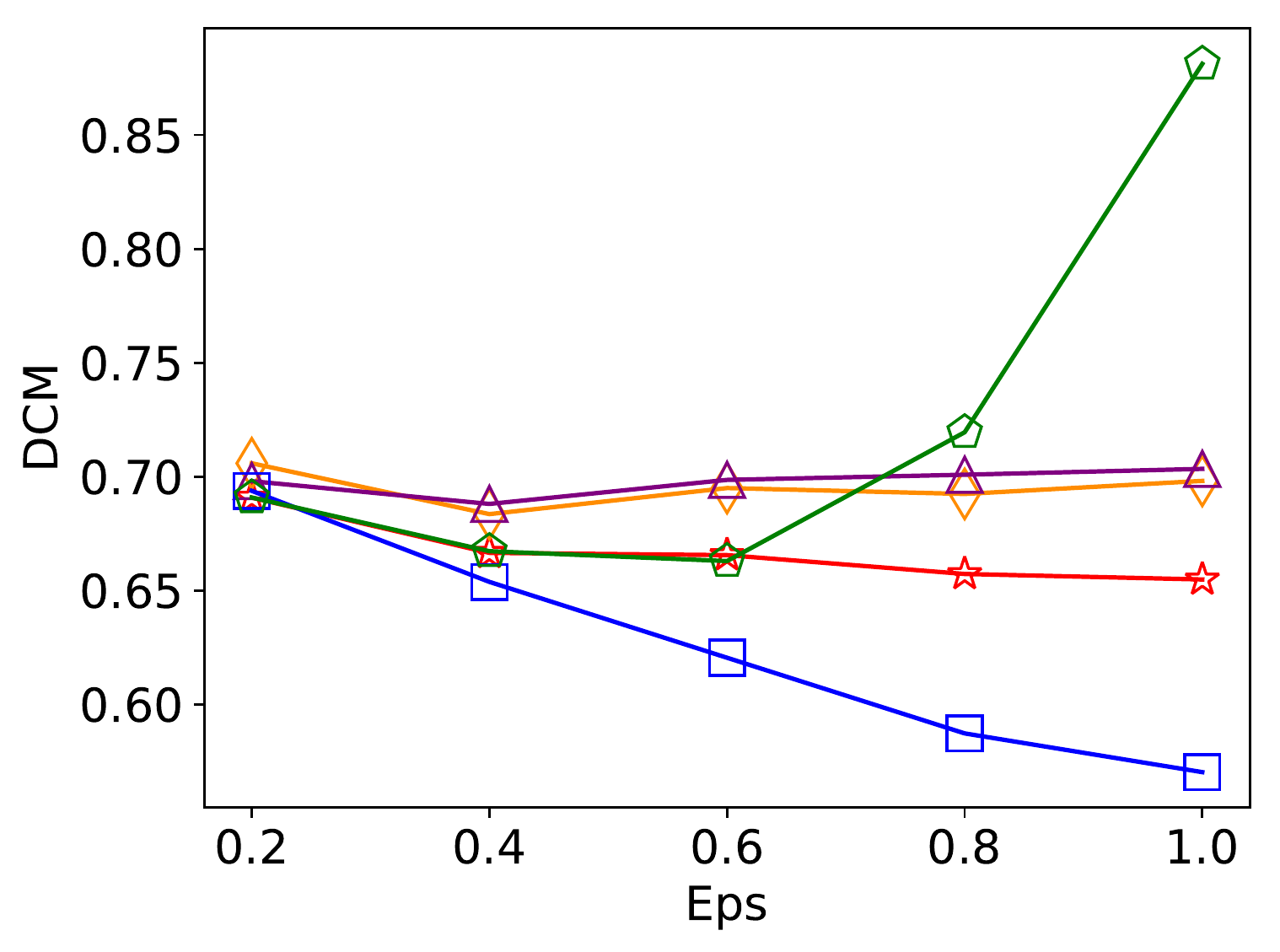}
	}
	\caption{Comparison between R-HT with different $r_{\rm loc}$ and G-GR\_hybrid on Gowalla}
	\label{fig:hybrid}
\end{figure*}
\subsection{Evaluation of LGR-Based Heuristics }

Experimental results illustrate that $LGR$ strategy improves the schemes as is mentioned in Section \ref{sec:LGR}. However, we find that the local radius $r_{\rm loc}$ computed by Algorithm \ref{alg:R} is sometimes too large. Indeed, in the $GR$ construction stage, the proportion that the whole $LGR$ cells are still not enough for reaching $EU$, for adopting $1.0\, r_{\rm loc} (0.0\%)$, $0.9\, r_{\rm loc} (2.1\%)$ and $0.88\, r_{\rm loc} (4.3\%)$ is very low. We perform the R-HT schemes with 0.9$r_{\rm loc}$ and 0.88$r_{\rm loc}$, respectively, to compare with the improved hybrid scheme of the G-GR. The experimental results on Gowalla are shown in Fig. \ref{fig:hybrid}.

When local radius $r_{\rm loc}$ is reduced, the ASR of R-HT  decreases naturally. Indeed, with the reduction of $r_{\rm loc}$, there are more tasks whose ASR can not achieve $EU$. In the case of $0.88\, r_{\rm loc}$, most ASR values for Gowalla can still reach $EU$, while the adoption of the G-GR\_hybrid scheme does not improve obviously the ASR of G-GR. $0.88$R-HT has a clear advantage in WTD, $12.4\%$ smaller than G-GR\_hybrid on average. As for HOP, although the two schemes are entangled to each other, $0.88$R-HT still leads $0.3\%$ overall. On the ANW, $0.88$R-HT is a bit higher due to the higher ASR. Further, On DCM $0.88$R-HT gradually loses the advantage, which is mainly related to the utility function of G-GR\_hybrid, defined by
\begin{equation}\label{eq:U_hybrid}
U_t^s=(1-\epsilon)\times u \times (1-\alpha)+ \epsilon \times Comp \times \alpha,
\end{equation}
where $\alpha=0.3$, $u$ represents the task acceptance probability of cell $s$, and $Comp$ is the DCM of $GR$ with cell $s$ included. As $\epsilon$ increases, the DCM weight in the utility function becomes larger. When $\epsilon$ is equal to 1.0, the utility function only depends on DCM.

By comparing with non-privacy algorithm, we can observe that non-privacy has obvious advantages in WTD and HOP because it has no grids. On ANW, we can see that 0.88R-HT, G-GR and G-GR\_hybrid are close to the non-privacy scheme when $\epsilon$ equals 1.0, which indicates that the three schemes behave well on reducing ANW, and among them the ASR of 0.88R-HT is the highest at this point.

Besides, we mention that To et al. \cite{TGF17} proposed a partial cell selection based on G-GR to deal with the ASR overflow problem when adding the last cell to $GR$, which reduces system overhead to some extent. However, due to the severe challenge that the actual ASR fails to reach $EU$, the partial cell selection undoubtedly aggravates this trouble, which is demonstrated by our various experiments. Therefore, the  partial cell strategy in R-HT is not skipped in this paper.

\subsection{Test on Running Time}\label{sec:runtime}

In real life, despite the success rate of task assignment and communication cost, the time costs from task request to GR release is also very important. We divided the entire assignment process of a single task into two major stages. The first stage is the domain partitioning (stage A), which includes the partitioning and adding noise to the number of real-time workers. Due to using historical data, this stage can be divided into the following two parts: Partitioning with noisy historical data (Stage A1); Then, updating real-time data and adding noise to the count of workers in level-2 cells (Stage A2). The second stage is the $GR$ construction stage (Stage B).

We consider the following three comparisons in terms of time consuming: First, the entire process (A+B), which means the running time of the entire algorithm; Second, updating real-time count in each  cell and constructing $GR$ (A2+B) with fixed grids, which is applicable to scenarios where the real-time data changes greatly in a period and we need only to upload real-time data for updating $GR$; Third, constructing $GR$ after updating real-time data (B).

We use Python 2.7 on Windows 10 (2.4 GHZ Intel i5 CPU, 8G RAM) to run 2,000 tasks on 3 datasets, and calculate each time cost by taking the average of 10 cycles, see Table \ref{tlb:consumption}.

\begin{table}[htbp]\renewcommand{\arraystretch}{1.5}
	\centering
	\caption{Time Consuming of R-HT and G-GR}\label{tlb:consumption}
	\begin{tabular}{|c|c|c|c|c|c|c|}
		\hline
		\multirow{2}*{Stage}&\multicolumn{2}{c|}{\tabincell{c}{Yelp\\( worker count: \\363330 + 17730)}}&\multicolumn{2}{c|}{\tabincell{c}{Gowalla\\( worker count:\\133771 + 6736)}}&\multicolumn{2}{c|}{\tabincell{c}{NYTaxi\\( worker count:\\ 841080 + 27165)}}\\
		\cline{2-7}
		~&R-HT&G-GR&R-HT&G-GR&R-HT&G-GR\\
		\hline
		B&20.4 ms&0.5 ms&9.8 ms&0.5 ms&16.9 ms&0.6 ms\\
		\hline
		A2+B&21.3 ms&$\setminus$&10.1 ms&$\setminus$&17.5 ms&$\setminus$\\
		\hline
		A+B&1.60 s&62 ms&0.56 s&22 ms&2.83 s&0.1 s\\
		\hline
	\end{tabular}
\end{table}

It can be seen in the table that in stage B, R-HT takes a relatively long time, because R-HT needs to calculate $r_{\rm loc}$, cell's utility and cell's score. For Stage A2+B, compared with stage B, updating data and adding noise only takes a very small period of time (less than 1 ms), and there is no data updates in G-GR after partitioning. In Stage A+B, The time consumption in partitioning is far greater than that in $GR$ construction, mainly due to the fact that the process of historical data prediction costs much time in R-HT.

However, our scheme is very suitable for the case that workers move rapidly, as is mentioned in Section \ref{sec:model}. Fix the grids and update real-time locations. This saves much on running time (A2+B) compared with G-GR on A+B.
%If the G-GR scheme fixes also the PSD, it means the PSD relies on the data collected only at a historical time point, but not a period like R-HT.

In a word, the running time of a single task on the three datasets is within 3 seconds, which forms an approximate direct proportion with the total count of real-time data, which guarantees the timeliness and practicability of task release in real life.

\section{Conclusion}\label{sec:conclusion}
In this paper, we proposed a location protection model for worker dataset in SC based on DP, which ensures that the privacy of worker locations is not disclosed at the task assignment stage. As far as we know, we are the first to introduce historical data learning and sampling simulation into domain partitioning
and then we achieved efficient allocation of privacy budget. This significantly reduces scales of random noises and enables the real ASR to reach the expected utility threshold stably. Moreover, We introduced several optimization techniques for constructing $GR$, particularly the newly designed quality scoring function. Our experimental results on real data demonstrated that the proposed R-HT scheme reduces the system overhead, and the time cost is practical.

Currently, we analyze a single-task (single-worker only) framework for privacy protection of worker locations based on historical data learning. If there is only real-time data, data segmentation (one part for domain partitioning and the other for $GR$ construction) can achieve parallel composition of privacy budget, which will be discussed in a forthcoming paper. As
future work, we aim to extend the privacy framework for the scenario of multi-task parallel assignments in SC and explore new cell selection strategies.

\section*{Acknowledgments}\label{sec:acknowledgments}

The authors thank Professor Chi Zhang for his excellent
comments which greatly helped us to improve this paper.
%S. Zhang is partially supported by the National Natural Science Foundation of China (No. 11301002) and Anhui Provincial Natural Science Foundation (2008085MF187). Z. Chen is partially supported by the National Natural Science Foundation of China (No. 61572031). H. Zhong is partially supported by the National Natural Science Foundation of China (No. 61572001).

\bibliographystyle{IEEEtran}
\bibliography{DPRH}

% Generated by IEEEtran.bst, version: 1.13 (2008/09/30)
\begin{thebibliography}{10}
\providecommand{\url}[1]{#1}
\csname url@samestyle\endcsname
\providecommand{\newblock}{\relax}
\providecommand{\bibinfo}[2]{#2}
\providecommand{\BIBentrySTDinterwordspacing}{\spaceskip=0pt\relax}
\providecommand{\BIBentryALTinterwordstretchfactor}{4}
\providecommand{\BIBentryALTinterwordspacing}{\spaceskip=\fontdimen2\font plus
\BIBentryALTinterwordstretchfactor\fontdimen3\font minus
  \fontdimen4\font\relax}
\providecommand{\BIBforeignlanguage}[2]{{%
\expandafter\ifx\csname l@#1\endcsname\relax
\typeout{** WARNING: IEEEtran.bst: No hyphenation pattern has been}%
\typeout{** loaded for the language `#1'. Using the pattern for}%
\typeout{** the default language instead.}%
\else
\language=\csname l@#1\endcsname
\fi
#2}}
\providecommand{\BIBdecl}{\relax}
\BIBdecl

\bibitem{TS18}
H.~To and C.~Shahabi, \emph{Location Privacy in Spatial Crowdsourcing}.\hskip
  1em plus 0.5em minus 0.4em\relax Cham: Springer, 2018.

\bibitem{TZZ20}
Y.~Tong, Z.~Zhou, Y.~Zeng, L.~Chen, and C.~Shahabi, ``Spatial crowdsourcing: A
  survey,'' \emph{The VLDB Journal}, vol.~29, no.~1, pp. 217--250, 2020.

\bibitem{XXY10}
Y.~Xiao, L.~Xiong, and C.~Yuan, ``Differentially private data release through
  multidimensional partitioning,'' in \emph{Workshop on Secure Data
  Management}.\hskip 1em plus 0.5em minus 0.4em\relax Springer, 2010, pp.
  150--168.

\bibitem{QYL13}
W.~Qardaji, W.~Yang, and N.~Li, ``Differentially private grids for geospatial
  data,'' in \emph{2013 IEEE 29th International Conference on Data Engineering
  (ICDE)}, 2013, pp. 757--768.

\bibitem{TGS14}
H.~To, G.~Ghinita, and C.~Shahabi, ``A framework for protecting worker location
  privacy in spatial crowdsourcing,'' \emph{Proceedings of the VLDB Endowment},
  vol.~7, no.~10, pp. 919--930, 2014.

\bibitem{TGF17}
H.~To, G.~Ghinita, L.~Fan, and C.~Shahabi, ``Differentially private location
  protection for worker datasets in spatial crowdsourcing,'' \emph{IEEE
  Transactions on Mobile Computing}, vol.~16, no.~4, pp. 934--949, 2017.

\bibitem{WLT17}
X.~Wang, Z.~Liu, X.~Tian, X.~Gan, Y.~Guan, and X.~Wang, ``Incentivizing
  crowdsensing with location-privacy preserving,'' \emph{IEEE Transactions on
  Wireless Communications}, vol.~16, no.~10, pp. 6940--6952, 2017.

\bibitem{YXS18}
C.~Yin, J.~Xi, R.~Sun, and J.~Wang, ``Location privacy protection based on
  differential privacy strategy for big data in industrial internet of
  things,'' \emph{IEEE Transactions on Industrial Informatics}, vol.~14, no.~8,
  pp. 3628--3636, 2018.

\bibitem{ZLZ17}
T.~Zhu, G.~Li, W.~Zhou, and S.~Y. Philip, \emph{Differential Privacy and
  Applications}.\hskip 1em plus 0.5em minus 0.4em\relax Springer, 2017,
  vol.~69.

\bibitem{STT12}
R.~Shokri, G.~Theodorakopoulos, C.~Troncoso, J.~P. Hubaux, and J.~Y. Le~Boudec,
  ``Protecting location privacy: Optimal strategy against localization
  attacks,'' in \emph{Proceedings of the 2012 ACM Conference on Computer and
  Communications Security}, 2012, pp. 617--627.

\bibitem{ABC13}
M.~E. Andr{\'e}s, N.~E. Bordenabe, K.~Chatzikokolakis, and C.~Palamidessi,
  ``Geo-indistinguishability: Differential privacy for location-based
  systems,'' in \emph{Proceedings of the 2013 ACM SIGSAC Conference on Computer
  $\&$ Communications Security}, 2013, pp. 901--914.

\bibitem{LCZ17}
B.~Liu, L.~Chen, X.~Zhu, Y.~Zhang, C.~Zhang, and W.~Qiu, ``Protecting location
  privacy in spatial crowdsourcing using encrypted data,'' \emph{Advances in
  Database Technology-EDBT}, pp. 478--481, 2017.

\bibitem{XZZ17}
P.~Xiong, L.~Zhang, and T.~Zhu, ``Reward-based spatial crowdsourcing with
  differential privacy preservation,'' \emph{Enterprise Information Systems},
  vol.~11, no.~10, pp. 1500--1517, 2017.

\bibitem{TSX18}
H.~To, C.~Shahabi, and L.~Xiong, ``Privacy-preserving online task assignment in
  spatial crowdsourcing with untrusted server,'' in \emph{2018 IEEE 34th
  International Conference on Data Engineering (ICDE)}.\hskip 1em plus 0.5em
  minus 0.4em\relax IEEE, 2018, pp. 833--844.

\bibitem{WPC19}
Z.~Wang, X.~Pang, Y.~Chen, H.~Shao, Q.~Wang, L.~Wu, H.~Chen, and H.~Qi,
  ``Privacy-preserving crowd-sourced statistical data publishing with an
  untrusted server,'' \emph{IEEE Transactions on Mobile Computing}, vol.~18,
  no.~6, pp. 1356--1367, 2019.

\bibitem{WLY19}
J.~Wei, Y.~Lin, X.~Yao, and J.~Zhang, ``Differential privacy-based location
  protection in spatial crowdsourcing,'' \emph{IEEE Transactions on Services
  Computing}, 2019, in press.

\bibitem{LWZ18}
X.~Li, Y.~Wang, X.~Zhang, K.~Zhou, and C.~Li, ``A more secure spatial
  decompositions algorithm via indefeasible laplace noise in differential
  privacy,'' in \emph{International Conference on Advanced Data Mining and
  Applications}.\hskip 1em plus 0.5em minus 0.4em\relax Springer, 2018, pp.
  211--223.

\bibitem{GZF18}
Y.~Gong, C.~Zhang, Y.~Fang, and J.~Sun, ``Protecting location privacy for task
  allocation in ad hoc mobile cloud computing,'' \emph{IEEE Transactions on
  Emerging Topics in Computing}, vol.~6, no.~1, pp. 110--121, 2018.

\bibitem{XWJ17}
D.~Xu, Y.~Wang, L.~Jia, Y.~Qin, and H.~Dong, ``Real-time road traffic state
  prediction based on {ARIMA} and {Kalman} filter,'' \emph{Frontiers of
  Information Technology $\&$ Electronic Engineering}, vol.~18, no.~2, pp.
  287--302, 2017.

\bibitem{LCZ18}
R.~Liu, G.~Cong, B.~Zheng, K.~Zheng, and H.~Su, ``Location prediction in social
  networks,'' in \emph{Asia-Pacific Web (APWeb) and Web-Age Information
  Management (WAIM) Joint International Conference on Web and Big Data}.\hskip
  1em plus 0.5em minus 0.4em\relax Springer, 2018, pp. 151--165.

\bibitem{JHC18}
Y.~Jiang, W.~He, L.~Cui, and Q.~Yang, ``User location prediction in mobile
  crowdsourcing services,'' in \emph{International Conference on
  Service-Oriented Computing}.\hskip 1em plus 0.5em minus 0.4em\relax Springer,
  2018, pp. 515--523.

\bibitem{CKZ19}
Z.~Chen, X.~Kan, S.~Zhang, L.~Chen, Y.~Xu, and H.~Zhong, ``Differentially
  private aggregated mobility data publication using moving characteristics,''
  \emph{arXiv preprint arXiv:1908.03715}, 2019.

\bibitem{KS12}
L.~Kazemi and C.~Shahabi, ``Geocrowd: Enabling query answering with spatial
  crowdsourcing,'' in \emph{Proceedings of the 20th International Conference on
  Advances in Geographic Information Systems}, 2012, pp. 189--198.

\bibitem{KSC13}
L.~Kazemi, C.~Shahabi, and L.~Chen, ``Geotrucrowd: Trustworthy query answering
  with spatial crowdsourcing,'' in \emph{Proceedings of the 21st ACM Sigspatial
  International Conference on Advances in Geographic Information Systems},
  2013, pp. 314--323.

\bibitem{ZYL16}
X.~Zhang, Z.~Yang, Y.~Liu, and S.~Tang, ``On reliable task assignment for
  spatial crowdsourcing,'' \emph{IEEE Transactions on Emerging Topics in
  Computing}, vol.~7, no.~1, pp. 174--186, 2016.

\bibitem{Dwork06}
C.~Dwork, ``Differential privacy,'' in \emph{Proceedings of the 33rd
  International Conference on Automata, Languages and Programming - Volume Part
  II}, 2006, pp. 1--12.

\bibitem{DR14}
C.~Dwork and A.~Roth, ``The algorithmic foundations of differential privacy,''
  \emph{Foundations and Trends{\textregistered} in Theoretical Computer
  Science}, vol.~9, no. 3--4, pp. 211--407, 2014.

\bibitem{LLS16}
N.~Li, M.~Lyu, D.~Su, and W.~Yang, ``Differential privacy: From theory to
  practice,'' \emph{Synthesis Lectures on Information Security, Privacy, $\&$
  Trust}, vol.~8, no.~4, pp. 1--138, 2016.

\bibitem{HG10}
B.~J. Hecht and D.~Gergle, ``On the `localness` of user-generated content,'' in
  \emph{Proceedings of the 2010 ACM Conference on Computer Supported
  Cooperative Work}, 2010, pp. 229--232.

\bibitem{MG13}
M.~Musthag and D.~Ganesan, ``Labor dynamics in a mobile micro-task market,'' in
  \emph{Proceedings of the SIGCHI Conference on Human Factors in Computing
  Systems}, 2013, pp. 641--650.

\end{thebibliography}

\end{document}